\documentclass{aa}  

\usepackage{graphicx}
\usepackage{txfonts}
\usepackage{textcomp}
\newcommand{\re}{$R_e$}
\newcommand{\Lsig}{$L-\sigma$}

\newcommand{\Lsigb}{$L=L'_0\sigma^{\beta}$}
\newcommand{\Lsigbtempo}{$L=L'_{0}(t) \sigma^{\beta(t)}$}                   

\newcommand{\MRa}{$R_e$-$M_s$}
\newcommand{\Rsigma}{$R_e - \sigma$}

\newcommand{\Ie}{$I_e$}

\newcommand{\IeRe}{$I_e - R_e$}
\newcommand{\IeSig}{$I_e - \sigma$}

\newcommand{\ie}{{i.e.}}
\newcommand{\eg}{{e.g.}}

\begin{document}

   \title{Galaxies' properties in the Fundamental Plane across time}


   \author{M. D'Onofrio
          \inst{1}
          \and
          C. Chiosi\inst{1}
          }

   \institute{Department of Physics and Astronomy, University of Padova,
              Vicolo Osservatorio 3, I-35122 Padova\\
              \email{mauro.donofrio@unipd.it};
             \email{cesare.chiosi@unipd.it}
             }

   \date{Received February 5, 2024; accepted ...}

 
  \abstract
   {Using the Illustris-1 and IllustrisTNG-100 simulations we investigate the properties of the Fundamental Plane (FP), that is the correlation between the effective radius \re, the effective surface intensity \Ie\ and the central stellar velocity dispersion $\sigma$ of galaxies, at different cosmic epochs.}
   {Our aim is to study the properties of galaxies in the FP and its projections across time, adopting samples covering different intervals of mass. We would like to demonstrate that the position of a galaxy in the FP space strongly depends on its degree of evolution, that might be represented by the $\beta$ and $ L'_0$ parameters entering the \Lsigbtempo\ law.}
   {Starting from the comparison of the basic relations among the structural parameters of artificial and real galaxies at low redshift, we obtain the fit of the FP and its coefficients at different cosmic epochs for samples of different mass limits.  Then, we analyze the dependence of the galaxy position in the FP space as a function of the $\beta$ parameter and the star formation rate (SFR).}
   {We find that: 1) the coefficients of the FP change with the mass range of the galaxy sample; 2) the low luminous and less massive galaxies do not share the same FP of the bright massive galaxies; 3) the scatter around the fitted FP is quite small at any epoch and increases when the mass interval increases; 4) the distribution of galaxies in the FP space strongly depends on the $\beta$ values (\ie\ on the degree of virialization and the star formation rate).}
   {The FP is a complex surface that is well approximated by a plane only when galaxies share similar masses and condition of virialization.}

   \maketitle
%

\section{Introduction}

The Fundamental Plane (FP), \ie\ the mutual correlation between the effective radius \re, the effective surface intensity \Ie\ and the central stellar velocity dispersion $\sigma$ of galaxies ($\log R_e = a \log\sigma + b \log I_e + c$), has been recognized long ago as an important tool for understanding the evolution of galaxies
\cite[see \eg,][]{Dressleretal1987,DjorgovskiDavis1987}.
In particular, when the FP was studied using the massive early-type galaxies (ETGs), it was recognized as a good distance indicator \citep[see \eg,][]{Dressler1987}, and as a useful tool for testing the expansion of the Universe \citep[see \eg,][]{Pahreetal1996}, for mapping the velocity fields of galaxies \citep[see \eg,][]{DresslerFaber1990,Courteauetal1993}, and for measuring the mass-to-light ratio variations across time \cite[see \eg,][]{PrugnielSimien1996,vanDokkumFranx1996,Busarelloetal1998,Franxetal2008}. In general the FP and its projections {($I_e$ vs $R_e$, $I_e$ vs $\sigma$, $R_e$ vs $\sigma$) } are diagnostic tools of galaxy evolution.

The main characteristics of the FP of nearby galaxies, essentially all ETGs of large masses {($ M_s \ge 10^{9} - 10^{10} M_\odot$)}, are the tilt of the best-fitted plane with respect to the prediction of the virial theorem  (VT) and its small scatter ($\approx0.05$ dex in the V-band).
The origin of the tilt has been discussed in several studies
\citep[see among many others,][]{Faberetal1987,Ciotti1991,Jorgensenetal1996,Cappellarietal2006,Donofrioetal2006,Boltonetal2007}, invoking different physical mechanisms: i) the systematic change of the stellar mass-to-light ratio ($M_s/L$) \citep[see \eg,][]{Faberetal1987,vanDokkumFranx1996,Cappellarietal2006,vanDokkumvanderMarel2007, Holdenetal2010,deugenio2021,deGraafetal2021}; ii) the structural and dynamical non-homology of ETGs \citep[see \eg,][]{PrugnielSimien1997,Busarelloetal1998,Trujilloetal2004,Donofrioetal2008,Oldhametal2017}; iii) the dark matter content and distribution \citep[see \eg,][]{Ciottietal1996,Borrielloetal2003,Tortoraetal2009, Taranuetal2015,deGraafetal2021,deGraaff2023};  iv) the star formation history (SFH) and initial mass function (IMF) \citep[see \eg,][]{RenziniCiotti1993,Chiosietal1998,Chiosi_Carraro_2002,Allansonetal2009}; v) the effects of environment \citep[see \eg,][]{Luceyetal1991,deCarvalhoDjorgovski1992,Bernardietal2003, Donofrioetal2008, LaBarberaetal2010, Ibarra-MedelLopez-Cruz2011, Samiretal2016}; vi) the effects of dissipation-less mergers \citep{Nipotietal2003}; vii) the gas dissipation \citep{robertsonetal06}; viii) the non regular sequence of mergers with progressively decreasing mass ratios \citep{Novak2008}; ix) the multiple dry mergers of spiral galaxies \citep{Taranuetal2015}. Many authors have confirmed that the observed tilt disappears when the mass FP is considered, \ie\ when the parameters used for the fit are marginally dependent on luminosity \citep[see e.g.][for more infos]{Bezanson2013,Zahidetal2016}.

Similarly, the small intrinsic scatter of the plane has never found a clear and definitive explanation. The invoked possible effects at play are: 1) the variation of the formation epoch; 2) the dark matter content; 3) the metallicity/age trends; 4) the variations of the mass-to-light ratio $M/L$; 5) the mixing of morphological types \citep[see e.g.,][]{Faberetal1987,Gregg1992,GuzmanLuceyBower1993,Forbesetal1998,Bernardietal2003,Redaetal2005,Cappellarietal2006,Boltonetal2008,Augeretal2010,Magoulasetal2012,Bernardietal2020,deugenio2021}. 

Previous studies have primarily focused on the properties of the FP derived from  massive ETGs at low redshifts. These studies often treated the analysis of the galaxy distribution in the FP-space, considering the projections of the plane \IeRe, \IeSig, and \Rsigma\ as independent topics for which different explanations were often advanced. Within these FP projections, distinct structural patterns emerge, including areas with pronounced clusters of objects and significant scatter, as well as regions devoid of galaxies (referred to as the 'Zone of Exclusions,' or ZOE), and regions displaying non-linear distribution patterns.

Such a wealth of information has never been used to infer a unitary view of the paths followed by galaxies in the various projection planes of the FP-space during  their evolution\footnote{Therein after we call FP-space the space of parameters characterizing a galaxy in which the FP is defined.}. Consequently, a comprehensive and coherent explanation  accounting for the interdependence of structural scaling relations, the peculiar shapes of the observational galaxy distributions in the different planes, and the connections between the various FP projections has remained elusive. This gap in understanding has made it difficult to fully grasp the causes of the tilt and scatter of the FP.

The use of the FP-space as cosmological tool for understanding galaxy evolution requires a global view of the properties of the FP and its projections at different cosmic epochs.
Unfortunately, the exploration of the FP at high redshift and for low-mass galaxies has been notably limited. This limitation arises from the considerable effort required by the  observations of faint, low-mass galaxies at high-redshift. These observations require long lasting campaigns and expensive instrumentation. Existing observational surveys at high redshift typically offer sparse data, primarily focused on the largest galaxies. Nevertheless, some empirical evidence has emerged, suggesting the possible  variation of the tilt of the FP with the redshift \citep[e.g., see the studies by][]{DiSeregoetal2005, Beifiorietal2017, Lindsayetal2017, deGraaffetal2021}, as well as deviations of faint galaxies from the FP of their more massive counterparts \citep{Heldetal1997, Bettonietal2016}.

Today, thanks to the large efforts in producing reliable cosmological simulations, it is possible to guess the properties of galaxies at different times and follow their evolutionary paths. In this context, the FP and its associated projections become valuable diagnostic diagrams, offering crucial insights into a galaxy's evolutionary state and history.

By exploiting the database of model galaxies of Illustris-1 and IllustrisTNG-100, we aim at infering some useful indication about galaxies evolution looking at the FP and its projections at different cosmic epochs. These simulations offer the best investigation tool currently available to us for a successful comparison between theory and observations, despite the fact that some problems still affect the analysis.

The first version of cosmological simulations named Illustris-1 appeared on 2014 \citep[see e.g.,][]{Vogelsberger_2014b,Genel_etal_2014,Nelsonetal2015}. Several works have later shown that there are some problems unsolved by this simulation: it yields an unrealistic population of ETGs with no correct colours, it lacks morphological information, the sizes of the less massive galaxies are too large, and the star formation rates are not always comparable with observations \citep[see \eg,][]{Snyderetal2015, Bottrelletal2017,Nelsonetal2018, Rodriguez-Gomezetal2019,Huertas-Companyetal2019,Donofrio_Chiosi_2023a}. 
Illustris-1 does not seem to produce a realistic red sequence of galaxies due to insufficient quenching of the star formation with too few red galaxies \citep{Snyderetal2015, Bottrelletal2017,Nelsonetal2018,Rodriguez-Gomezetal2019}, the number of red galaxies is in tension with respect to the observed population of ETGs and, for what concern the internal structure of galaxies, the measured S\'ersic index, the axis ratio and the radii, are in marginal agreement with observations \citep{Bottrelletal2017}.

Some years later, in 2018, IllustrisTNG \citep{Springeletal2018,Nelsonetal2018,Pillepichetal2018b} was released, a new version of the simulation that seems to produce much better results \citep{Nelsonetal2018, Rodriguez-Gomezetal2019},  in better agreement with observations. In IllustrisTNG galaxies have much better internal structural parameters \citep{Rodriguez-Gomezetal2019}, better colors and radii in closer agreement with available data at low redshift.

In our previous works we convincingly demonstrated that these simulations, although not perfect, do reproduce important features observed in the projections of the FP, and even the tilt of the FP and the small scatter around it \citep{Donofrio_Chiosi_2022,Donofrio_Chiosi_2023a}. 

In the present study, using both libraries of galaxy models, we systematically addressed the FP and its projections at different cosmic times. The study of the FP up to z=2 was addressed by \cite{Luetal2020} using the TNG simulations. {We will compare our results with theirs}.

{However, it is worth clarifying that the main focus of the present study is not to determine the most correct fit of the FP across time, but to investigate the distribution on the FP's at different redshifts of galaxies with different $I_e$, $R_e$, $L$, $M_s$, $\sigma$, and associated parameters  $\beta$ and  $L'_0$ defined by \citet{Donofrioetal2017}. Since in each galaxy, $I_e$, $R_e$, $L$, $M_s$, and $\sigma$ vary with time (redshift), and so do $\beta$ and  $L'_0$.  }

{The ultimate goal is to show that it is possible to infer some useful indications on the degree of evolution of a galaxy by looking at the observed position of it in the FP-space. The goal can be reached thanks to the new method developed by \citet{Donofrioetal2017}. In brief, } they were inspired by the classical Faber \& Jackson relation (FJ; \cite{FaberJackson1976}) linking the luminosity of $L$ with the stellar velocity dispersion $\sigma$ of galaxies in the nearby Universe at redshift z$\simeq$0. The novelty, however, was to consider that galaxies in the course of their evolution first may change both luminosity and  stellar velocity dispersion  for many reasons (star formation, mass acquisition by mergers or stripping events, natural aging of their stellar populations, and so forth), and second for most of time they are in mechanical equilibrium, that is they obey the VT. Based on these grounds, 
\citet{Donofrioetal2017} first generalized the luminosity-velocity dispersion relation by explicitly including the temporal dependence 

\begin{equation}
 L(t)  = L'_0(t) \sigma(t)^{\beta(t)}.
 \label{eq1}
\end{equation}
where $t$ is the time, $\sigma$ the velocity dispersion, and the proportionality coefficient $L'_0$ and the exponent $\beta$ functions of time.  Second, they coupled it to the VT, which is a function of $M_s$, $R_e$ and $\sigma$,  to obtain a system of two equations in the unknowns $\log L'_0$ and $\beta$ whose coefficients are function of the variables characterizing a galaxy ($M_s$, $R_e$, $L$, $\sigma$, and $I_e$). This system of equations is discussed in detail {in Appendix \ref{Appendix_A} below.}

The new empirical relation (\ref{eq1}), although formally equivalent to the FJ relation for ETGs, has a profoundly different physical meaning: $\beta$ and $L'_0$ are time-dependent parameters that can vary considerably from galaxy to galaxy, according to the mass assembly history and the evolution of the stellar content of each object. $L'_0$ is the most important variable parameter of this equation that empirically mirrors the effects of the evolution while $\beta$ gives important information on the paths followed by each galaxy in the FP-space in the course of time.
These parameters are found to be good indicators of a galaxy's historical mass accretion, star formation, and evolutionary processes, offering an immediate insight into its current stage of evolution. Indeed the $\beta$ parameter determines the direction of motion of a galaxy in the FP space.
Adopting this new perspective, it was possible to simultaneously explain the tilt of the FP and the observed distributions of galaxies in the FP projections and, at the same time, to understand the real nature of the FJ relation. 

In a series of papers on this subject, we called attention on some of the advantages offered by the joint use of the VT and the \Lsigbtempo\ law \citep{Donofrioetal2017,Donofrioetal2019,Donofrioetal2020,DonofrioChiosi2021,Donofrio_Chiosi_2022,Donofrio_Chiosi_2023a,Donofrio_Chiosi_2023b}. Accepting the idea of a variable $\beta$ parameter, taking either positive or negative values, yields an explanation of the movement of galaxies in the FP-space. This approach allows us to simultaneously account for: i) the tilt of the FP, ii) the existence of the ZoE,  and iii) the flip of galaxies in the FP projections. 

In the present study, taking advantage of what we learned from  galaxies at redshfit z$\simeq 0$, we extended the same analysis of the FP and its projections to galaxies at high redshift. Adopting the data of Illustris-1 \citep{Vogelsberger_2014a,Vogelsberger_2014b} and IllustrisTNG-100 \citep{Springeletal2018,Nelsonetal2018,Pillepichetal2018a} from z=0 up to z=4, we attempted to reconstruct the history of galaxies across time, shading some light on the main mechanisms at work during the evolution.

The paper is organized as follows: In Sec. \ref{sec:2} we briefly describe the samples of galaxies (both real and simulated) used in our work. In Sec. \ref{sec:3} we present the FP-space at z=0 and we discuss the results for real and simulated galaxies. In Sec. \ref{sec:4} we analyze the fits of the FP at high redshift for samples including galaxies of different masses. Here we compare our results with those obtained by \cite{Luetal2020}. In Sec. \ref{sec:5} we show how the FP-space changes when galaxies with different $\beta$ values are considered. In Sec. \ref{sec:6} we discuss the problem of the FP scatter and its relation with the $\beta$ parameter as a thermometer of the virialization condition. {In Sec. \ref{sec:7} we look at the FP-space distribution of the galaxies as a function of their star formation rate (SFR). In Sec. \ref{sec:8} we address the problem of the determination of the errors that should be associated to the $\beta$ parameter. In Sec. \ref{sec:9} we present our conclusions.  Finally, in Appendix \ref{Appendix_A} we summarize the key equations of the new theory of the FP and scale relations,  while in Appendix \ref{Appendix_B} we present all the calculations relative to the errors discussed in Sec. \ref{sec:8}. }
  
For the sake of internal consistency with the previous studies of this series, in our calculations with the Illustris-1 database we adopted the same values of the $\Lambda$-CDM cosmology used by  \citep{Vogelsberger_2014a,Vogelsberger_2014b}:
$\Omega_m$ = 0.2726, $\Omega_{\Lambda}$= 0.7274, $\Omega_b$ = 0.0456, $\sigma_8$ = 0.809, $n_s$ = 0.963, $H_0 = 70.4\,km\, s^{-1}\, Mpc^{-1}$. Slightly different cosmological  parameters are used by for the IllustrisTNG-100 simulations: $\Omega_m$ = 0.3089, $\Omega_{\Lambda}$= 0.6911, $\Omega_b$ = 0.0486, $\sigma_8$ = 0.816, $n_s$ = 0.967, $H_0 = 67.74\,km\, s^{-1}\, Mpc^{-1}$ \citep{Springeletal2018,Nelsonetal2018,Pillepichetal2018a}. Since the systematic differences  in $M_s$, $R_e$, $L$, $I_e$, and $\sigma$ are either small or nearly irrelevant to the aims of this study, no re-scaling of the data has been applied. 

\section{Observational data and model galaxies}
\label{sec:2}

\textsf{\bf{ Observational Data}}. The observational data for the real galaxies are the same adopted in our previous works on this subject \citep[see,][]{Donofrio_Chiosi_2022,Donofrio_Chiosi_2023a}. The data at redshift z$\sim 0$ have been extracted from the WINGS and Omega-WINGS databases
\citep{Fasanoetal2006,Varela2009,Cava2009,Valentinuzzi2009,Moretti2014,Donofrio2014,Gullieuszik2015,Morettietal2017,Cariddietal2018,Bivianoetal2017}. This sample contains only ETGs, generally with stellar mass $M_s > 10^9 \, M_\odot$.

The combination of the WINGS spectroscopic and photometric samples used here contains $\sim 1690$ galaxies with measured \re, \Ie\ and $\sigma$. A sub-sample of this, made of 270 galaxies, containing the stellar mass, the star formation rates measured by \citet{Fritzetal2007} and the morphological types given by MORPHOT \citep{Fasanoetal2012} is also used for some more detailed analyses when necessary. The ETGs sample used here takes also advantage of the data for 24 dwarf galaxies studied by \citet{Bettonietal2016} including objects of lower masses 
{($M_s \sim 10^7 - 10^9 \, M_\odot$)}. We explicitly mention the use of this dwarf sample in the work.

   \begin{figure*}
   \centering
   \includegraphics[scale=0.9]{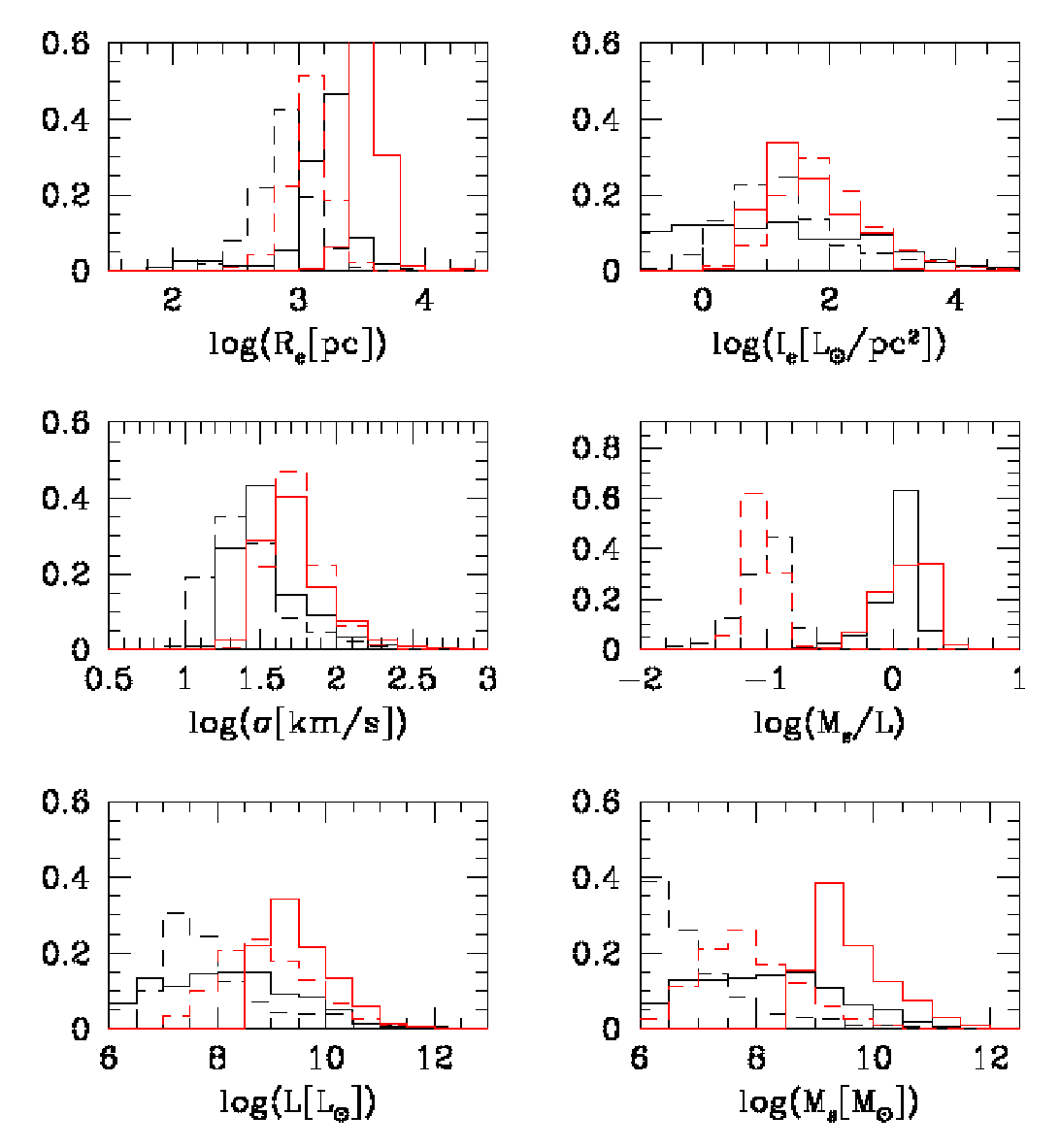}
   \caption{Comparison between the data of Illustris-1 and IllustrisTNG-100. The black lines mark the TNG data, while the red ones the Illustris-1. The histograms are normalized to the total number of galaxies. The dashed lines refer to z=4, while the solid line to z=0.
   }
              \label{fig:1}
    \end{figure*}
    
The maximum error on the measured parameters is $\simeq 20\%$. These are not shown in our plots, because they are much lower than the observed range of variation of the structural parameters in the FP-space. The small size of the errors does not  affect the whole distribution of galaxies. The errors are not considered in our fits of the FP, because the range spanned by the coefficients of the fitted planes, when different samples are used, is greater than the uncertainties on the fit induced by them.

\textsf{\bf{ Theoretical Models}}. {We adopt the galaxy models of the Illustris-1 and IllustrisTNG-100 spanning ample ranges of masses and redshifts. Some preliminary comments on the two are databases are necessary here. }

The sample of model galaxies in Illustris-1 contains $\sim 2400$ objects of all morphological types with masses at z=0 larger than $10^9\, M_\odot$. We used the data in the V-band photometry, the mass and half-mass radii of the stellar particles (i.e., integrated stellar populations), for which Cartesian comoving coordinates are available. This sample is well discussed in all our previous works \citep[see \eg,][]{Cariddietal2018,Donofrioetal2020}. For this sample
the values of \re\ are  calculated considering the luminosity growth curves of the galaxies that are members of clusters, following the same technique as in the case of real objects. {For the galaxies of Illustris-1 sample, {the family-tree} of each object has been reconstructed so that it is possible to follow each object along the the cosmic time (redshift). } 

From the IllustrisTNG-100 dataset we extracted four samples with 1000 objects at increasing redshift from z=0 to z=4, ordered with decreasing stellar masses. The samples have been obtained using the online Search Galaxy/Subhalo Catalog\footnote{see https://www.tng-project.org/data/.}. In this case the half-mass stellar radius is used instead of the effective radius \re. This radius is not so different from the effective radius and its use does not change  the conclusions reached here at all.  {The list of progenitors of each galaxy across the cosmic time is in progress.  }

The choice of using both Illustris-1 and IllustrisTNG-100 has the following motivations: i) we want to be consistent with our previous works on this subject; ii)  the differences in $M_s$, $R_e$, $I_e$, $L$, and $\sigma$  of Illustris-1 and IllustrisTNG do not bias significantly the results  for  the FP fit  and the values of  $\beta$ and $L'_0$ in the \Lsigb\ law \citep[see][]{Donofrio_Chiosi_2023a}; iii)  the two data samples are in some way complementary: IllustrisTNG-100 has better measurements of the half-mass radii of the less massive galaxies, while Illustris-1 is  richer in massive objects; iv) the two simulations agree in the physical parameters of the massive objects and produce very similar distribution of galaxies in the FP-space, apart from the small differences due to the lower \re\ of the dwarf galaxies in IllustrisTNG-100; v) the use of both samples gives an idea of the degree of uncertainty present in our analysis of the FP-space.

It is important to bear in mind the selection criteria applied to model galaxies when analyzing the FP-space. Specifically, it is noteworthy that the Illustris-1 sample gradually excludes low-mass objects (with masses $\leq 10^9 \, M_\odot$) as we approach redshift z=0. In contrast, the IllustrisTNG-100 sample, which is approximately half the size of the Illustris-1 sample, maintains the mass range of galaxies across all redshifts due to a consistent mass-based selection criteria. Therefore, the theoretical deficiency of low mass objects does not reflect a real  scarcity of these galaxies, but rather a consequence of our galaxy selection criteria.

{The mass resolution of the models is about $10^6 M_\odot$. Therefore galaxies in the low mass intervals (say up to $10^7  M_\odot$ or so)  are described by a handful of mass points, consequently their radius $R_e$ and velocity dispersion $\sigma$ are poorly determined. Galaxies with masses $M_s \leq 10^7  M_\odot$ are excluded from our analsysis. }

The detailed analysis of the differences between Illustris-1 and IllustrisTNG-100 has not been addressed here because it has already been presented in other studies on this subject \citep[see \eg,][]{Pillepichetal2018a,Pillepichetal2018b,Rodriguez-Gomezetal2019, Huertas-Companyetal2019}. One of the major issues of tension that may be  relevant for the present study is the radius size of the low mass galaxies ($M_s \leq 5.0\times 10^{10}$), where the IllustrisTNG-100 radii are about a factor of 2 smaller that those of Illustris-1  while above it they are nearly equal \citep{Pillepichetal2018a,Pillepichetal2018b,Rodriguez-Gomezetal2019,Huertas-Companyetal2019}. We will see later that for the aims of the present analysis, the effects of such differences do not introduce significant biases in our interpretation of the FP properties.

In any case, in order to understand the range of the structural parameters that are involved in this work, in Fig. \ref{fig:1} we present the comparison between the data of Illustris-1 (red lines) and IllustrisTNG-100 (black lines) at the redshifts z=0 (solid lines) and z=4 (dashed lines).
It is clear from the figure that the effective radii of Illustris-1 are systematically larger than those of IllustrisTNG-100. Another significant difference is found in the distribution of the total luminosity and total stellar mass. As already explained, Illustris-1 does not contain objects with mass lower than $10^9\, M_\odot$ at z=0. It follows that the distributions in mass and luminosity are different for the two samples.
{As far as the other parameters are concerned, there are also some differences between the two samples. We will see later that such differences in radii, masses,  luminosities, and velocity dispersions  do not compromise the analysis of the FP  as well as the main conclusions. Despite the small differences in the distribution of galaxies in the parameter space, the global behavior is the same for the two simulations and suggest the same conclusions about the physical effects at play in shaping the FP-space. }

Other points of weakness in the two libraries of model galaxies  are of little relevance for our analysis because: i) we do not make use of the galaxy colors; ii) we have demonstrated \citep[see,][]{Donofrio_Chiosi_2023a} that  Illustris-1 and IllustrisTNG-100 samples produce very similar distributions of the $\beta$ and $L'_0$ parameters of the \Lsigb\ law; iii) the FP-space at each redshift for the two samples is very similar, not in the sense that the FP coefficients are identical, but in the general distribution of the galaxies in this space; iv) the point mass view of the galaxies adopted here secures that our analysis is not too much affected by the problems affecting  the inner structure of the model galaxies of the two libraries.
In fact  both for Illustris-1 and IllustrisTNG-100 we did not use information on the  morphology of  galaxies. ETGs and late-type galaxies (LTGs) are mixed together. This choice originates from the observation that the FP projections of ETGs and LTGs are almost identical \citep[see,][]{Donofrio_Chiosi_2023b}. In addition, we will see that the FP defined by  massive ETGs is only  part of a more general distribution of galaxies in the FP-space. In this sense the presence of LTGs in the samples does not alter none of the conclusions previously reached for the massive ETGs.

{The last remark concerns the completeness of the samples. This is not critical for the conclusions drawn here. We will see that the fit of the FP is always performed with a large number of galaxies and that the observed differences can be clearly attributed to the characteristics of the sample under analysis.}
This is somehow independent of the level of precision reached by the model galaxies of the two  samples. In fact what we want to emphasize  is not the precision reached in the derivation of the FP, but the general behavior of galaxies in the FP-space as a function of their mass and the dependence of the observed distributions on the $\beta$ parameter.

\section{Fit of the FP at z=0}
\label{sec:3}

   \begin{figure*}
   \centering
   \includegraphics[scale=0.4]{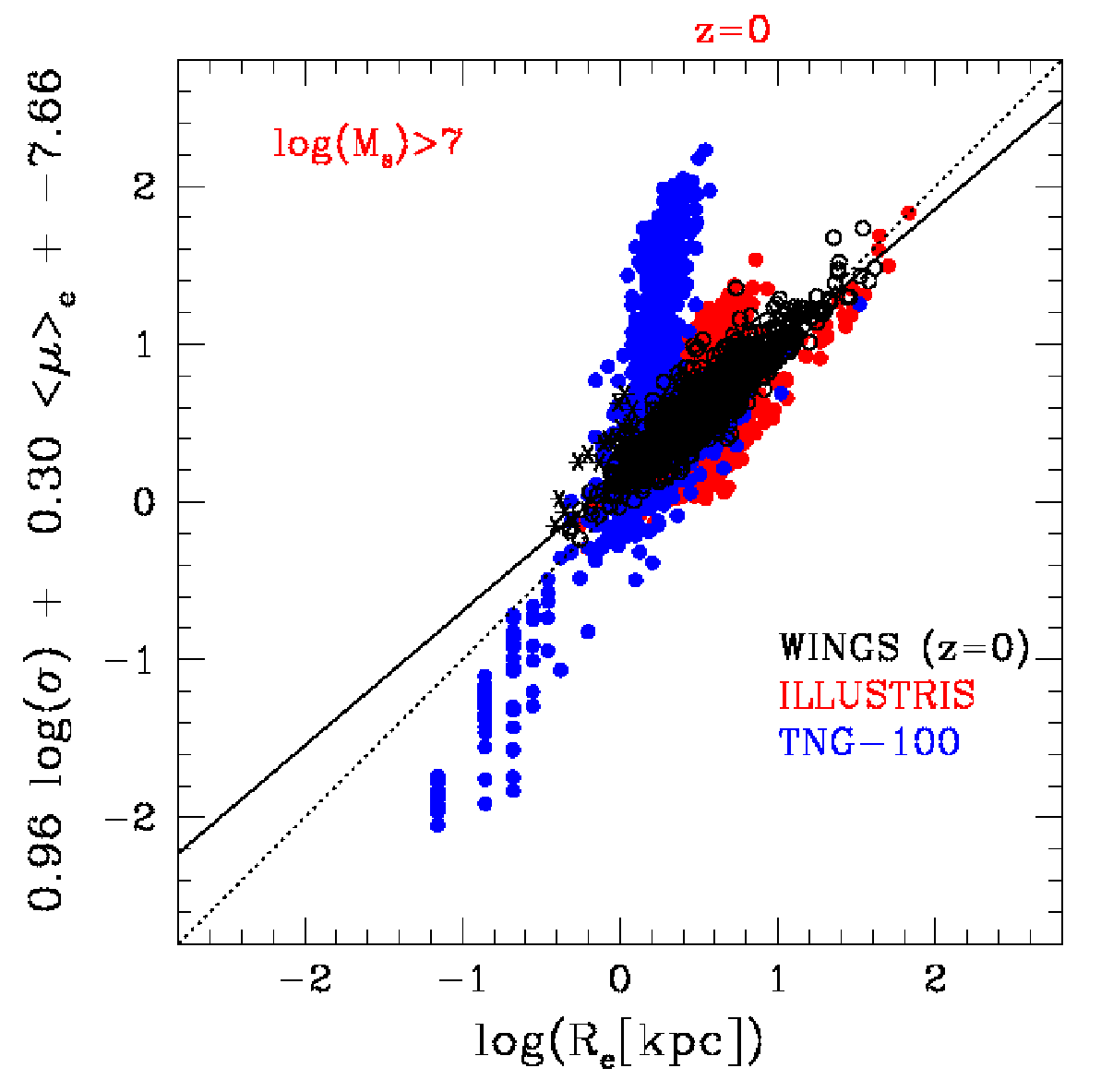}
   \includegraphics[scale=0.4]{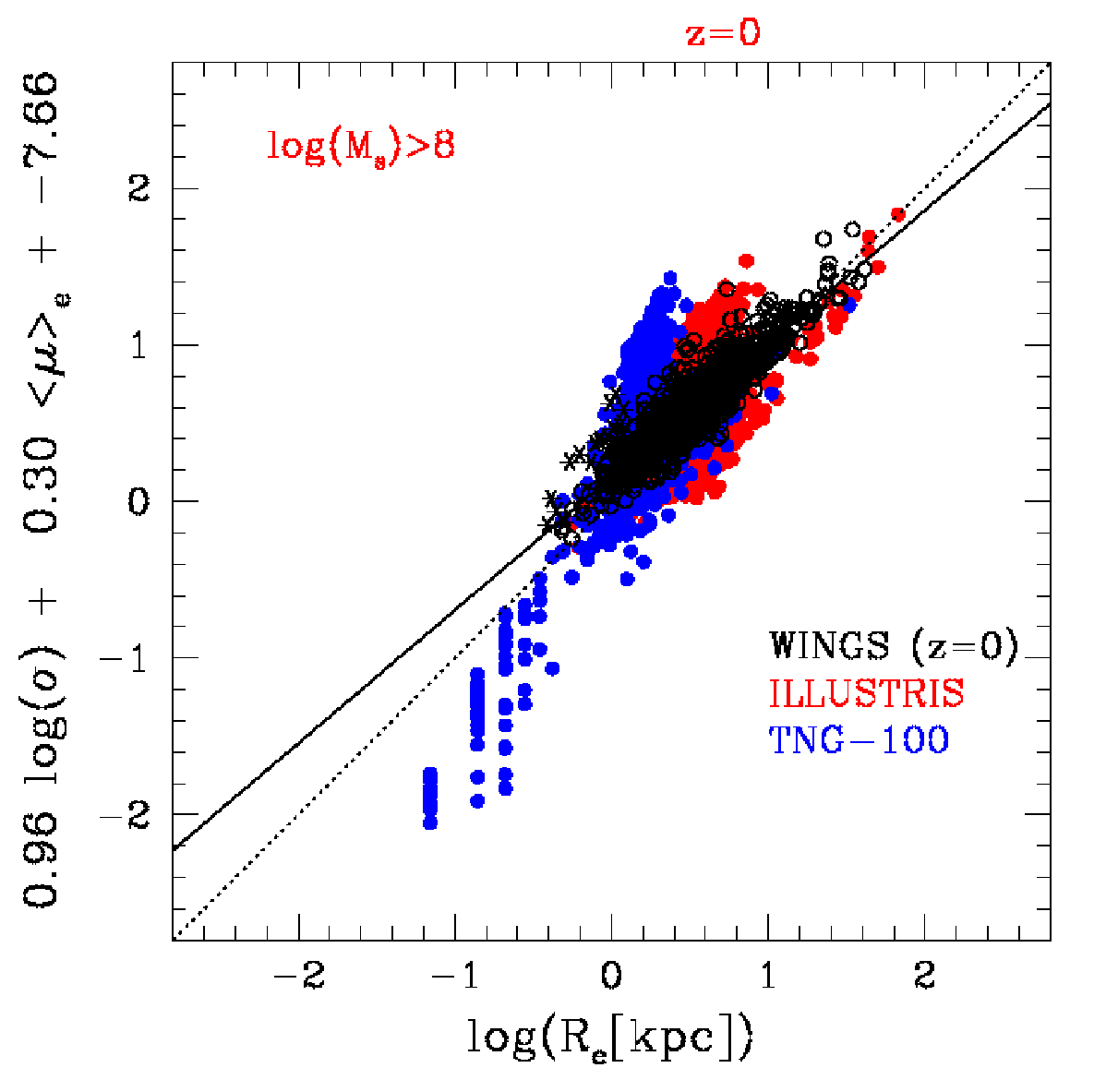}
   \includegraphics[scale=0.4]{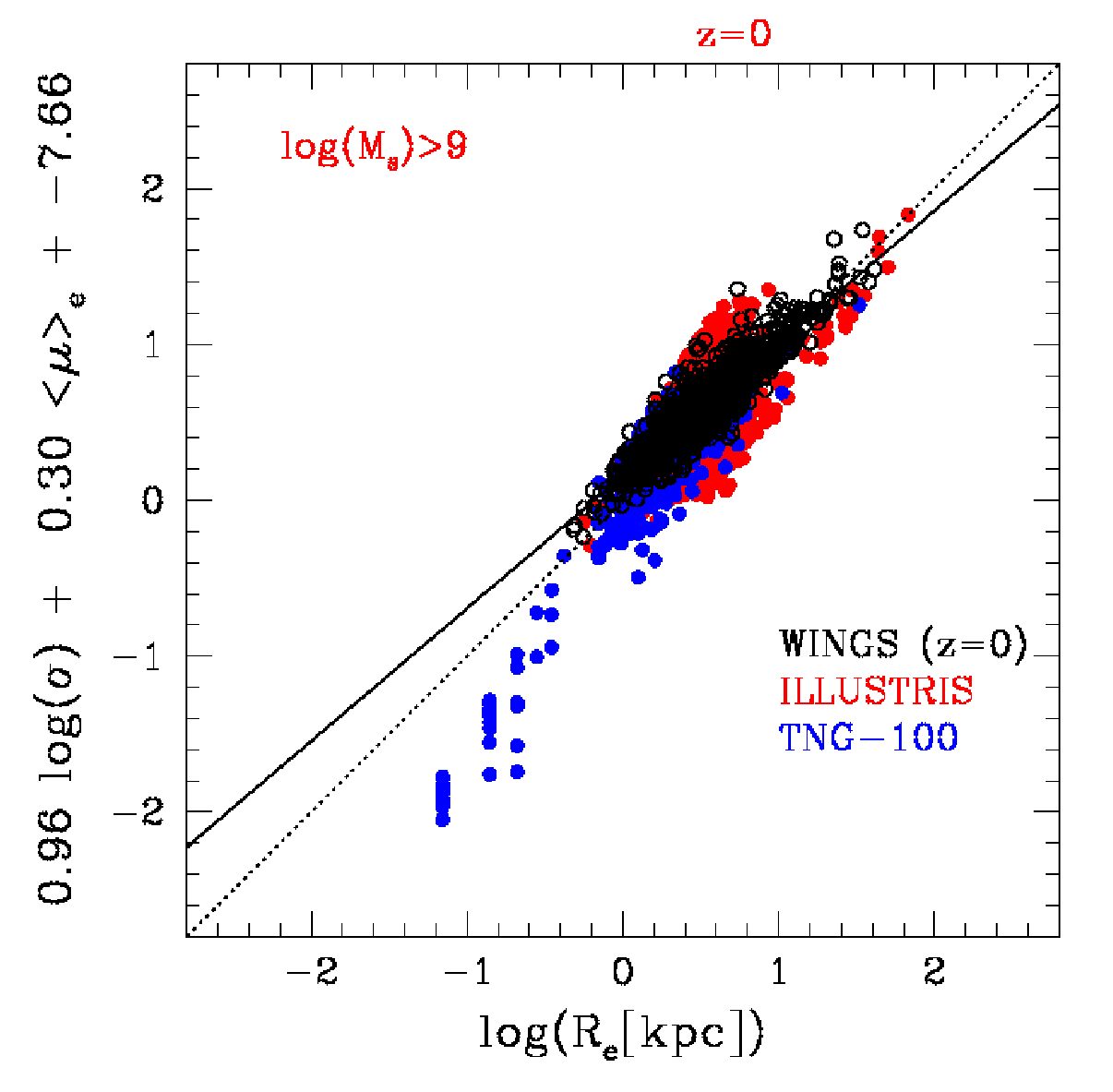}
   \includegraphics[scale=0.4]{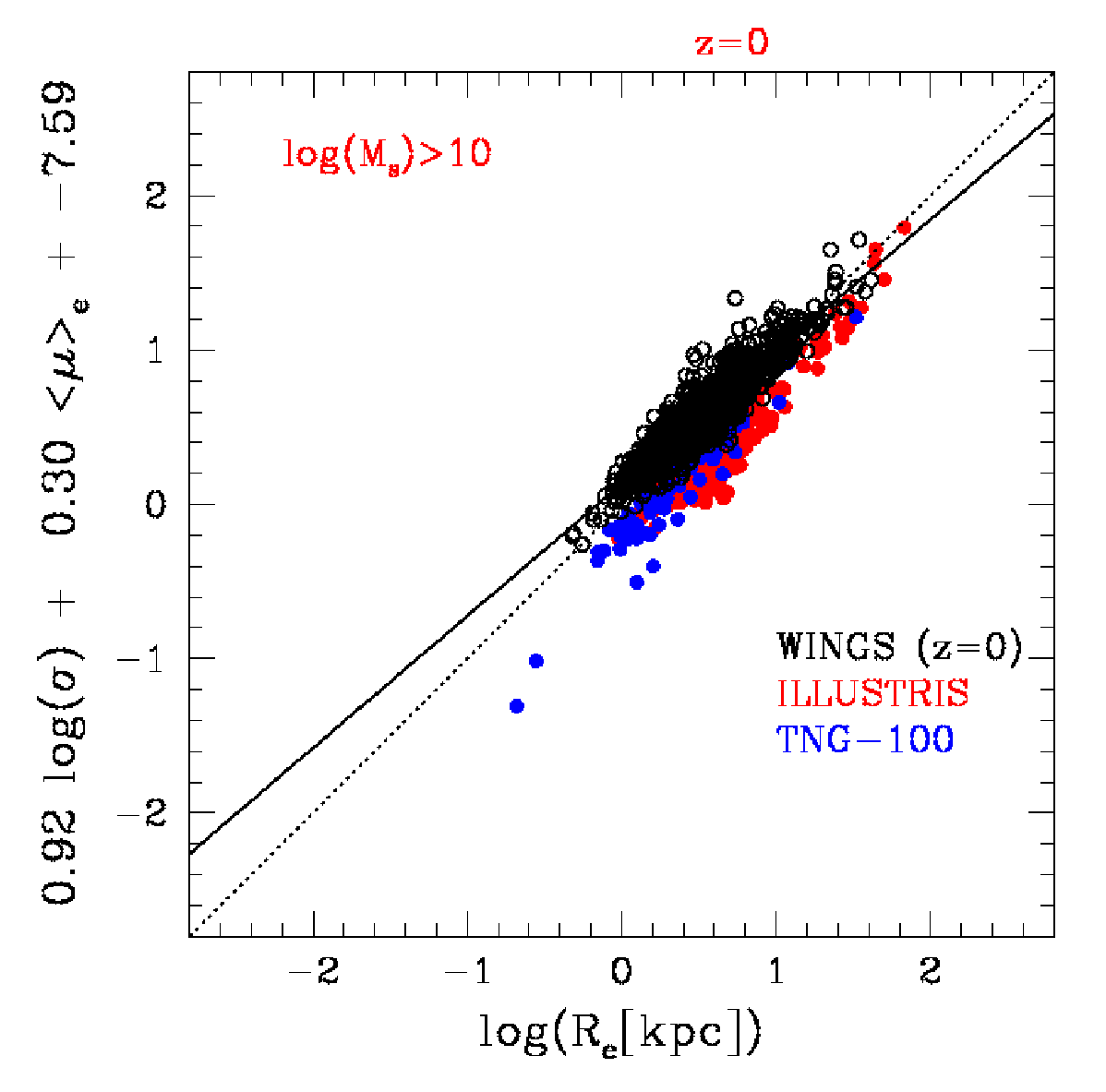}   
   \caption{The FP seen edge-on for the full WINGS sample of $\sim1600$ ETGs (black open circles), Illustris-1 (red filled circles) and TNG-100 (blue filled circles) at z=0. The four panels show the distribution of the galaxies in the FP when different range of mass are considered: $M_s \ge 10^7 \, M_\odot$ (upper left panel), $M_s   \ge 10^8\, M_\odot$ (upper right panel), $M_s \ge 10^9 \, M_\odot$ (lower left panel) and $M_s \ge 10^{10}\, M_\odot$ (lower right panel). The full black solid line is the best fit of the galaxies of the whole WINGS sample (no lower mass limit). The dashed line is the one-to-one correlation.}
    \label{fig:2}
    \end{figure*}
 
 {The first step to undertake is to set up the FP and its projections at z=0 for the galaxies of our observational sample and the Illustris-1 and IllustrisTNG-100 galaxies to compare them with other FP's in literature. }

{There are different analytical expressions of the FP, we adopt here the following one
 
 \begin{equation}
  \log R_e = a \log (\sigma) + b < \mu_e > + c
  \label{expr_fp}
 \end{equation}
 
\noindent
where   $< \mu_e >$ is the mean effective surface brightness expressed in mag/arcsec$^2$ and all other symbols have their usual meaning and are expressed in suitable units, see also appendix \ref{Appendix_A} and \citet{Donofrio_Chiosi_2023a,Donofrio_Chiosi_2023b}. 
\footnote{Usually the effective specific intensity $I_e$ is expressed in $L_\odot/pc^2$. However, limited to the FP, we express $I_e$ as $<\mu_e>$, the mean surface brightness, in mag/arcsec$^2$. }
The fit of the WINGS data (mainly massive galaxies with $\log (M_s/M_\odot) > 10$) yields the following coefficients of the FP:  $a=0.96$, $b=0.30$, and $c=-7.66$. 
All our 3D fits are based on the statistical regression of the $R$ program (the free software environment for statistical computing and graphics). }

{Figure \ref{fig:2} shows the FP at z=0 (solid black line) for the WINGS sample to which the 24  dwarf ellipticals of \citet{Bettonietal2016} are added (black dots and asterisks, respectively) and compare it with the data of Illustris-1 (red dots) and IllustrisTNG-100 (blue dots). The comparison  is made by correlating the $R_e$ of a galaxy (either model or real) with the $R_e$ we would obtain from eqn. (\ref{expr_fp}) representing the FP. The dotted line is the  one-to-one correlation (i.e. perfect coincidence between the two values of $R_e$ for each object). } 
The four panels plots the distributions for different ranges of the stellar mass. It is clear from the plots that if the coefficients of the FP  derived from the whole WINGS database  are used also for the fainter, lower mass galaxies, these latter deviate from the main trend. This occurs in particular with the IllustrisTNG-100 dataset. The plume-like features pointing toward directions above the plane are well evident and  quite similar in both samples of model galaxies (there is however a small offsets  due to the different values of \re\ in the two databases). For IllustrisTNG-100 there are also objects that deviate in the opposite direction: these are likely objects with very small \re\ and high effective intensity \Ie. The opposite for the upward plume-like features (larger \re\ and lower \Ie). This point will be much more clear when the projections of the FP are shown. These features are nearly absent in the last panel of Fig.\ref{fig:2} at the bottom right where only galaxies more massive than $ 10^{10}\, M_\odot$ are displayed.

It is also important to note that the deviation observed for the low mass model galaxies is present even in the sample of real galaxies. Here we used the small sample of faint ETGs studied by \cite{Bettonietal2016} (black asterisks). These objects starts to show the progressive shift from the FP of massive galaxies along the same direction suggested by both libraries of theoretical models, that is toward lower values of \Ie\ (upward shift). The smaller shift shown by real objects with respect to that seen for models is likely due to the small number of dwarfs (24 only) available in our sample and to selection effects: only the brightest dwarf ETGs with similar \Ie\ have been detected and studied (i.e., have measured $\sigma$).
Their smaller range of \Ie\ is at the origin of the observed deviations.

In summary, the models suggest that the galaxies with masses lower than $\sim 10^9\, M_\odot$ do not share the same FP of {the more massive ETGs}. The galaxies with low \Ie\ extend toward larger \re, while those with high \Ie, should correspond to those with short \re. This confirms previous results already obtained during the study of the FP \citep[see e.g.][as an example]{HydeBernardi2009}.

Furthermore, these plume-like features should not depend on the galaxy morphology; we cannot prove this statement because the information on morphology is missing, but we guess that even for a sample of pure ETGs the trend should be  the same. It is only the range spanned by the specific intensity \Ie\ the ultimate cause of the plume-like features. Since ETGs and LTGs span approximately the same range of \Ie\ \citep[see,]{Capacciolietal1992}, they should also exhibit the same features.

\section{The FP at high redshift}
\label{sec:4}

\citet{Donofrio_Chiosi_2023b} demonstrated that Illustris-1 and IllustrisTNG-100 are in quite good agreement with the observational data of the FP-space at z=0. The same is also true for the FP at redshift z=1, at least looking at the data of \citet{DiSeregoetal2005} at z$\sim 0.8$, and those at z=2 according to \cite{Luetal2020}.

{We need to clarify that passing from the FP at z=0, for which we have used a sample of galaxies selected according to some criteria, the same galaxies are not  traced back to high redshifts, but galaxies that are present  at high redshift are reselected according to the same criteria. In other words we do not use the family-tree becasuse it is not avalable for both galaxy samples (Illustris-1 and IllustrisTNG-100). Fortunately, this is less of a problem because we are not interested in reconstructing the FP for objects belonging to the same family-tree,  but only in discussing  the statistical properties of the FP at different redshifts. In other words, the main results of our analsysis are not signicantly affected by this type of selection. }

{Based on the two libraries of model galaxies to our disposal, we examined the FP at different redshifts, using different  mass intervals of galaxies. In addition to this, we compared the edge-on distribution of galaxies using two representations of the FP: i) at all redshifts the FP is supposed to be the one derived at z=0, that is no evolution of the FP; ii) at each redshift a new FP is derived using galaxies at that redshift. The FP changes with redshift (time). 
Also here, all 3D fits are based on the $R$ method (we have already for the case of z=0). }

   \begin{figure*}
   \centering
  { \includegraphics[scale=0.72, angle=-90]{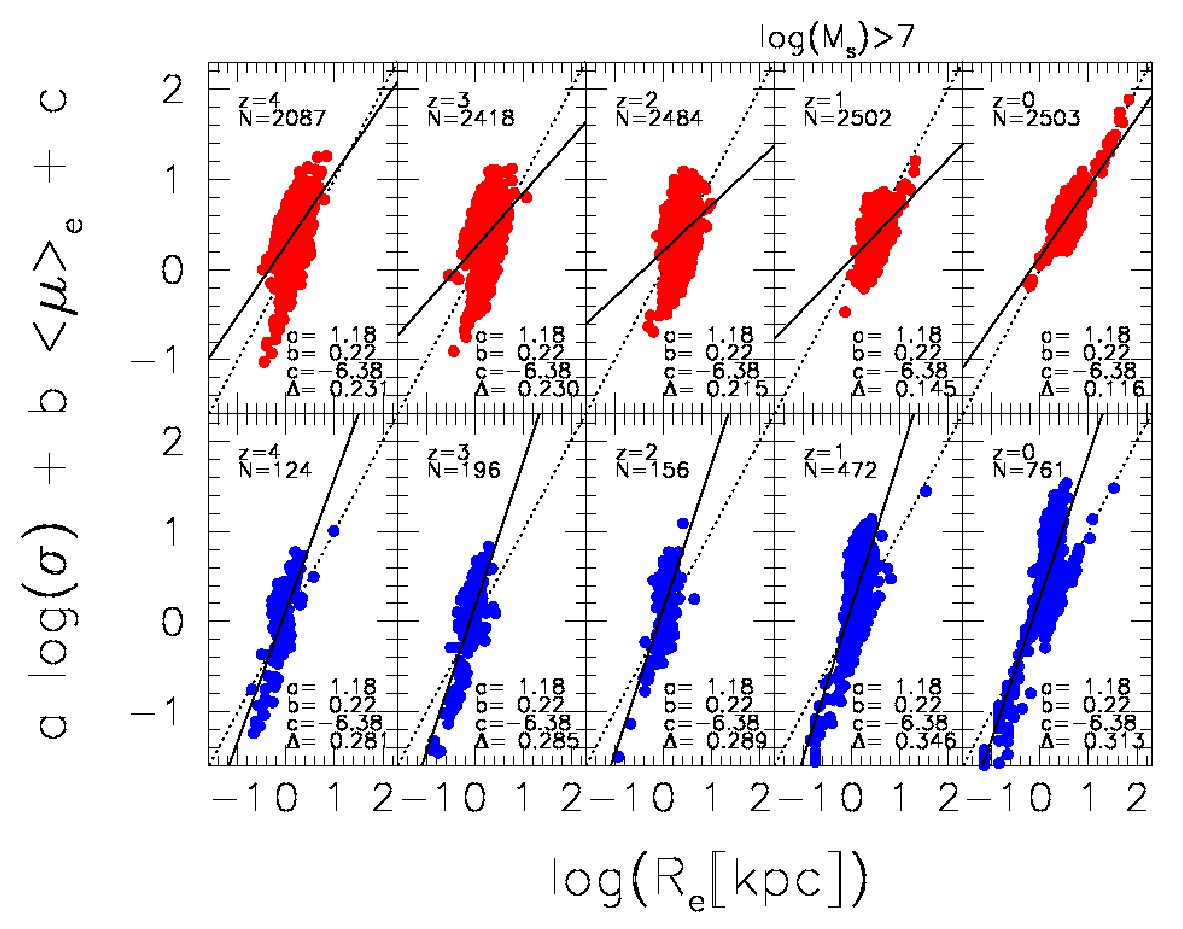}      
   \includegraphics[scale=0.58, angle=-90]{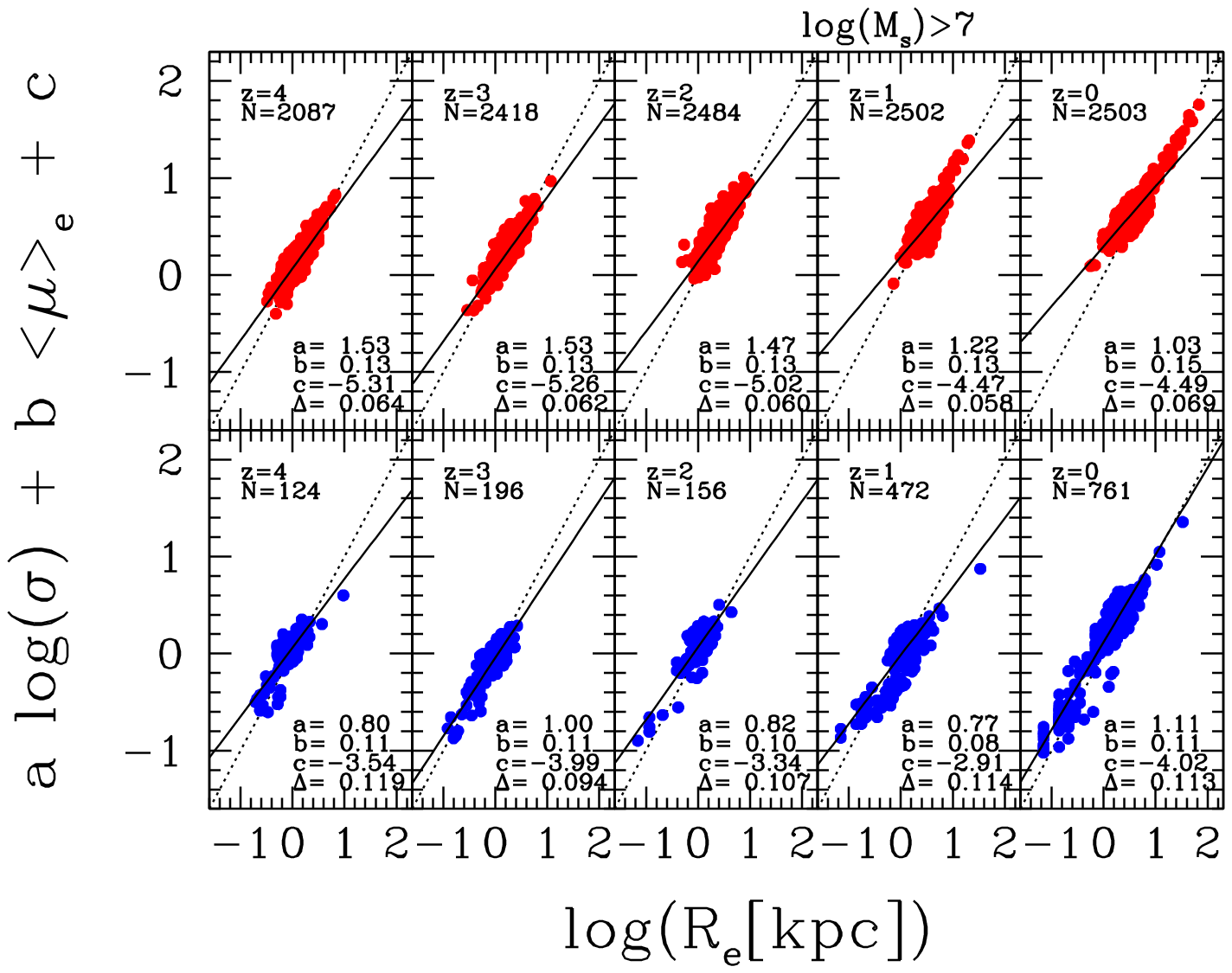}   }    
   \caption{The edge-on FP at different redshifts for the Illustris-1 and IllustrisTNG-100 datasets. 
   All panels with red dots are for Illustris-1, and all panels with blue dots are for IllustrisTNG-100.  In each panel we show the redshift z, the total number $N$ of galaxies, and the coefficients $a,b,c$ of the FP and the dispersion $\Delta$ around it.  The low  mass limit of the galaxies is $\log(M_s/M_\odot)>7$. The dotted lines mark the one-to-one relations. The black solid line (when present) is the fit of the FP. In the group of panels at the top side, the FP at z=0 is supposed to hold also at all other redshifts, while in the group of panels at the bottom side the FP varies according to the redshift. See the text for all other details. }
    \label{fig:3}
    \end{figure*}

   \begin{figure*}
   \centering
 {  \includegraphics[scale=0.62, angle=-90]{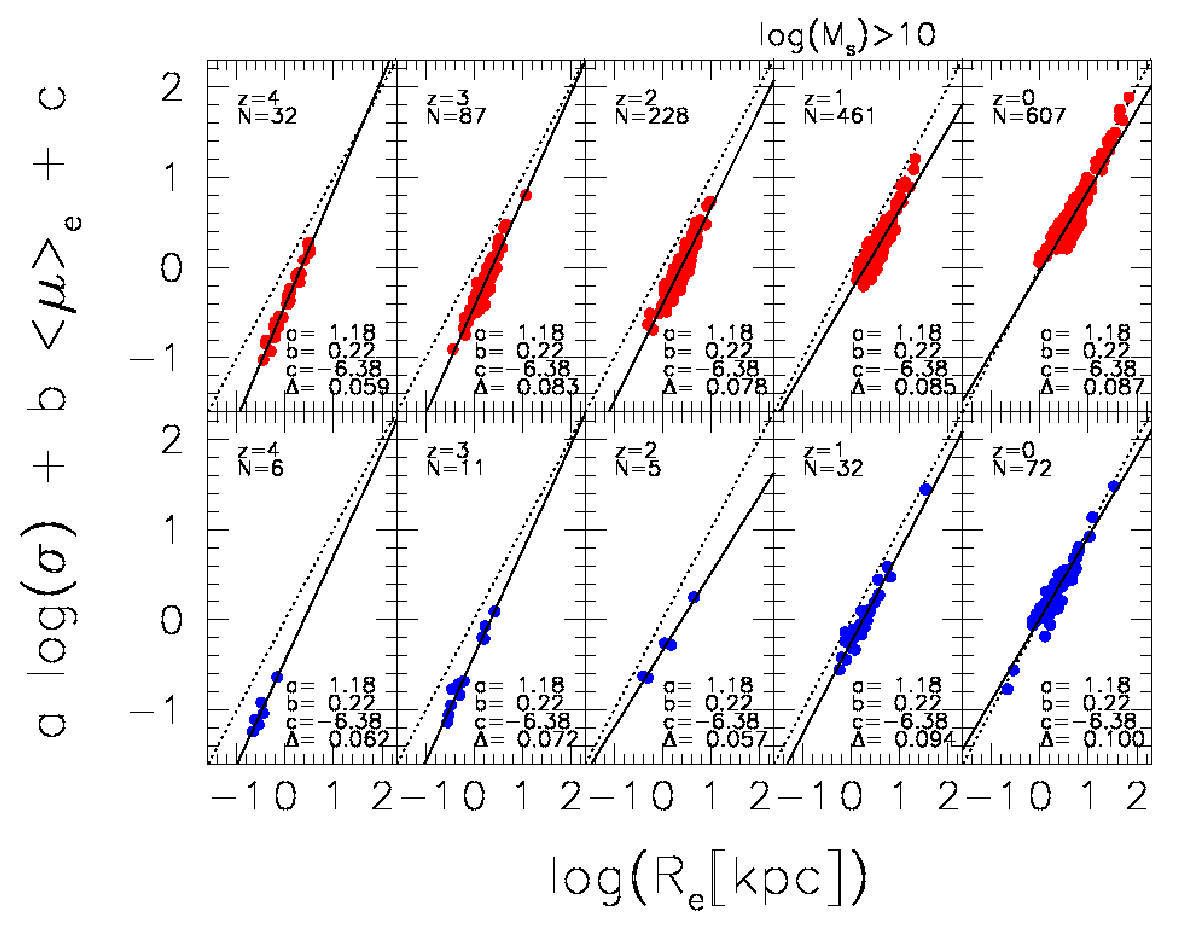}        
   \includegraphics[scale=0.52, angle=-90]{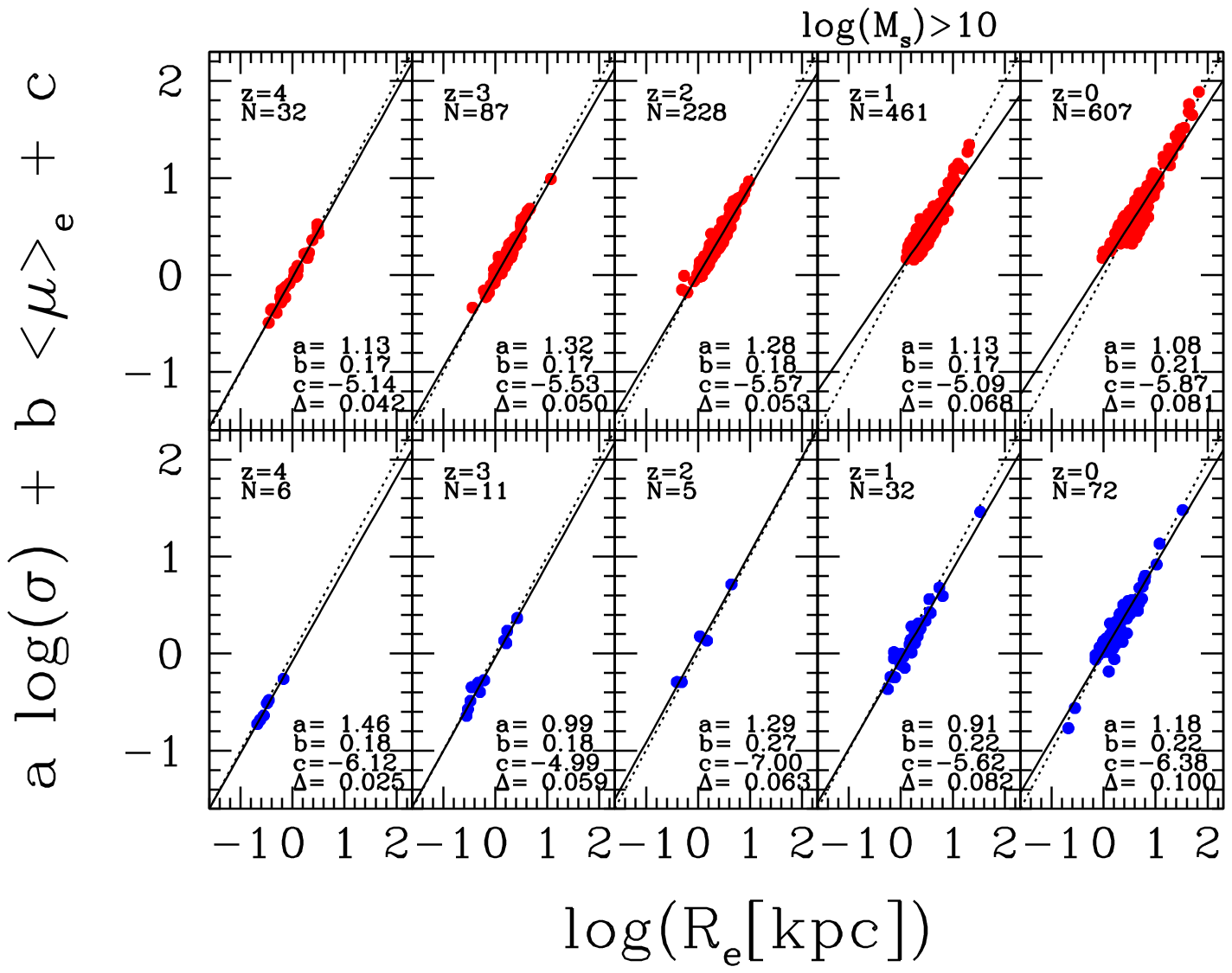}  } 
   \caption{The same as in Fig.\ref{fig:3}  but for galaxies with masses greater than $\log(M_s/ M_\odot)>10$. }
    \label{fig:4}
    \end{figure*}


{Figure \ref{fig:3} shows the edge-on view of the FP at different redshifts for all model galaxies of Illustris-1 and IllustrisTNG-100 with mass greater than $\log(M_s/M_\odot)>7$. 
More precisely, in Fig.\ref{fig:3} (and all following ones with the same layout) for each object we plot the correlation between the true radius $R_e$ (x-axis) and the $R_e$ we would obtain from the fit of the FP (y-axis). The dotted lines is the one-to-one correlation. The FP on display refers to massive galaxies ($ \log (M/M_\odot) \geq 10$ derived from the WINGs data (it is often referred to as the ``reference FP``). }

{Two groups of ten panels are displayed. In the top group, the FP is supposed not to change with redshift. The FP determined at  z=0 is supposed to hold at all redshifts (no FP evolution with time). In the bottom group the FP is let change with the redshift (FP evolves with time). In each group,  the top row  of panels shows the Illustris-1 data (red dots) while the bottom row of panels shows the IllustrisTNG-100 data (blue dots). In each panel we also display the redshift z, the total number $N$ of galaxies, the coefficients ($a$, $b$, and $c$) of the FP fit,  and the dispersion $\Delta$ of galaxies around it. In each panel, the dotted line is the one-to-one relationship, while the black solid  line  is the FP fit associated to the that panel. The redshift goes from z=4 to z=0 in steps of $-1$.  Note that the reference FP given in each panel is different for the Illustris-1 and IllustriTNG-100 datasets. }  

{A careful inspection of all panels reveals that: 
(i) When the FP is kept fixed to the z=0 values obtained with the massive galaxies ($\log(M_s)>10$), a progressive deviations from the one-to-one correlation is observed, in particular when the sample of less massive galaxies is considered. 
(ii) In the case of the Illustris-1 galaxies, the FP of massive galaxies is given by $a=1.08$, $b=0.21$, $c=-5.87$. The dispersion $\Delta$ slightly decreases with redshift passing from $\Delta = 0.08$ at z=0 to $\Delta =0.04$ at z=4. For IllustrisTNG-100, the FP is given by  $a=1.18$, $b=0.22$, $c=-6.38$, while the dispersion decreases from $\Delta = 0.1$ at  z=0 to $\Delta =0.02$ at z=4. The agreement of the two determinations of $R_e$ is not satisfactory in particular for the low number of massive IllustrisTNG-100 models at high redshift. 
(iii)  The situation improves when the FP is calculated independently for each galaxy sample at each redshift. The FP coefficients vary with redshift and the dispersion around it significantly smaller with respect to those calculated when the FP is fixed. Notably in this case the galaxies are always close to the fitted planes.
(iv) The analysis clarifies that the FP plane of the galaxy models varies with the redshift. The details depend on the  models in use and  the adopted selection criteria. In general, the agreement between the observed distribution and the FP at varying the redshift is poor if the FP holding at z=0 is supposed to hold at all redshift, while the agreement is good if the FP is left vary with redshift. }

The same analysis was repeated changing the lower mass limit for the galaxies. Now they are those with mass greater than $\log(M_s/M_\odot)>10$. The results are shown in the two groups of panels of Fig. \ref{fig:4}. All other assumptions and symbols meaning are the same of Fig. \ref{fig:3}. 

This analysis demonstrated that the coefficients of the fitted FPs vary both with the redshift and with the mass interval in use. As in the case of  Fig. \ref{fig:2}, when the FP coefficients are fixed, the less luminous and less massive galaxies deviate from the plane of the bright massive objects. On the other hand, when at each redshift the galaxy populations in place at that redshift are used to derive the FP pertinent to the epoch under consideration, a new FP is found with a fairly small scatter.
In addition we note that: i) the scatter around the plane decreases with increasing galaxy mass (it is smaller for the massive galaxies), and ii) the rotation coefficient $b$ of the plane changes by a factor of $\sim 2$ when high and low mass objects are separately analysed. The main driver for this change of the rotation coefficient is the effective surface intensity \Ie\ that varies by several order of magnitudes across the sample.

{Another point to keep in mind is that the number of galaxies of different masses varies with redshift. There are two sources of variations. One is real and it depends on the progressive increase of the galaxy mass at increasing cosmic time. The other is an artifact of the selection criteria we adopted to set up the sample of galaxies to examine and also the assumptions made for the mass distribution of the calculated models. All this  may somewhat bias the coefficients of the FP we have determined. The FP coefficients  may somewhat depend on the  size of the galaxy sample \citep[see e.g.,][]{Donofrioetal2008}. However, this effect is only marginally relevant here, because the fits for the FP coefficients are always performed using more than about 30 galaxies.}

To somehow cope with this uncertainty affecting the FP coefficients, we performed a series of fits of galaxy distributions in the FP-space by randomly extracting  a fixed number of galaxies from each data sample (each time 200 objects were extracted, and the whole fit procedure was repeated 500 times). In this way we can derive the mean values of the FP coefficients and their scatter, thus providing an estimate of the possible variations induced by the bias
\footnote{A problem arises when the number of galaxies is lower than 200. This may occur with the most massive objects. In this case, an acceptable way out is to count massive objects more than once.  Clearly, the final value of the coefficients turns out to be somewhat biased by it.}.

\begin{figure*}
   \centering
   { \includegraphics[scale=0.4, angle=0]{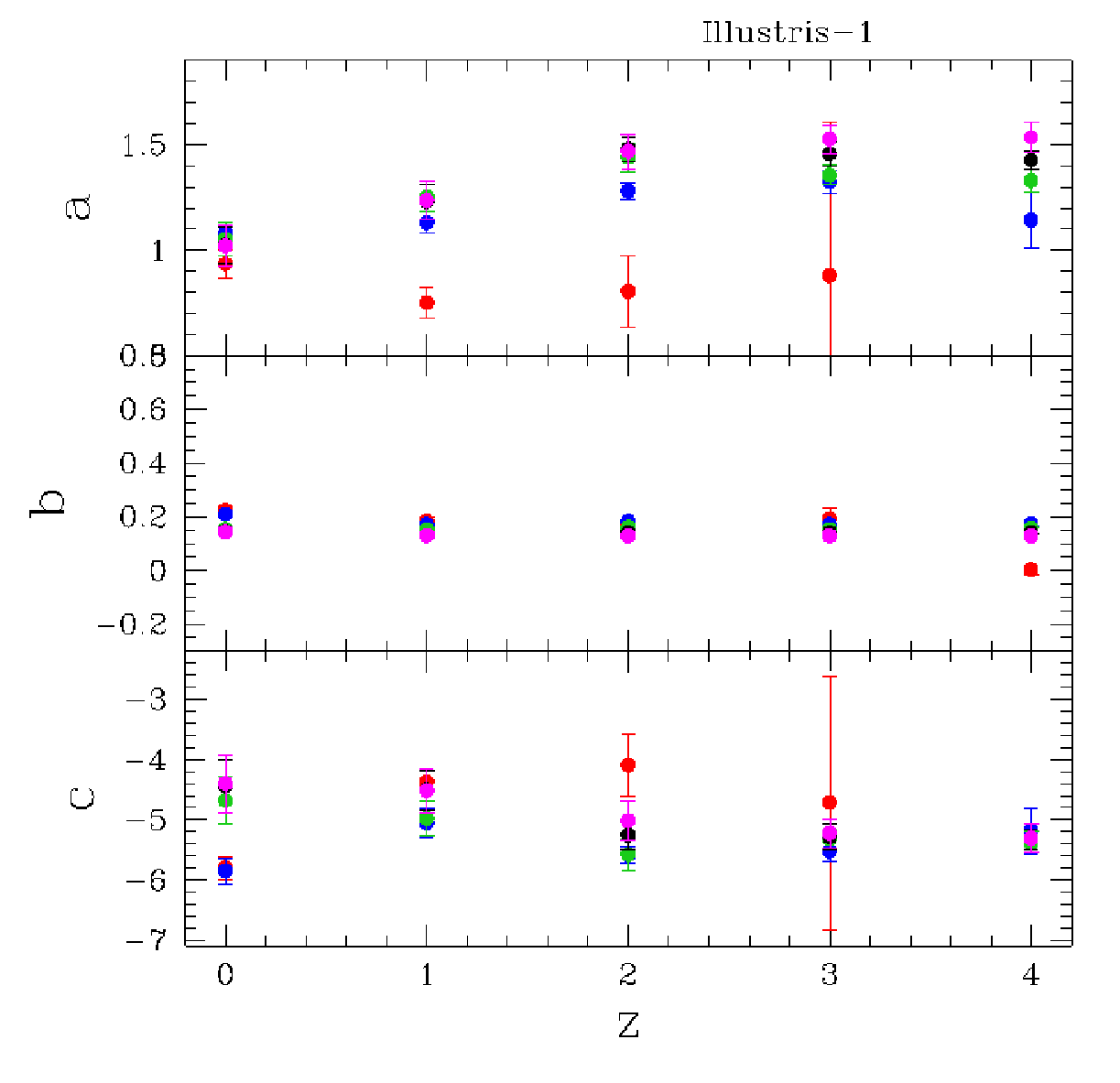}
     \includegraphics[scale=0.4, angle=0]{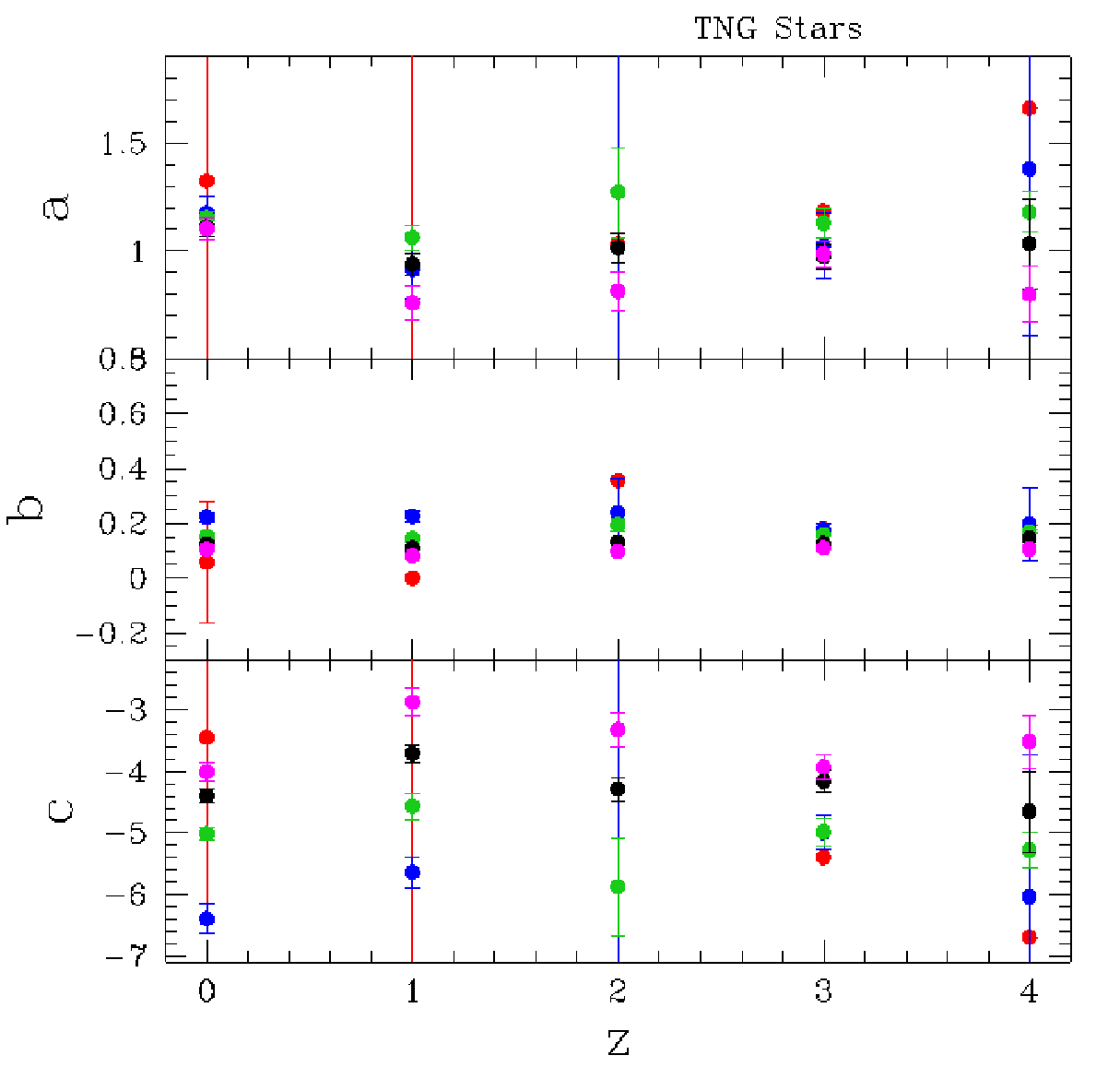} }      
   \caption{The coefficients of the FP at different redshift for Illutris-1 (left panel) and IllustrisTNG-100 (right panel) obtained from 500 fits of the FP created with a random sample of 200 galaxies. The color of the dots indicate the values for different range of masses: red dots ($\log(M_s/ M_\odot)>11$), blue dots ($\log(M_s/ M_\odot)>10$), green dots ($\log(M_s/M_\odot)>9$), black dots ($\log(M_s/M_\odot)>8$), magenta dots ($\log(M_s/M_\odot)>7$).}
    \label{fig:5}
    \end{figure*}

Figure \ref{fig:5} shows the average coefficients of the FP obtained for the Illustris-1 sample (left panel) and for the IllustrisTNG-100 sample (right panel) using the above procedure. For each redshift, we calculated the average values of the coefficients and their scatter around the mean. We also tested the effect introduced by varying the mass interval of the selected galaxies. This is indicated by the color code of the plotted dots.

It is soon evident that the variation with redshift of the FP coefficients is different for the two databases of galaxy models. In Illustris-1, the coefficient $a$ increases from z=0, where $a\sim 1.0$ at z=0 to $a\sim 1.4$ at z=2 and remains nearly constant thereafter. This is true for all the mass intervals, with exception of the highest masses (red dots), where $a$ slightly departs from the general trend, most likely because of the small number of galaxies at higher redshift. Conversely, in IllustrisTNG-100 the  coefficient $a$ has smaller variations: it is $\sim 1.2$ at z=0 and  a bit smaller at higher $z$. At z=4, the coefficient $a$ is very different for the different mass intervals, increasing progressively as the mass increases. 
In Illustris-1, the coefficient $b$  is very similar at all cosmic epochs ($b\sim 0.2$) with a small scatter around the mean value. This is not the case for IllustrisTNG-100, where the scatter of the coefficient $b$ is much larger, reaching a variation of a factor of $\sim 2$  when the low and high mass intervals are used. This difference is likely due to the selection of the galaxies that is different for the two samples. Illustris-1 progressively looses the low mass and faint objects that are characterized by low surface brightness. It follows that in this case the values of $b$, which strongly depend on \Ie, are always very similar.
The  coefficient  $c$ (the zero point of the plane) depends on the coefficient $a$ and shows much larger variations in both datasets. The values of the  coefficients $c$ at each redshift are very different for the different samples with different masses. Notably at z=4, Illustris-1 predicts very similar values of $c$ for the samples, while IllustrisTNG-100 gives different values of $c$ for the different samples with different masses.

Considering that the Illustris-1 sample is biased toward large masses at z=0 and that the galaxy radii are systematically much larger, we are inclined to trust more on the results obtained with the IllustrisTNG-100 data. However, the situation is not fully clear with the present data and it requires a deeper analysis in which the bias introduced by the selection effects is taken into account.

In any case, independently on the reliability of the FP coefficients at each redshift, the basic result obtained here from both databases is that at any epoch massive and dwarf galaxies do not share the same FP. This is well clear in particular for the rotation of the plane that is implicit in the different values of the coefficient $b$ when galaxies of different masses are used.

However, it is worth noting that at any redshift there is a mean FP for all mass intervals that is always characterized by a  small scatter. This issue is discussed in great detail in the next sections.

\subsection{Comparison with literature data}
We compare here our results with those obtained by \cite{Luetal2020}. {In order to frame the comparison in the right context we should remind that: 1) The equation of the FP they adopted is }

\begin{equation}
    \log(R_e[kpc])=a+b\log(\sigma[km/s])+c\log(I_e[L_\odot/kpc^2])
\end{equation}

   \begin{figure*}
   \centering
   { \includegraphics[scale=0.4, angle=0]{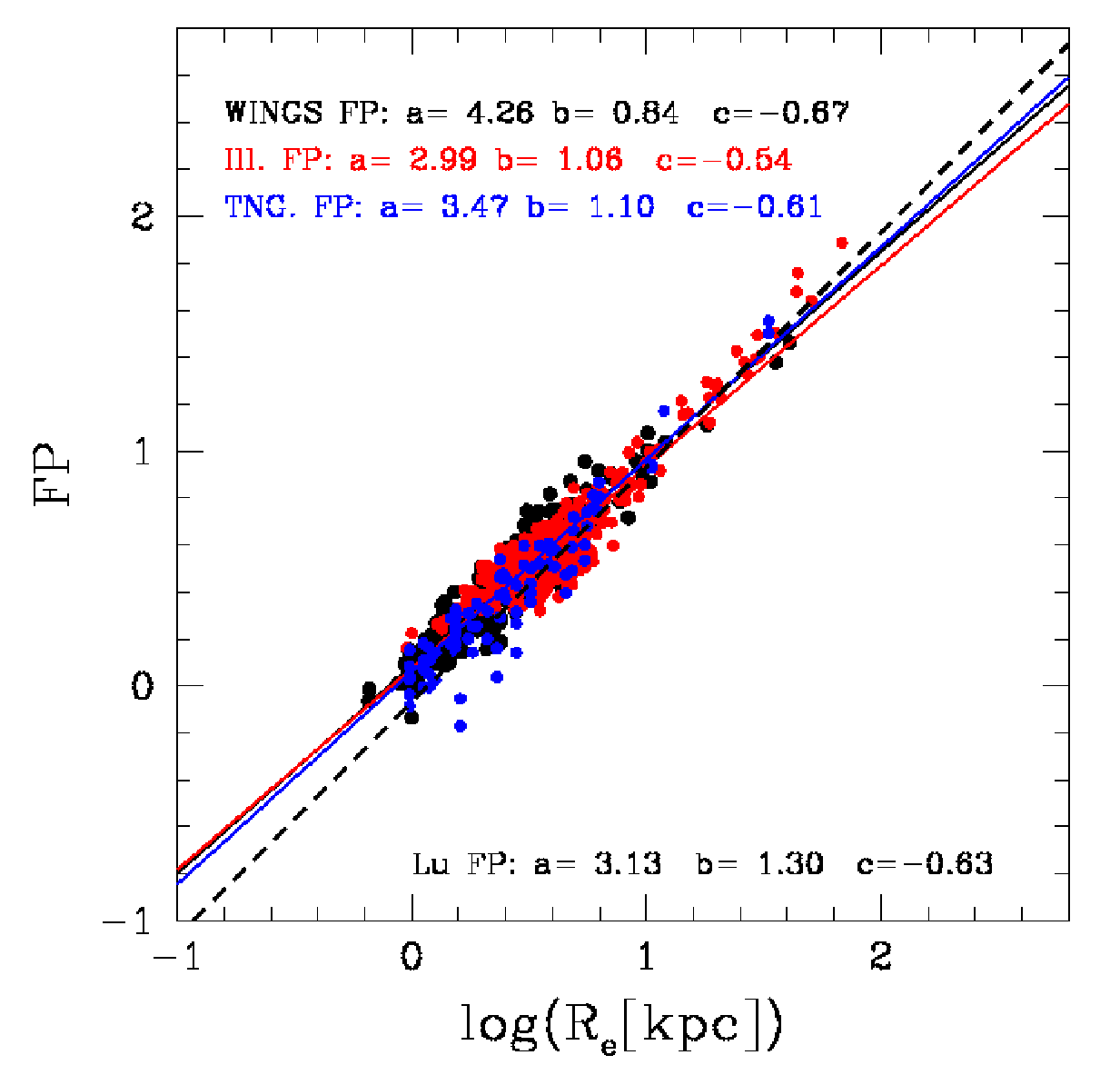}
     \includegraphics[scale=0.4, angle=0]{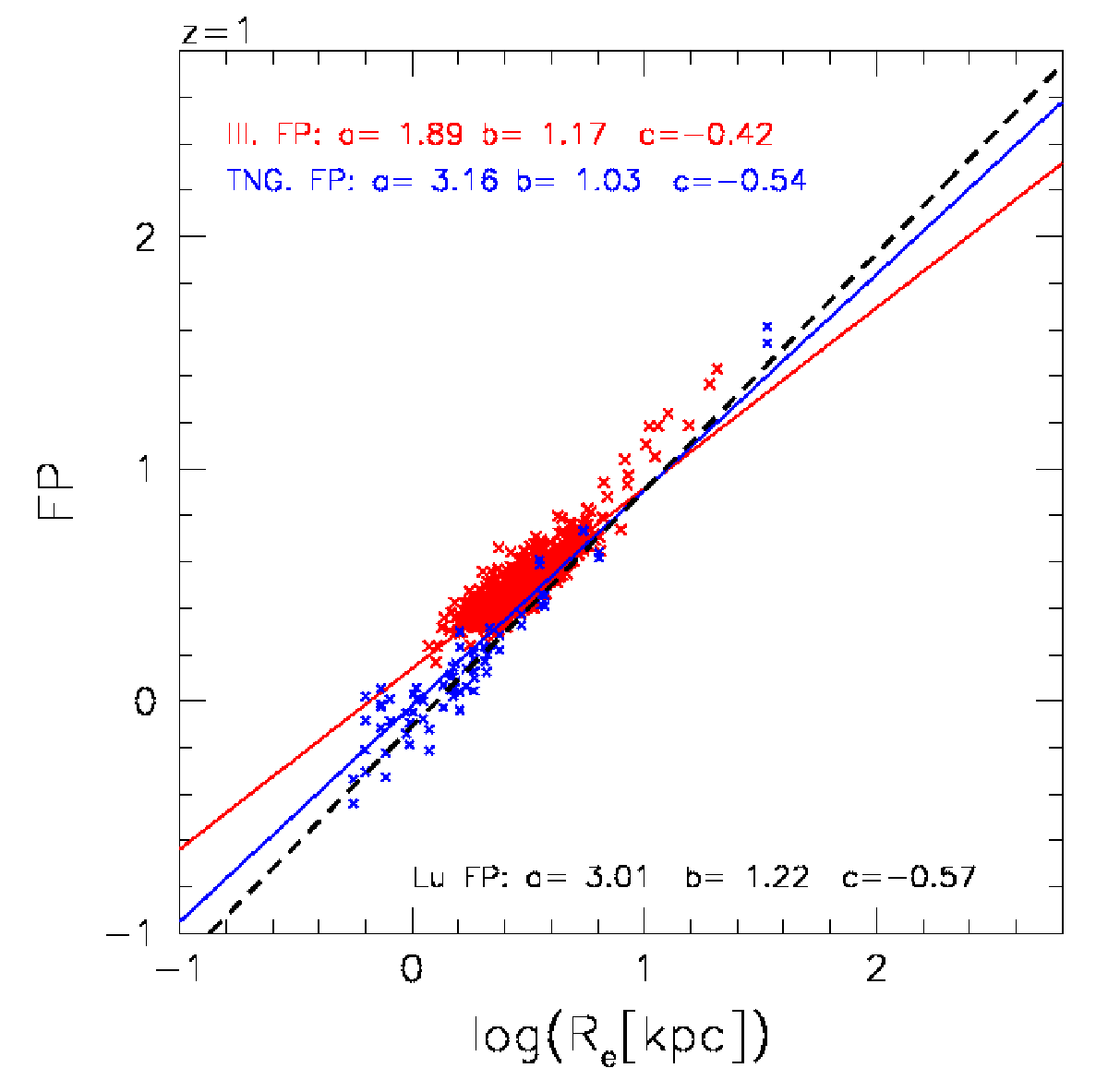} }      
   \caption{{The fits of the FP for our data compared with those of \cite{Luetal2020}. Left panel: The FP coefficients obtained at z=0 for the WINGS and Illustris data fitted in the same mass interval chosen by \cite{Luetal2020}. The black dots mark the WINGS data, the red ones the Illutris-1 and the blue ones the IllustrisTNG-100 data. The FP coefficients obtained by \cite{Luetal2020} are listed at the bottom of the panel and  the dashed line visualizes the relation. Right panel: 
   The same as in the left panel but for  the data and relationsips at z=1.  The dashed line is the FP of \cite{Luetal2020} FP whose coefficients are also displayed in the panel for the sake of clarity.} }
    \label{fig:new}
    \end{figure*}

\noindent
2) The IllustrisTNG-100 data they consider have stellar masses $\log(M_s)>9.7$. This limit was chosen to take into account the mass resolution of the simulation ($1.4\times10^6 M_\odot$).
3) {Their photometric data are in the $r$-band.}

{The comparison is shown in the two panels of Fig. \ref{fig:new} for different values of the redshift, that is  z=0 (left panel) and z=1 (right panel).  In the left panel we show the FP's for the WINGS data (black dots), the Illustris-1 simulations (red dots) and the IllustrisTNG-100 simulations  (blue dots) at z=0. The solid lines are the FP's for the three cases whose coefficientes ({\it a, b,} and {\it c}) are indicated at the top of the panel according to the adopted color code. {Finally}, at the bottom of the panel we display the coefficients of the FP of \cite{Luetal2020}.  The same comparison is also made in the right panel of Fig. \ref{fig:new} for models and observational data at redshift z=1 (same color code). WINGS data are of course missing. The result of this comparison is that 
in the case of  the IllustrisTNG-100 data, the coefficients of the fitted distributions (FP's) are quite similar. The small differences can be easily explained in terms of the different number of galaxies, the different bands ($V$ vs $r$), and the different strategy used for the fit. We conclude that our analysis confirms the results of \cite{Luetal2020}. }

\section{The scatter around the FP}
\label{sec:5}
    
At any redshift, the scatter around the FP is always small when the fit of the galaxy sample is made for galaxies at that redshift and no particular mass interval is selected. In other words, the  classical FP discovered and derived for galaxies at z=0 is not the same at all epochs, but most likely it changes with time \citep[see e.g.][]{Luetal2020}. Extending and using it as it is at z=0 back in time is not correct. {Each epoch has its own FP and the scatter around it is small. In this sense only, the FP is a universal property, whose physical origin is still elusive. }

{In Figs. \ref{fig:2}, \ref{fig:3}, an \ref{fig:4}  we have presented the FPs,  the comparisons of these with observational data and model galaxies, and quoted the dispersion $\Delta$ of data and models around the FPs without discussing the definition of dispersion we have adopted. Given the  FP  represented the straight line of eq. \ref{expr_fp}  (formally the line R:aX +bY+c), we measure the dispersion of data with respect to this line by the mean value of the distance $d_p$ of all data  points with  coordinates $X_p$, $Y_p$ from the line R:aX +bY+c. 
For any arbitrary point the distance $d_p$ is:

\begin{equation}
     d_p =  \frac{|a X_p + Y_p + c| }{\sqrt{a^2 +b^2}} 
\end{equation}

\noindent
and the mean dispersion $< \Delta >$ for all the $N$ data points is 
\begin{equation}
 < \Delta > = \frac{1}{N} \sum d_p 
 \label{delta_pos}
\end{equation}
In our analysis we adopt this definition.

Another way of defining the mean dispersion is to look for any data point at the difference between the observational position $O$  on the FP and the one indicated by $P$ it would take if eq. \ref{expr_fp} is applied. Therefore we have

\begin{equation}
 < \Delta >  = \frac{1}{N} \sum \sqrt{(O-P)^2} 
 \label{delta_ops}
\end{equation}
This definition is used to compare our results with those in literature.

Finally, in order to highlight the side with respect to the reference FP the deviation occurs, in the discussion below we occasionally  use the simple expression:

\begin{equation}
 < \Delta >  = \frac{1}{N} \sum (O-P) 
 \label{delta_neg}
\end{equation}
which may take both positive and negative values. }

\begin{figure*}
   \centering
   \includegraphics[scale=0.38, angle=0]{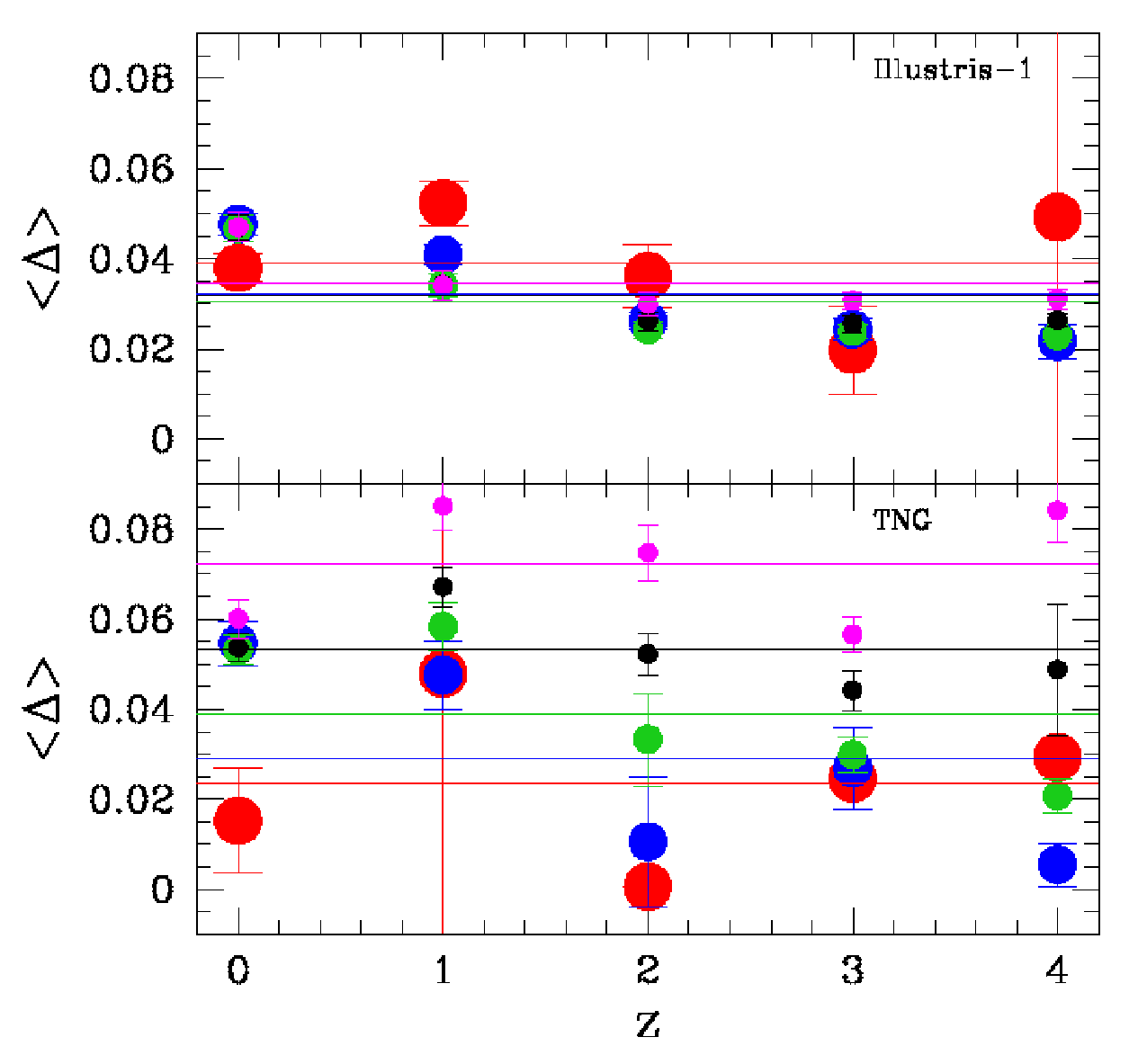}    
   \includegraphics[scale=0.4, angle=0]{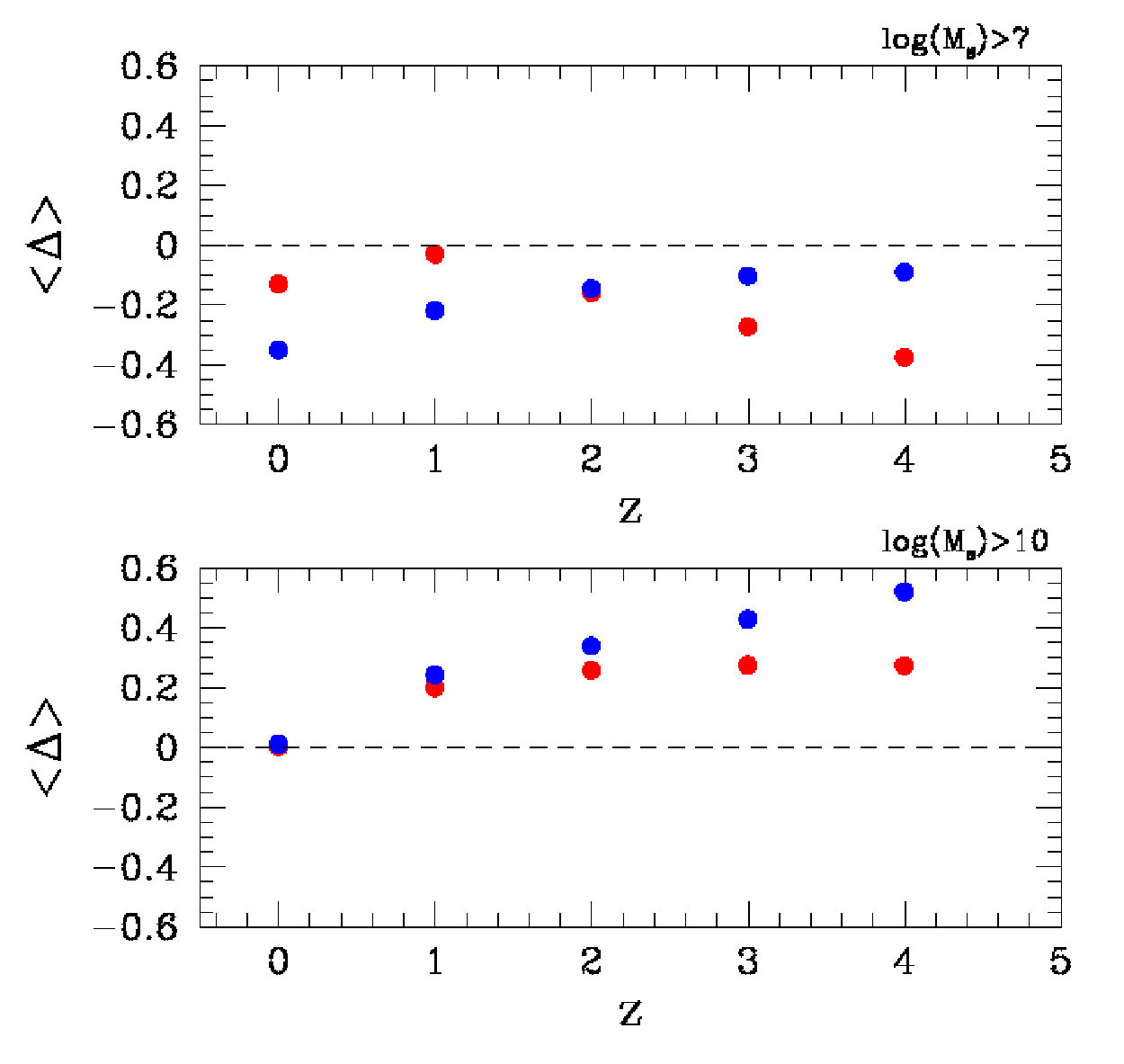}
   \caption{Left panel: The scatter around the FP for Illutris-1 (upper panel) and IllustrisTNG-100 (lower panel) obtained from simulations. The dots of different colors mark the values of the scatter obtained for different range of masses: red dots ($\log(M_s/M_\odot)>11$), blue dots ($\log(M_s/M_\odot)>10$), green dots ($\log(M_s/M_\odot)>9$), black dots ($\log(M_s/M_\odot)>8$), magenta dots ($\log(M_/M_\odot)>7$). The colored lines give the average scatter across time for each mass sample. {The $\Delta$'s are calculated according eq.
   \ref{delta_pos}. }   Right panel: The scatter around the FP for Illustris-1 (red dots) and IllustrisTNG-100 (blue dots) at different redshift. In this case the scatter is obtained from the difference of the measured $\log(R_e)$ and the value calculated using the expression of the FP with fixed coefficients given by the sample at z=0 with masses larger than $10^{10} M_\odot$. {The $\Delta$'s are calculated according eq.    \ref{delta_neg}.
   The upper and lower panels show the scatter for the two samples of models with different low mass limit.} }
    \label{fig:6}
    \end{figure*}

{Figure \ref{fig:6} shows, with different colors for the different mass interval of the galaxies from which the FP is derived, the scatter $\Delta$ expected at  different redshifts for models galaxies taken from the Illustris-1 (top panel) and IllustrisTNG-100 (bottom panel) libraries. The two data-sets  yield  slightly different results. In the left panels the  $\Delta$'s are calculated according eq. \ref{delta_pos}, while in the right panels according to eq.    \ref{delta_neg} to better show that the FP holding at z=0 cannot be extended to higher redshifts. }

If we take IllustrisTNG-100 as the reference sample, because it is less biased in mass, the scatter around the FP is systematically lower for cases with a high mass limit. In any case, the mean value of the scatter is quite small (maximum value $\Delta \leq 0.1$) and similar for both libraries of model galaxies. A possible cause of systematic differences, in this case, could be that the number of galaxies of large mass progressively decreases at increasing  redshift.

{The situation changes completely if the FP determined at z=0 is  supposed to hold  at all redshifts and it is used to infer the scatter of galaxies around it.  Figure \ref{fig:6} (right panel) {shows the values of} $\Delta$ we would expect in such a case. The results are presented both for Illustris-1 (red dots) and IllustrisTNG-100 (blue dots) and limited to two values for the galaxy mass limits, namely $\log (M_s /M_\odot) \geq 7$ (all galaxies are used, top panel) and $\log (M_s /M_\odot) \geq 10$ (only the massive ones, bottom panel). The scatter is measured by the difference of the observational $\log(R_e)$ and the value calculated using the expression of the FP whose coefficients are derived from the galaxies at z=0 with masses larger than $10^{10} M_\odot$ (the classical FP). We note that now the scatter is much larger than before, both at changing the redshift and varying the mass limit. Furthermore, in the top panel, the mean scatter is negative, because too many objects would acquire too large $R_e$ if calculated with the expression of FP derived for z=0. 
The negative values for the scatter in this case can be explained by the presence of a large number of objects with mass smaller than $10^{10} M_\odot$ which have radii smaller  than predicted by the adopted FP of reference (for galaxies more massive than  $10^{10} M_\odot$ at z=0, the classical FP).
In contrast, in the lower panel, when the sample includes only galaxies with mass $M>10^{10}\, M_\odot$, the scatter becomes positive going toward high redshifts. }

{The scatter we have estimated is comparable with the results found by \citet{Luetal2020} and \citep{Cappellari_etal_2013}. Our values are, however, somewhat smaller than those found those authors. The reason resides in the different definitions of scatter that are used. We adopt the distance method given by relation (\ref{delta_pos}) while they used difference method expressed by relation (\ref{delta_ops}).

In any case, it is worth recalling that the scatter we are talking about cannot be mistaken with the scatter deriving from uncertainties affecting both the theoretical simulations \citep{Luetal2020} and observational results \citep{Cappellari_etal_2013}. The scatter in question here is caused by the uncorrect use of the FP for massive galaxies at z=0 all over the whole mass range and at all redshifts. }

In summary, the analysis of the FP at different redshifts done with the two libraries of model galaxies suggests that: 1) at each cosmic epoch there is a best fitting plane in the FP-space; 2) this plane is not the same across the whole range of masses, but slightly changes with the mass interval; 3) the scatter around the plane is always quite small (likely a bit larger if the galaxy mass interval is large) when the coefficients of the plane are those derived from galaxies at the redshift under consideration; 4) on the contrary, the scatter can be  significantly large if coefficients of the FP  are those derived from galaxies at z=0, in other words the classical FP is used at any redshift. These conclusions agree with and confirm similar results presented in other recent studies \citep[e.g.][]{HydeBernardi2009,Luetal2020}.

\section{The FP-space and the $\beta$ parameter}
\label{sec:6}

\citet{Donofrio_Chiosi_2023a} and \citet{Donofrio_Chiosi_2023b} already demonstrated that the behavior of galaxies in the FP-space is tightly related to the value of $\beta$  in the \Lsigbtempo\ law. More specifically,  the direction of motion of galaxies in this space varies with $\beta$, because this parameter encodes the effects of galaxy evolution in terms of merging events, star formation, and many other possible physical processes. Aim of this section is to quantify {the above statement} that the distribution of galaxies in the FP-space changes  with $\beta$. \\

\textsf{The reference case: the WINGS galaxies at redshift z=0 and $M_s > 10^9\, M_\odot$}.  The equation of the FP for this case is 
\begin{equation}
  \log R_e = 0.92 \log \sigma + 0.28 <\mu_e> - 7.15\, .  \nonumber
\end{equation}

\begin{figure*}
   \centering
  { \includegraphics[scale=0.45, angle=0]{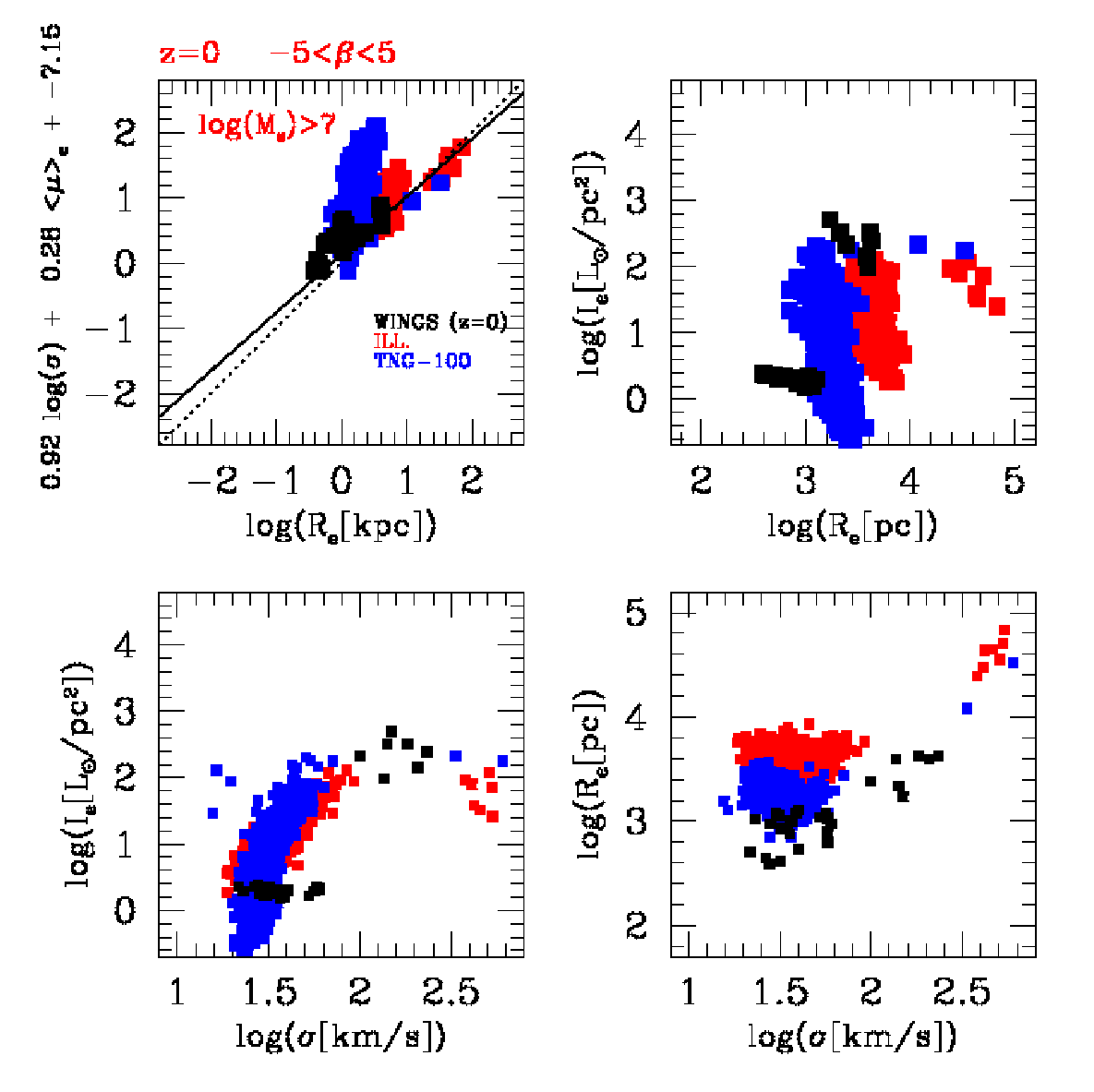}
    \includegraphics[scale=0.45, angle=0]{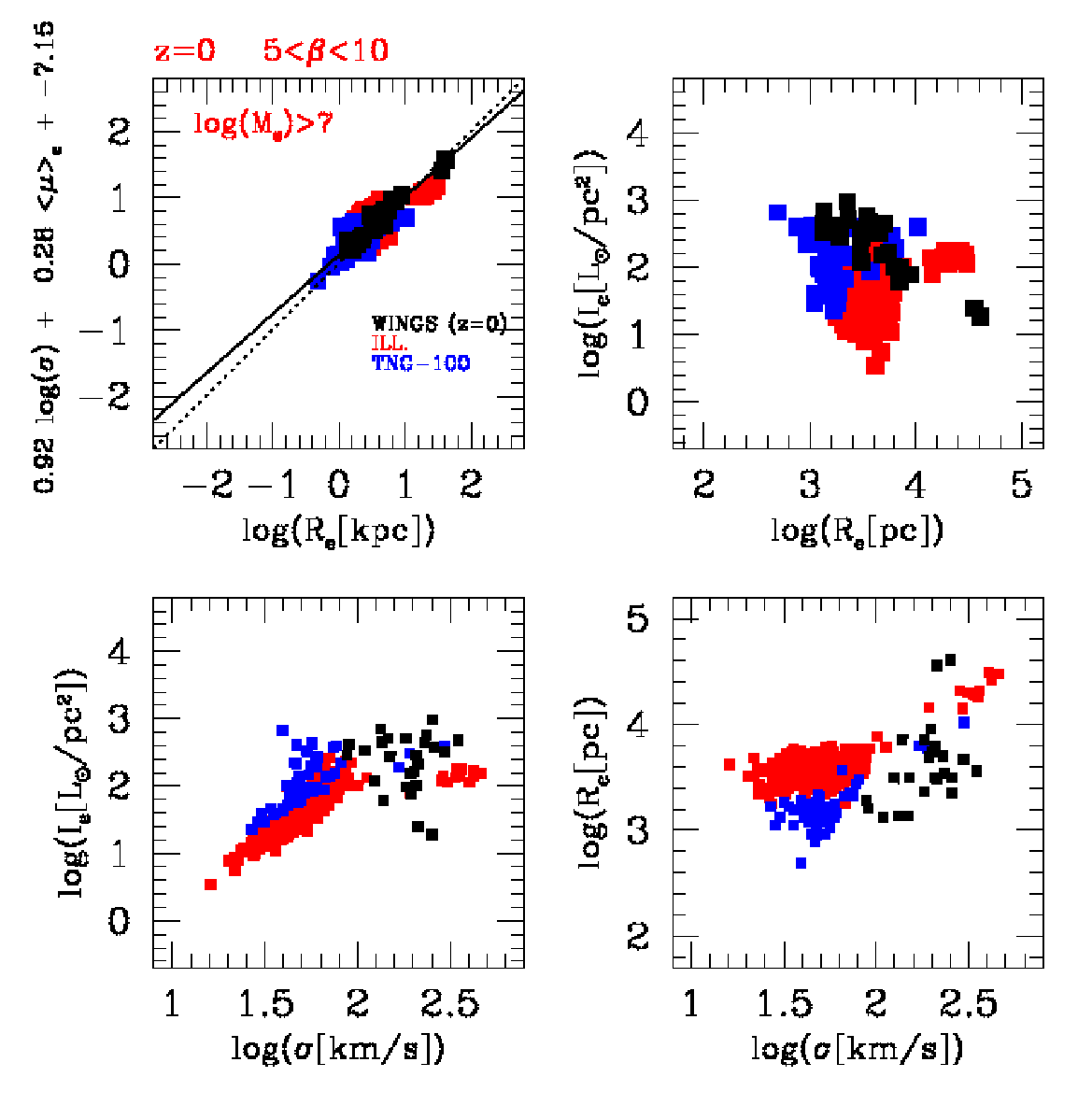}   }
   \caption{The FP edge-on and its projections for galaxies with masses greater than 
   $\log (M_s/M_\odot) > 7$. \textsf{Left Panel}: the sample at z=0 for WINGS (black dots), Illustris-1 (red dots) and Illustris TNG (blue dots), and $\beta$ in the interval 
   $-5 < \beta < 5$. \textsf{Right Panel}: the same as in the left panel but for $5 < \beta < 10$. The solid line and the dotted line in  leader panel (top left) are the FP and the one-to-one relation, respectively. }
    \label{fig:7}
    \end{figure*}

\begin{figure*}
   \centering
  { \includegraphics[scale=0.45, angle=0]{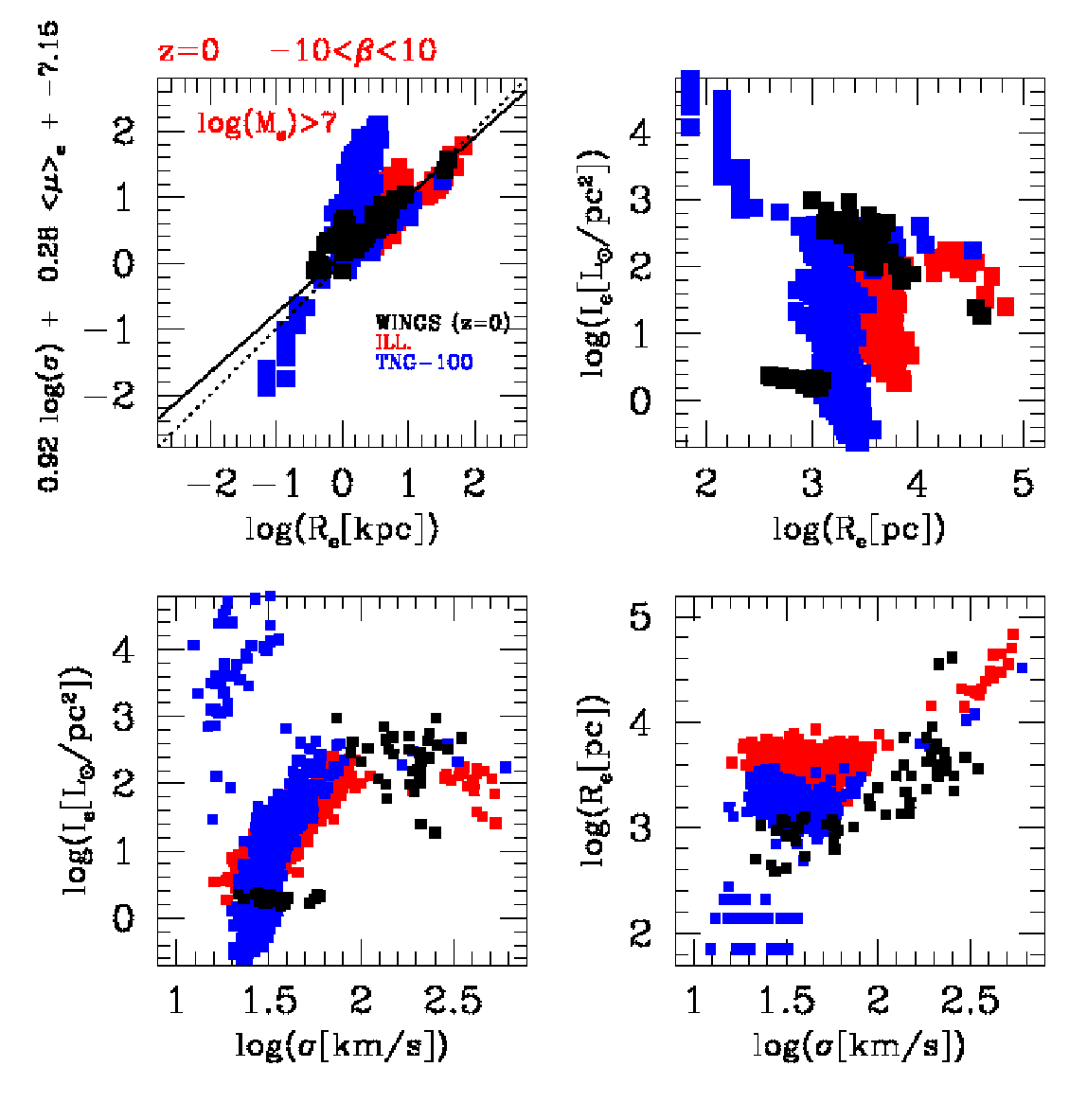}
   \includegraphics[scale=0.45, angle=0]{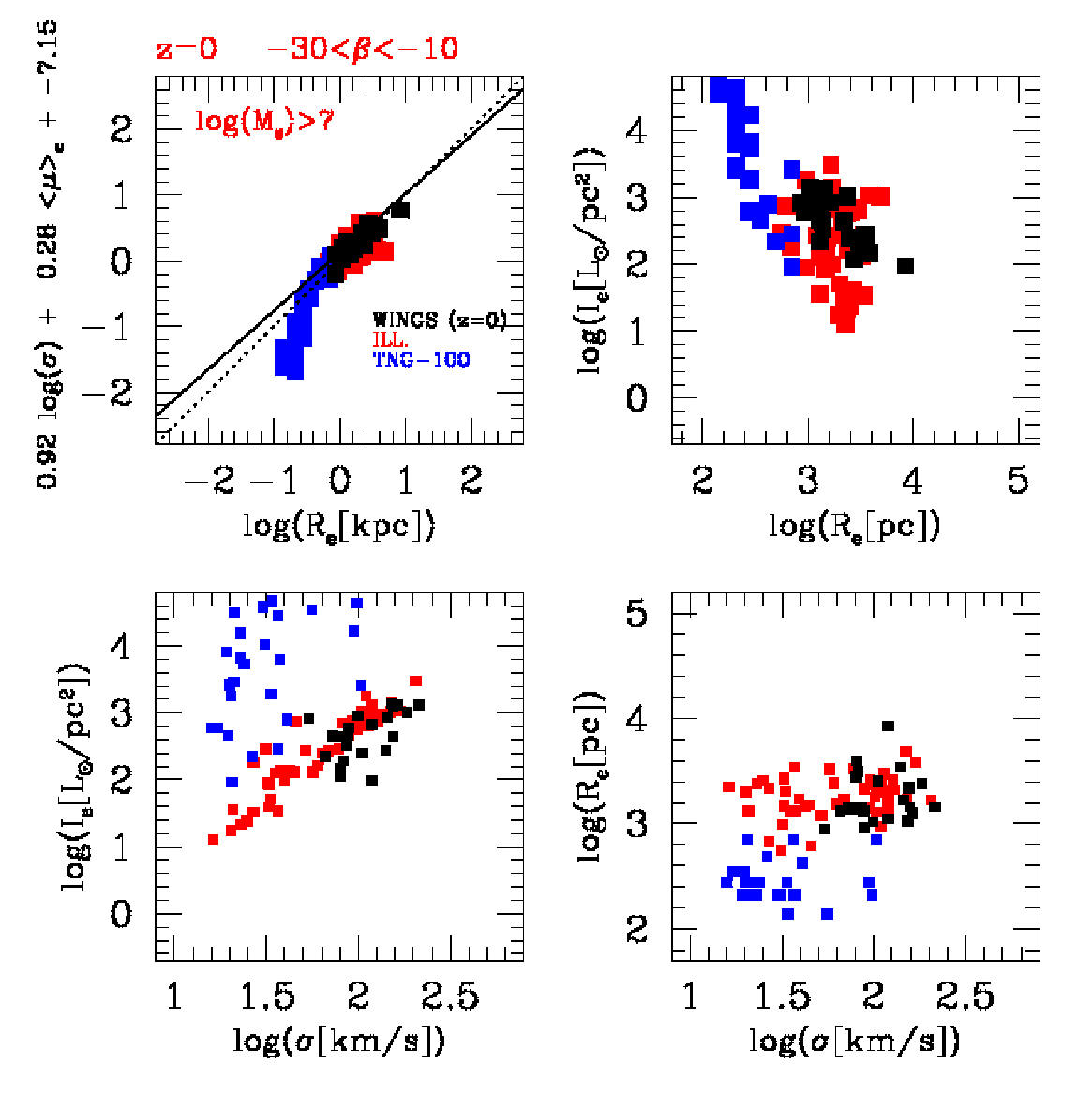}  } 
   \caption{The FP edge-on and its projections for galaxies with masses greater than $\log (M_s/M_\odot) > 7$. \textsf{Left Panel}: the sample at z=0 for WINGS (black dots), Illustris-1 (red dots) and Illustris TNG (blue dots) and $\beta$ in the interval 
   $-10 < \beta < 10$. \textsf{Right Panel}: the same as in the left panel but for $-30 < \beta < -10$. The solid line and the dotted line in  leader panel (top left) are the FP and the one-to-one relation, respectively.}
    \label{fig:8}
    \end{figure*}

\begin{figure*}
   \centering
  { \includegraphics[scale=0.45, angle=0]{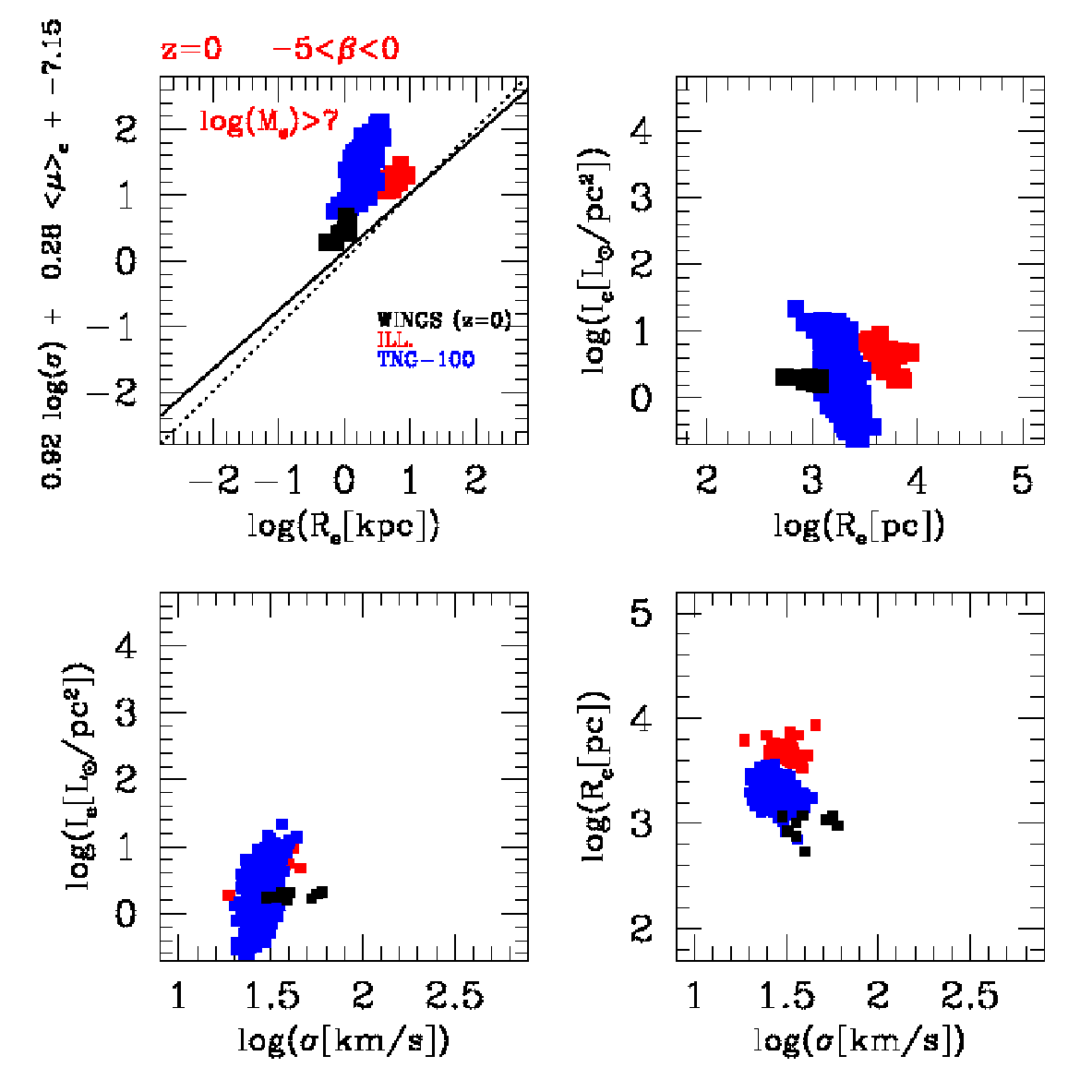}
   \includegraphics[scale=0.45, angle=0]{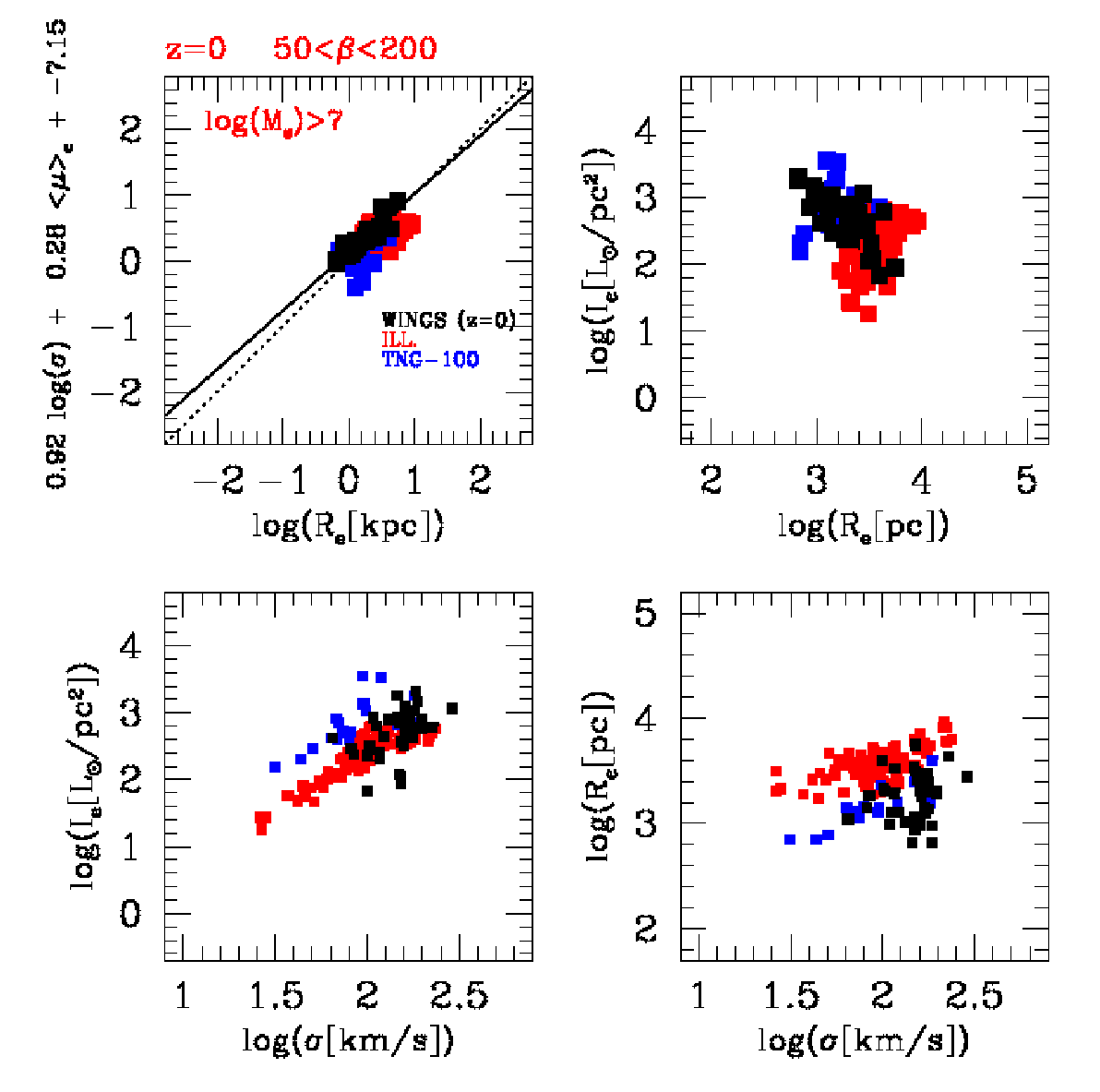}  } 
   \caption{The FP edge-on and its projections for galaxies with masses greater than $\log (M_s/M_\odot) > 7$.  \textsf{Left Panel}: the sample at z=0 for WINGS (black dots), Illustris-1 (red dots) and Illustris TNG (blue dots) and $\beta$ in the interval 
   $-5 < \beta < 0$. \textsf{Right Panel}: the same as in the left panel but for $50 < \beta < 200$.  }
    \label{fig:8bis}
    \end{figure*}

\noindent
{when the dwarf galaxies of  \cite{Bettonietal2016} are not included.  
Figures \ref{fig:7}, \ref{fig:8} and \ref{fig:8bis} display the FP-space, both in its edge-on view and its projections, for a sample of galaxies at redshift z=0 with masses exceeding $10^7 \,M_\odot$ and $\beta$ falling in different intervals.} The three figures  are organized in groups of four panels (or physical cases). {The group leader is the panel showing the FP on its ordinate (the solid black line)  } and is labeled with the value of redshift and range of $\beta$ of the model galaxies in use (e.g. z=0 and $-5 < \beta <5$, first top left panel in Fig.\ref{fig:7}). {The dotted line is the one-to-one relationship for comparison}. The other panels surrounding the leader (at the top right, immediate bottom left and bottom right)  show the projections of the FP, that is $I_e$ vs $R_e$, $I_e$ vs $\sigma$, and $R_e$ vs $\sigma$ in the order.  
In each group of four panels the intervals of  $\beta$ are different. Finally, in each panel we display the observational data of WINGS (black filled squares), {the Illustris-1 models (red filled squares)} and the IllustrisTNG-100 models (blue filled squares). 

The careful inspection and mutual comparison of the various groups of panels (physical cases) reveals that position and even the occurrence of galaxies in general and/or particular sub-groups of these on the FP and on its projections find tight correspondence in the values and/or range of values spanned by the $\beta$ parameter. In other words, we found that some features of the galaxy distribution in the afore mentioned planes come and go according to the underlying values of $\beta$. 

To illustrate the point let us look at the case of the faint galaxies. {In Fig. \ref{fig:7}, they are the objects with $\log(R_e) \simeq 0.5$  (in Kpc) and $-1 < \log I_e < 3$ {(in $L_\odot/pc^2$) } in the top right panel } of the first group with $-5 <\beta <5$ and corresponds to the plume-like structure above the FP shown in the top left panel of the same group.  They disappear moving to the second group of the same figure and same panel but in which $\beta$ falls in the range $5 <\beta <10$.  
Conversely, as $\beta$ becomes larger and more negative, it is well visible only the downward deviation from the FP of the bright galaxies.
The same faint objects deviate from the FP in the first panel of the first group but are absent in the same panel of the second group. 
Furthermore, the galaxies with $\beta$ in the range  $-5$ to $5$, show both objects along the classical edge-on view of the FP (black solid line), that is the upward deviation typical of the faint galaxies. This deviation is not evident for galaxies with $\beta$ ranging from 5 to 10, while it is the only feature visible when $-5<\beta<0$. As $\beta$ decreases further and reach large negative values, the IllustrisTNG-100 sample begins to exhibit a downward deviation in the edge-on view of the FP. Remarkably, when $\beta$ falls below $-10$, the upward deviation disappears.  

Continuing the inspection of the various panels {in Fig. \ref{fig:8} and \ref{fig:8bis}, }the list of features that come and go depending on the range of $\beta's$ under consideration gets longer and longer. In general we observe that the most massive galaxies are found along the classical FP quite independently from the value of $\beta$. The only difference we can note among all these possible situations is in the number of objects present in each interval of $\beta$. Notably for $-10< \beta <-5$ only one galaxy is found, while in the interval 
$-5<\beta<5$ there are  galaxies along the FP and the classical tail of the \IeRe\ plane. For large positive and negative $\beta$'s, we always find galaxies along the classic FP, but the distributions in the FP projections may change significantly. The tail in the \IeRe\ plane is better visible for low values of $\beta$ ($-10<\beta<10$).
The obvious conclusion is that at z=0,  galaxies more massive than  $10^7\, M_\odot$  are distributed in a different ways in the FP-space according to their mass and evolutionary conditions that are here indicated by  the $\beta$ parameter. Variations of $\beta$ correspond to including or excluding  galaxies of different mass and/or different evolutionary conditions. 

Looking for a plausible physical explanation of this otherwise odd behaviour of galaxies at varying $\beta$, we suggest here that it  could be ascribed to the effective surface brightness \Ie\ and its interplay with  \re. When \Ie\ is low and \re\ is large, $\beta$ gets close to 0. The opposite if \Ie\ is high and/or \re\ is small or normal.
{ Equation \ref{eq2} in appendix  \ref{Appendix_A} clarifies} why $\beta$ can assume different values, either positive or
negative, either small or large. The reason is that $L'_0$ depends on the quantity $1-2A'/A$ (see eq. \ref{eq3} in appendix  \ref{Appendix_A}), that is  on the ratio between $2\log(\sigma)$ and the combination of \Ie\ and \re\ in log units. When $1-2A'/A$ approaches zero,  $L'_0$ diverges and consequently $\beta$ does the same. Values of $\beta$ close to 0 means that $1-2A'/A$ is significantly different from 0. The brightest galaxies have in many cases large values of $\beta$ because $1-2A'/A \sim 0$. Many others, in particular the ETGs along the bright tail of the \IeRe\ relation, have values of $\beta\sim 0$. These objects are believed to continuously experience minor dry mergers that increase their radius and decrease their effective surface brightness. Such combination yields $1-2A'/A$ significantly different from zero.

\begin{figure}
   \centering
   \includegraphics[scale=0.4, angle=0]{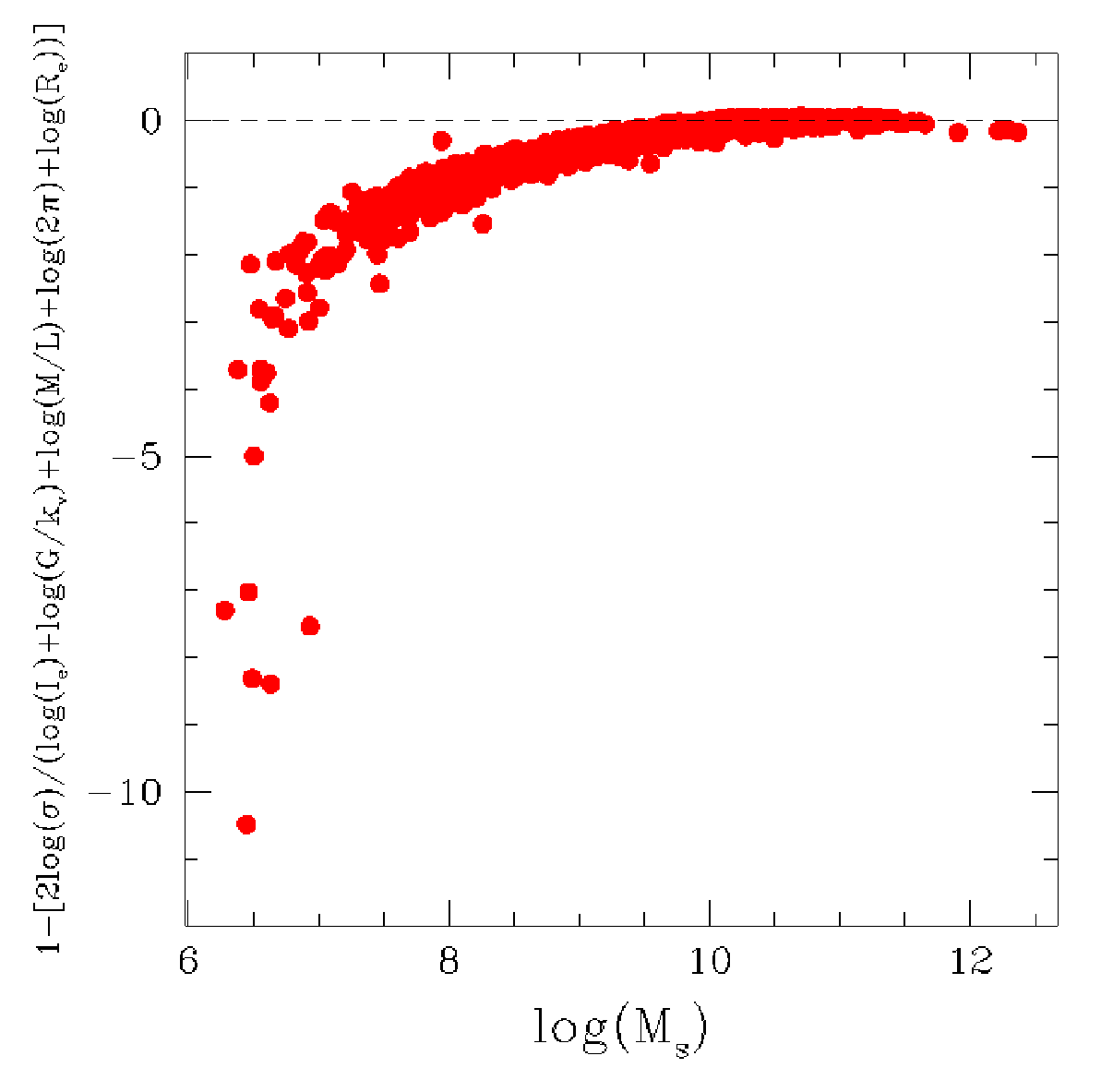} 
   \caption{The quantity $1-2A'/A$ versus $\log(M_s)$. The less massive galaxies progressively deviate from the "full virialization" condition.}
    \label{fig:MAA}
    \end{figure}

\citet{Donofrio_Chiosi_2023a} argued that  galaxies with $\beta$ close to 0 are far from  "full virialization", whereas those with large either positive or negative $\beta's$ are much closer to this condition. A brief comment on what we really meant with ``full virialization'', is mandatory here to avoid any misunderstanding. Indeed, galaxies are always close to the mechanical equilibrium (any momentary deviation from this condition by interaction with another is soon wiped off on a short dynamical timescale), that is for most of time they are very close to virialization. However,  when light is used to derive the galaxies' structural parameters, the main effect of it is that their distribution appears tilted in the FP-space with respect to the virial prediction. Several studies have already demonstrated that, when the half-mass radius and surface-mass density are used as parameters, the tilt disappears \citep[see e.g.,][]{Cappellarietal2006}. This means that light introduces an extra information, different from that provided by the mass.
When we observe galaxies with large values of $\beta$ (either positive and negative), it means that the combination of \Ie, \re, and $M_s/L$ nearly exactly matches $2\log(\sigma)$, the condition of virialization written using the light structural parameters, those derived from the luminous component of galaxies. The light parameters \Ie\ and \re\ do not trace perfectly the mass parameters. The term "full virialization" is used here to say that galaxies are in such peculiar condition, where $2\log(\sigma)\sim A$. Many galaxies are far from this condition and have $\beta\sim 0$. Indeed their combination of \Ie\ and \re\ does not match such condition (see Fig. \ref{fig:MAA}). \\

   \begin{figure}
   \centering
   \includegraphics[scale=0.45, angle=0]{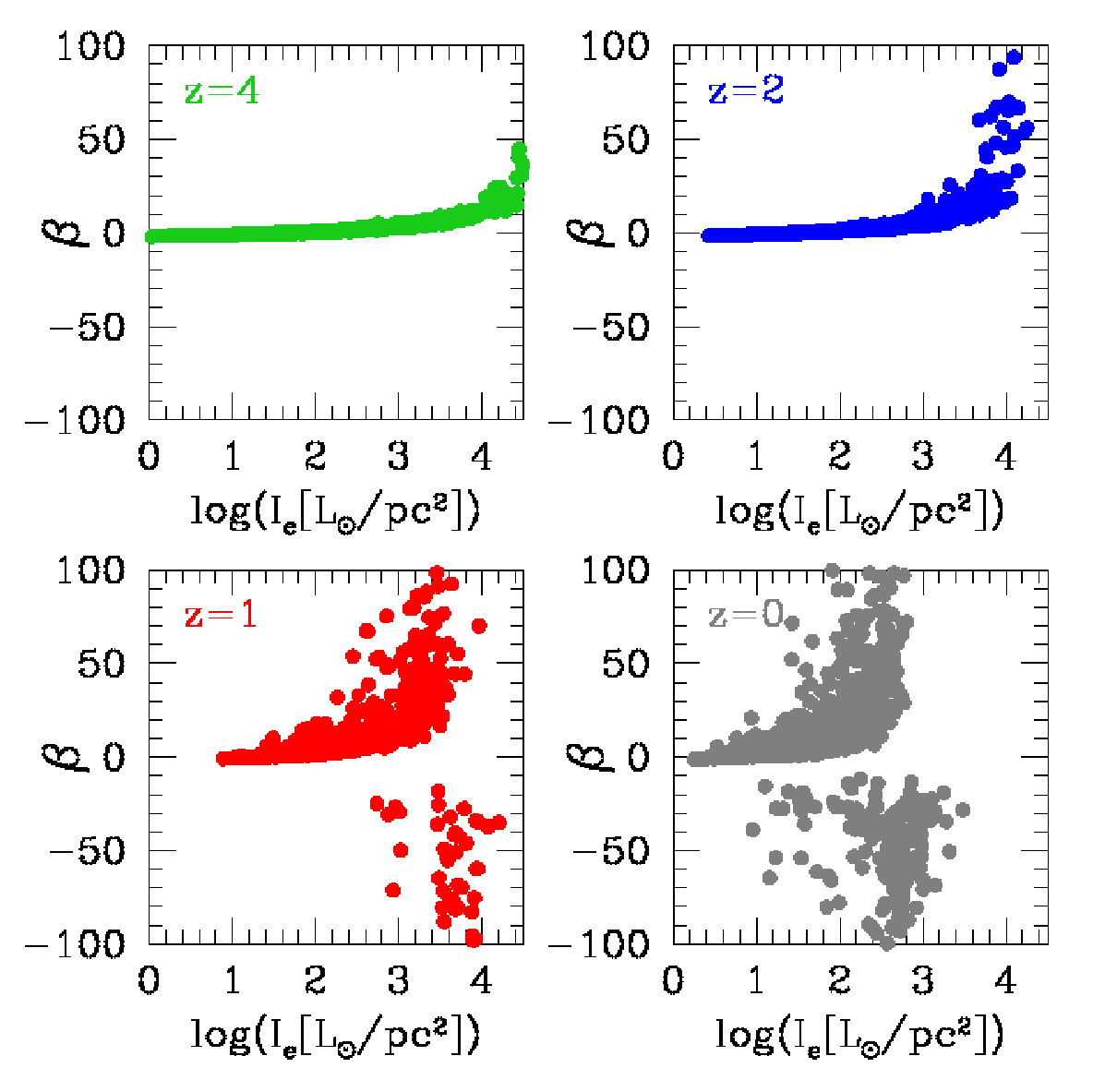} 
   \caption{The $\beta-I_e$ relation for the Illustris-1 galaxies at four different redshift. Note the progressive increase of $\beta$ and the average decrease of \Ie\ when z=0 is approached. }
    \label{fig:Iebeta}
    \end{figure}

Figures \ref{fig:Iebeta} and \ref{fig:MAA} provide an explanation of the increase of $\beta$ across time. Many galaxies of larger dimension and mass can reach the condition of "full virialization" (i.e. $1-2A'/A \sim 0$) and their $\beta$ increases on both negative and positive sides, depending on the sign of $L'_0$. The remaining  galaxies of similar dimensions and mass with $1-2A'/A \neq 0$ have $\beta\sim 0$. \\

   \begin{figure*}
   \centering
  { \includegraphics[scale=0.45, angle=0]{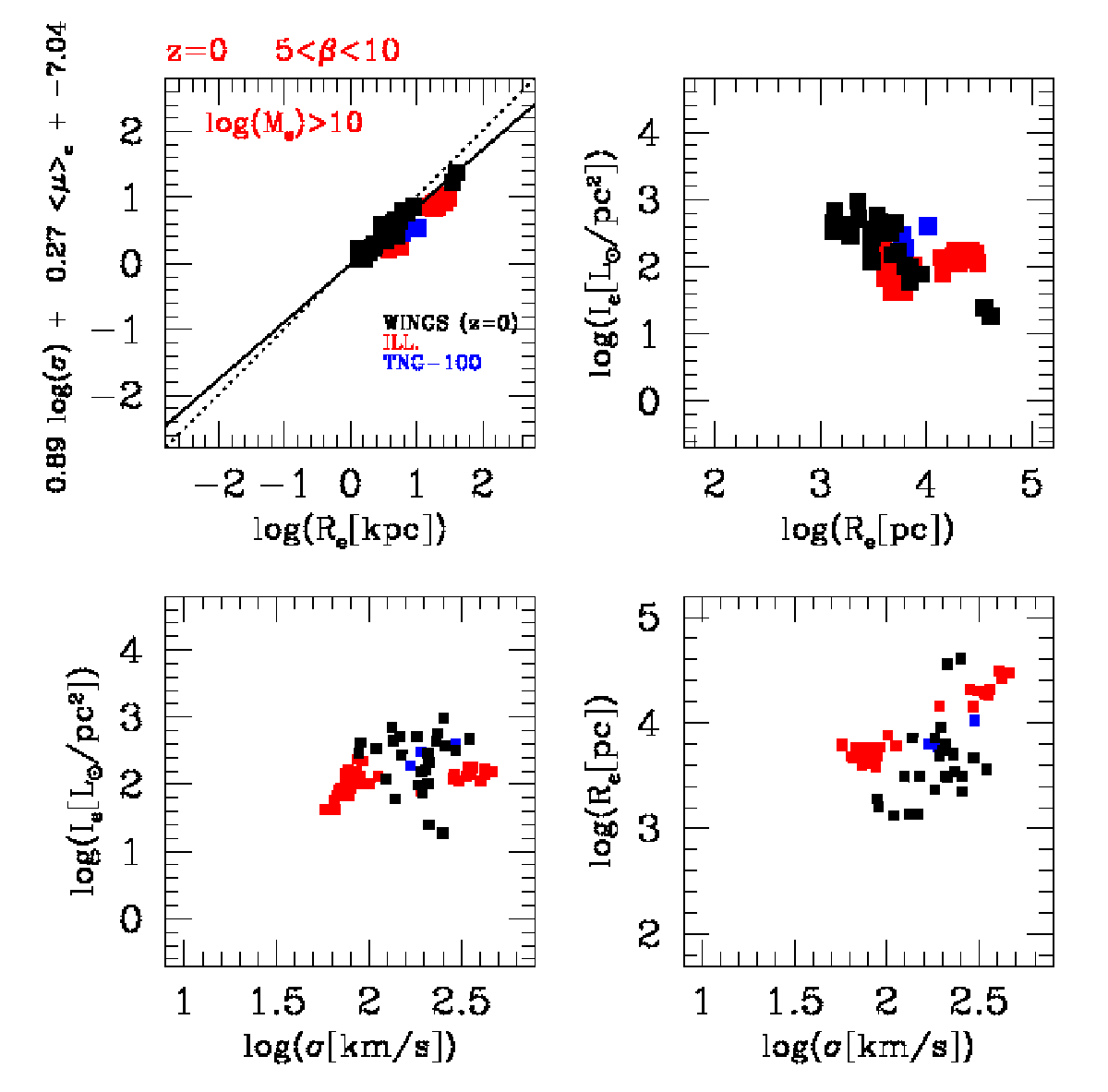}
   \includegraphics[scale=0.45, angle=0]{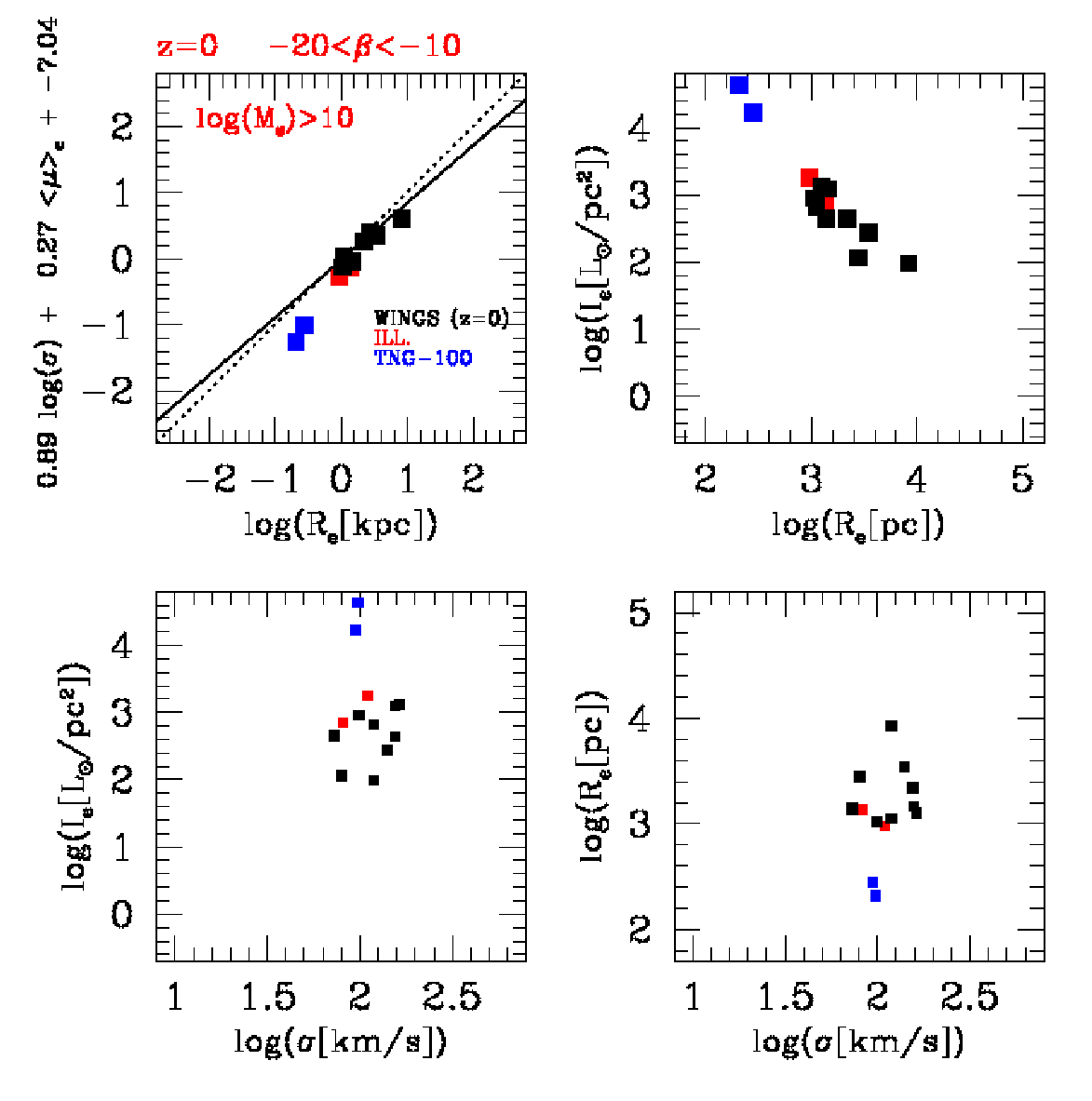}  }     
   \caption{The FP edge-on and its projections for galaxies with masses greater than $\log M_s/M_\odot > 10$. \textsf{Left Panel}: the sample at z=0 for WINGS (black dots), Illustris-1 (red dots) and Illustris TNG (blue dots) and $\beta$ in the interval $5< \beta < 10$. \textsf{Right Panel}: the same as in the left panel but for $ -20 \beta < -10$.  }
    \label{fig:9}
    \end{figure*}

  \begin{figure*}
   \centering
   { \includegraphics[scale=0.45, angle=0]{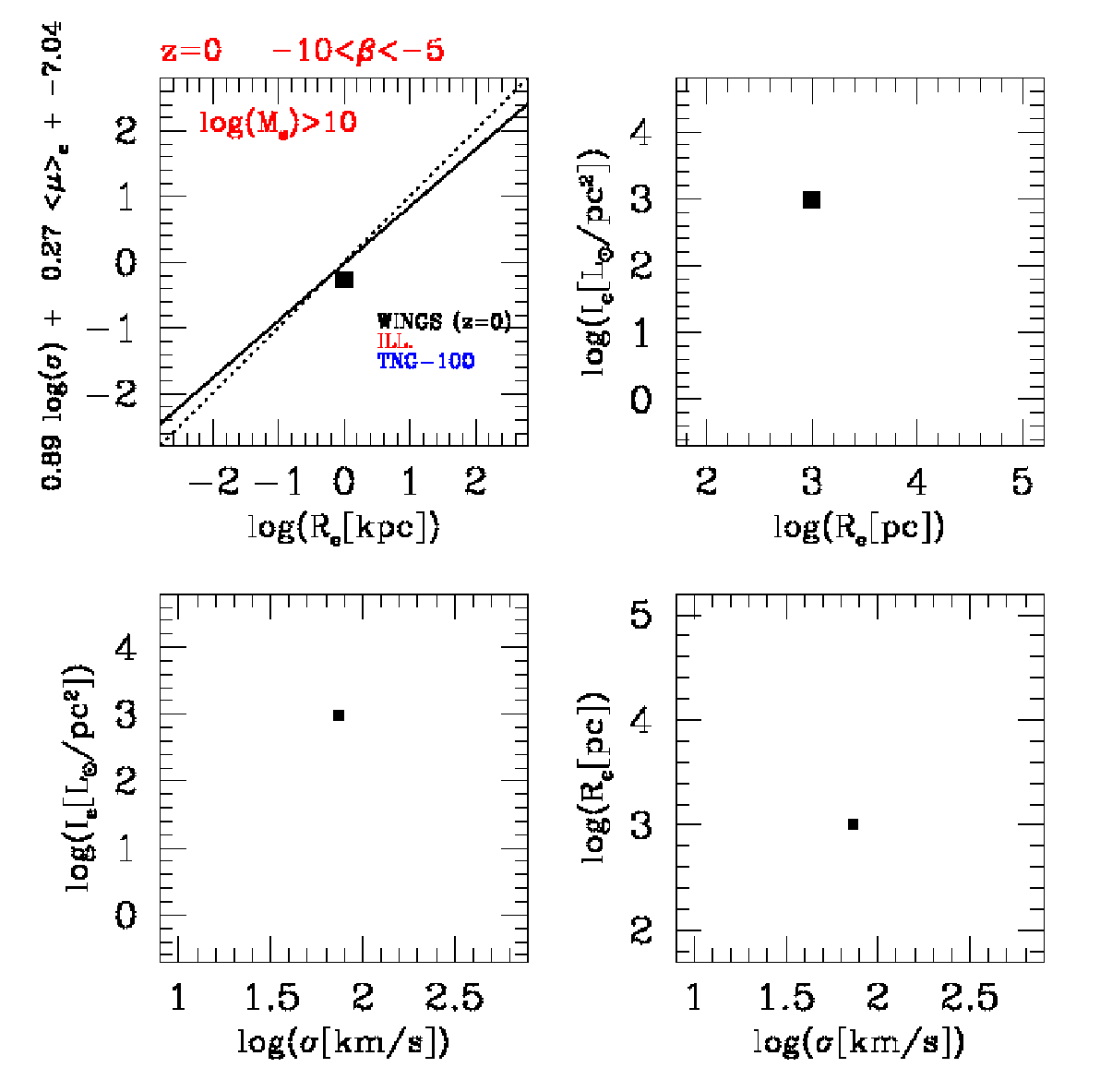}
     \includegraphics[scale=0.45, angle=0]{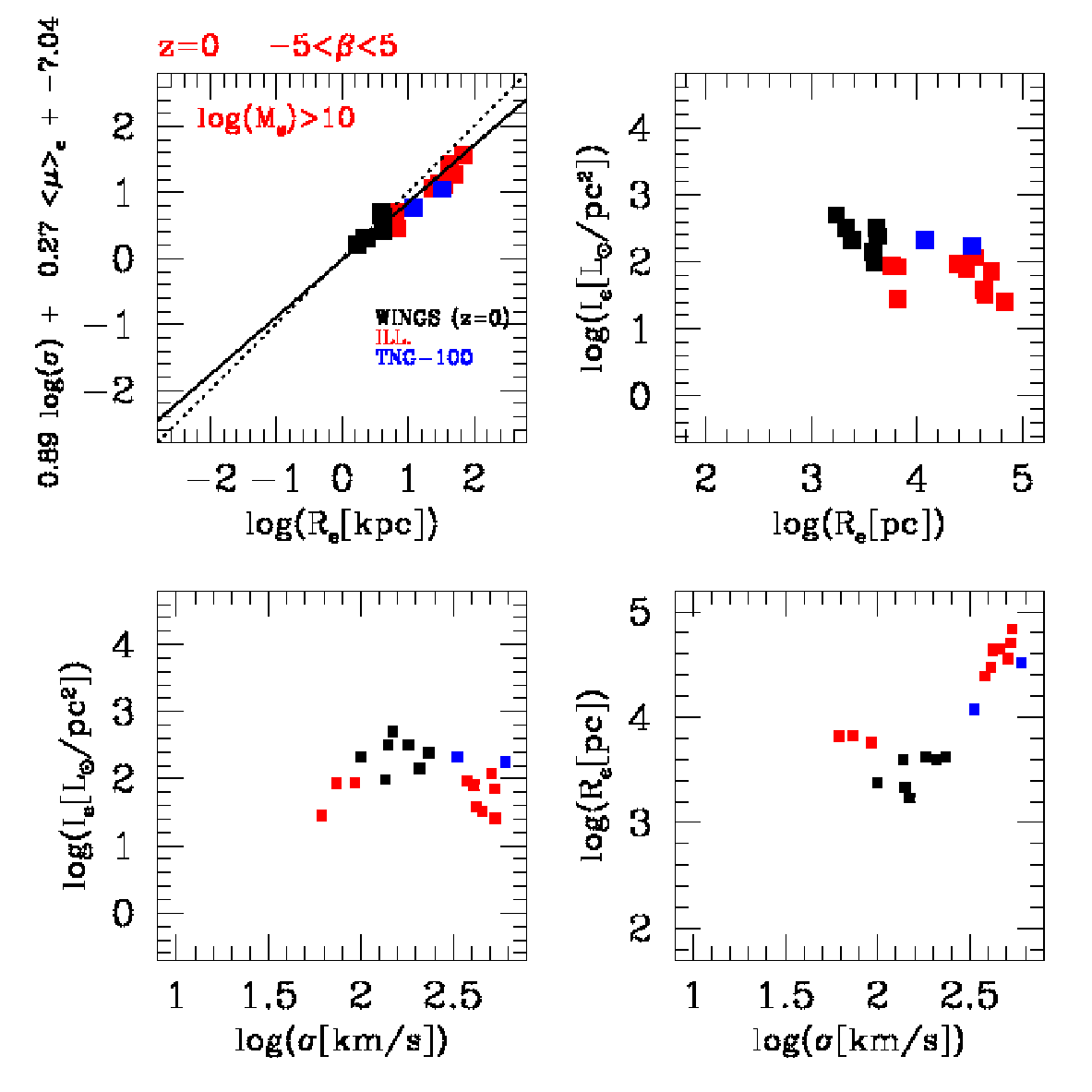}    }   
   \caption{The FP edge-on and its projections for galaxies with masses greater than $\log M_s/M_\odot > 10$. \textsf{Left Panel}: the sample at z=0 for WINGS (black dots), Illustris-1 (red dots) and Illustris TNG (blue dots) and $\beta$ in the interval $-10 < \beta < -5$. \textsf{Right Panel}: the same as in the left panel but for $ -5 \beta < 5$.}
    \label{fig:9bis}
    \end{figure*}

\textsf{Changing the lower mass limit}. For the sake of completeness, we briefly discuss the case in which the lower mass limit of galaxies is pushed up to $\log (M_s/M_\odot) = 10)$. It is obvious that the number of galaxies to disposal drastically decreases (both observational data and model galaxies). 
The equation of the FP for this case is 
\begin{equation}
 \log R_e = 0.89 \log \sigma + 0.27 <\mu_e> - 7.04\, ,  \nonumber
\end{equation}

\noindent
not much different from the previous one. Figures \ref{fig:9} and \ref{fig:9bis} illustrate the situation. The layout of the figures and their sub-panels is exactly the same as before. Compared to the reference case there is not much to say, but the overall conclusions we reached are similar although more difficult to defend because of the paucity of data. \\

\begin{figure*}
   \centering
  { \includegraphics[scale=0.42, angle=0]{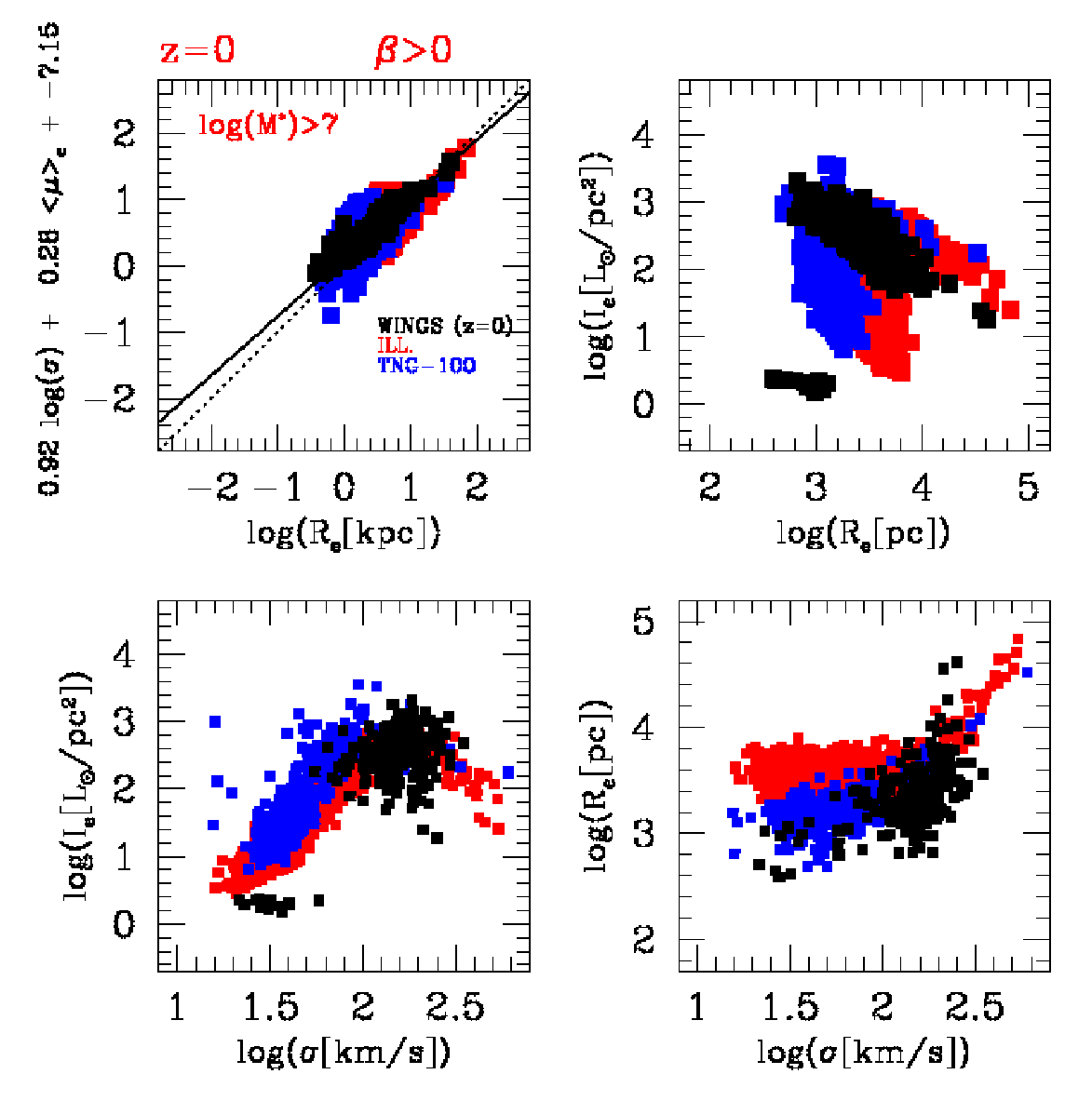}
    \includegraphics[scale=0.42, angle=0]{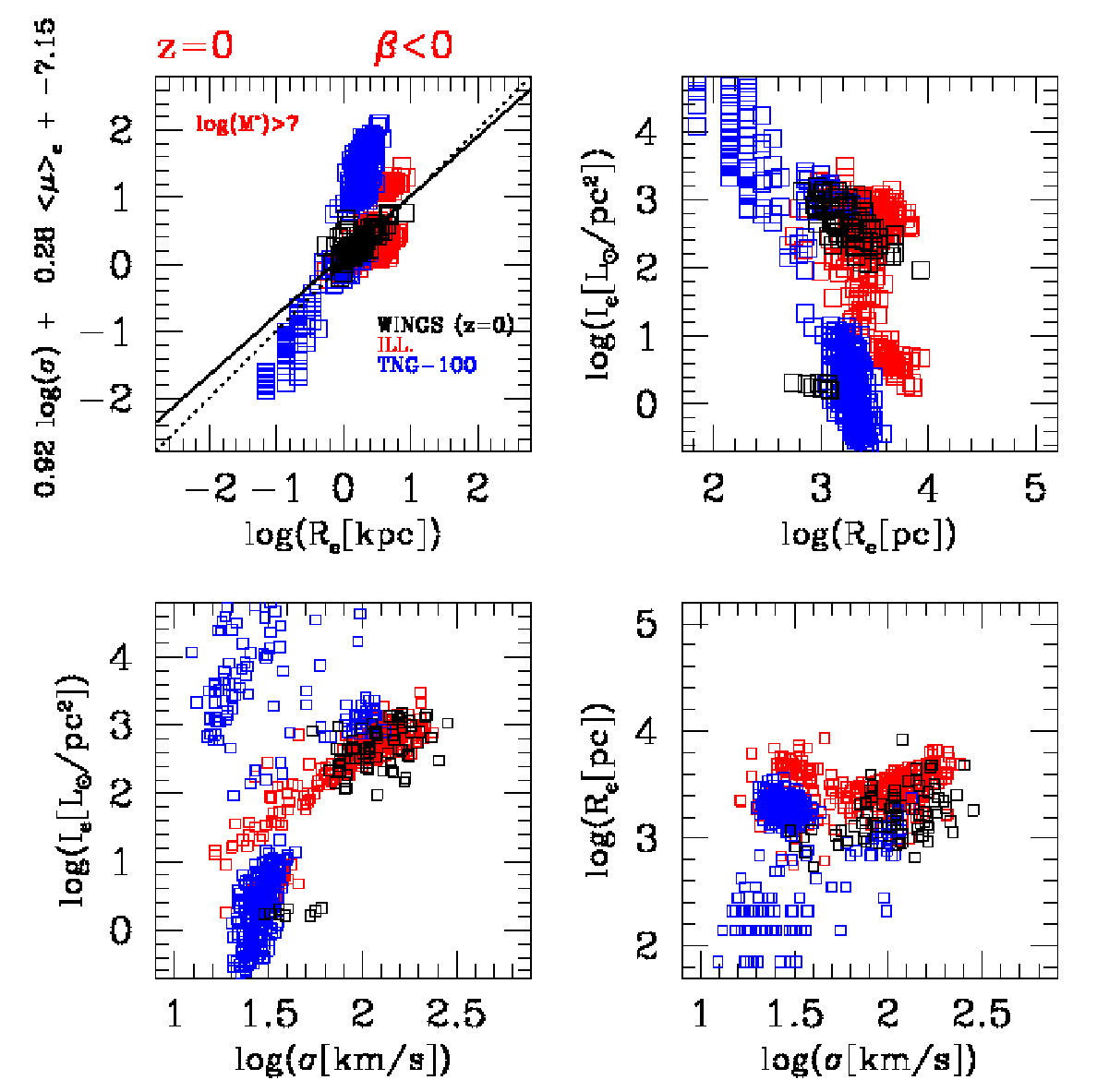}    }   
   \caption{The FP edge-on and its projections at z=0. \textsf{Left Panel}: the sample of WINGS  at z=0 (black symbols), and the samples of Illustris-1 (red symbols) and Illustris TNG (blue symbols) at z=0. All galaxies with mass $\log (M_s/M_\odot) > 7$ and $\beta > 0$ are considered. \textsf{Right Panel}: the same as in the left panel but for $\beta <0$; galaxies are indicated by open squares with the same color code. }
    \label{fig:10prebis}
    \end{figure*}

\begin{figure*}
   \centering
  { \includegraphics[scale=0.42, angle=0]{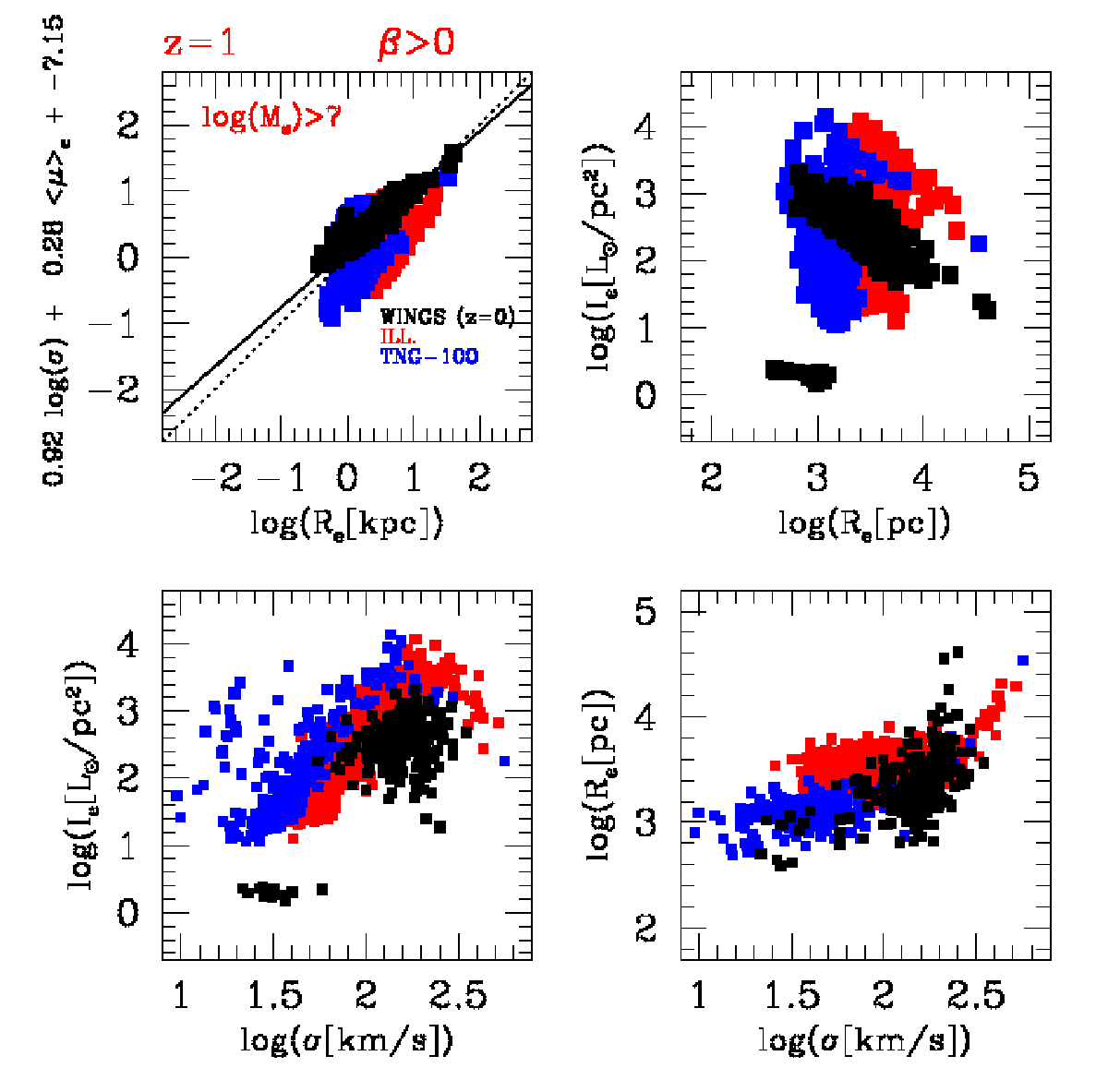}
    \includegraphics[scale=0.42, angle=0]{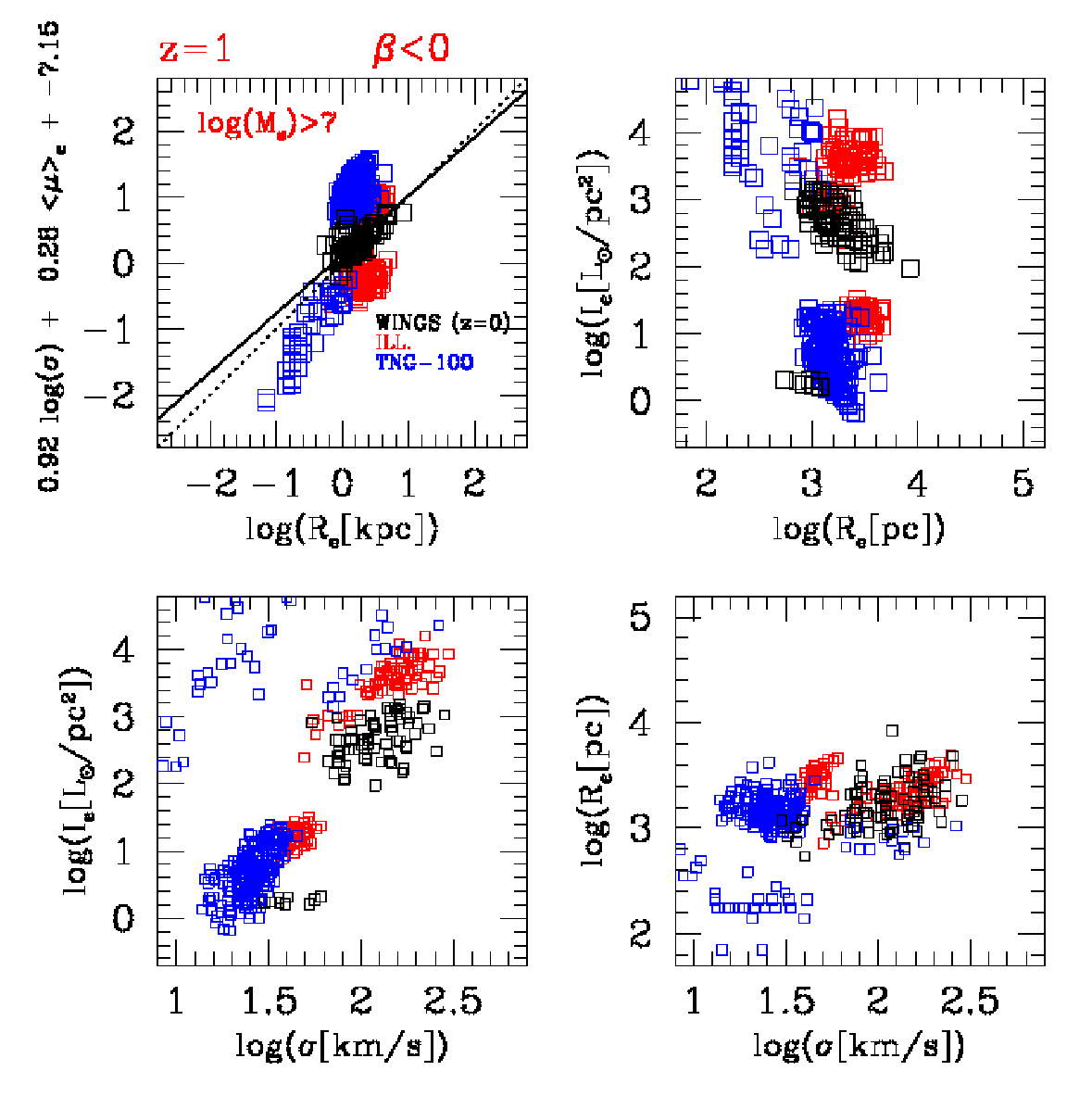}    }   
   \caption{The FP edge-on and its projections at z=1. \textsf{Left Panel}: the sample of WINGS  at z=0 (black symbols), and the samples of Illustris-1 (red symbols) and Illustris TNG (blue symbols) at z=1. All galaxies with mass $\log (M_s/M_\odot) > 7$ and $\beta > 0$ are considered. \textsf{Right Panel}: the same as in the left panel but for $\beta <0$; galaxies are indicated by open squares with the same color code. }
    \label{fig:10}
    \end{figure*}

\begin{figure*}
   \centering
  { \includegraphics[scale=0.45, angle=0]{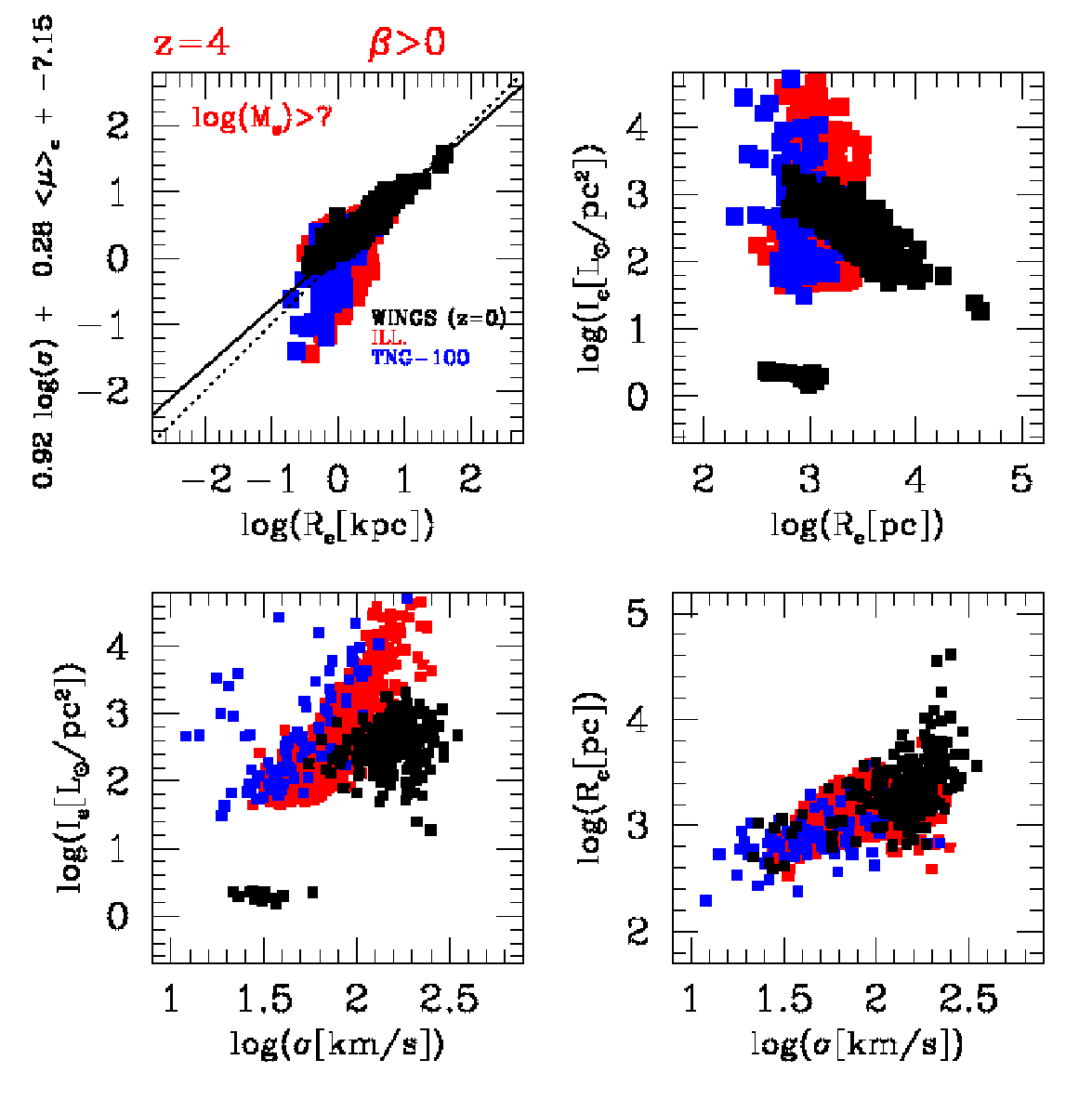}
    \includegraphics[scale=0.45, angle=0]{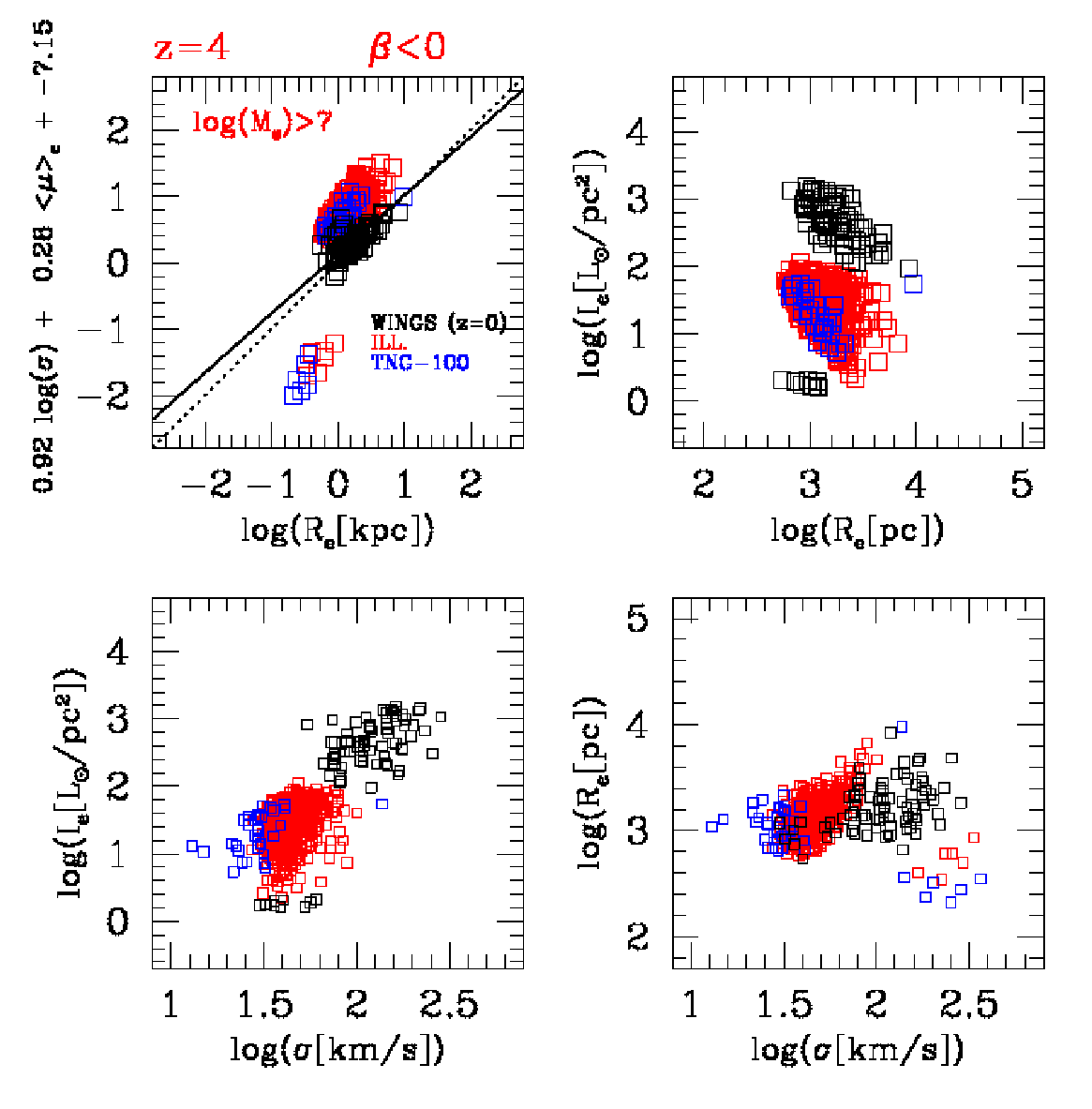}   }
   \caption{The FP edge-on and its projections at z=4. \textsf{Left Panel}: the sample of WINGS (black symbols) at z=0, and the samples of Illustris-1 (red symbols) and Illustris TNG (blue symbols) at z=4. All galaxies with mass $\log (M_s/M_\odot) > 7$ and $\beta > 0$ are considered.
   \textsf{Right Panel}: the same as in the left panel but for $\beta <0$; galaxies are indicated by open squares with the same color code.  }
    \label{fig:10bis}
    \end{figure*}

\textsf{Going to high redshift}. {The situation does not change going to high redshifts. For the sake of brevity we limit the discussion to redshifts z=1 and z=4 (and z=0, for comparison).  The lower mass limit of galaxies is now back to  $\log (M_s/M_\odot) = 7)$ while the FP is the one of the WINGS data at z=0. } Contrary to what amply discussed in Sec. \ref{sec:4}, instead of using the FP equation holding good for the new samples of data (observations and models  with fewer massive galaxies),  we kept also here the same equation for the FP of the reference case at z=0. The uncertainty implicit in this approximation is not relevant in the present qualitative context. Furthermore, to simplify the analysis and presentation of the results we split the data in two groups those with $\beta > 0$  and those with $\beta < 0$. {For the sake of comparison we also consider the case z=0,   Figures \ref{fig:10prebis}, \ref{fig:10}, and \ref{fig:10bis}  shows the FP and its projections at z=0, z=1, and z=4, respectively. } The layout of the figures is always the same. Galaxies with $\beta > 0$ are indicated with filled squares those with $\beta <0$  with open squares. The color code has remained the same (black: WINGS; red: Illustris-1, blue: IllustrisTNG-100). 
It is clear that objects with $\beta>0$ and $\beta<0$ do not share the same distribution.

The distribution of the galaxies changes considerably with respect to that of the WINGS galaxies at z$\sim 0$. Figure \ref{fig:10bis} shows that at z=4 the galaxies with mass greater than $10^7\, M_\odot$ and $\beta>0$ are below the FP, while those with $\beta<0$ are above the plane of the WINGS galaxies at z=0.

The $\beta$ parameter changes also with redshift, but does not seem to much affect the scatter around the FP. At z=0 only a modest scatter around the plane is indicated by the model galaxies of both Illustris-1 and IllustrisTNG-100 databases.

In order to check the relevance of the $\beta$ parameter as a proxy of the degree of evolution (either in terms of galaxy dynamics that in terms of mass accretion and stellar evolution), we present a  plot showing the stellar mass versus the SFR for the Illustris-1 galaxies at z=0.

    \begin{figure}
   \centering
   {\includegraphics[scale=0.45, angle=0]{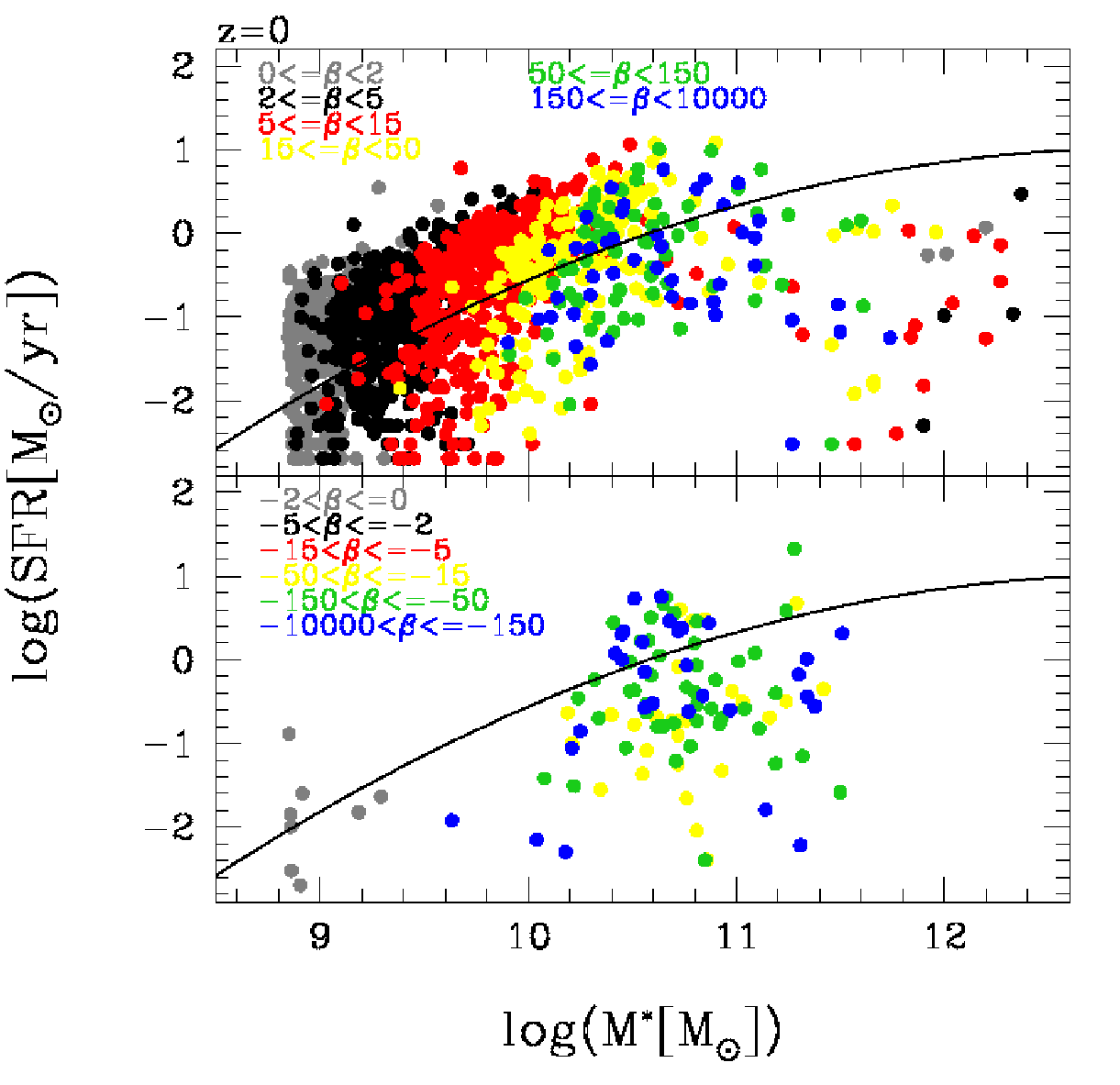}} 
   \caption{The Main Sequence (MS) relation at z=0 for the Illustris-1 galaxies. The different colors of the dots mark the different intervals of $\beta$ (listed in the figure). The upper panel considers only the positive values of $\beta$, while the bottom panel the negative ones. The black solid line is the fit of the data as proposed by \cite{Popessoetal2023}.}
    \label{fig:new2}
    \end{figure}

Figure \ref{fig:new2} shows the Main Sequence (MS) relation. Colors mark different $\beta$ values. It is apparent from the figure that in different intervals of $\beta$ we can find only galaxies of certain masses. In such intervals galaxies might have different values of the SFR. Note in particular: 1) the absence of several negative $\beta$'s along the MS at small masses; 2) the small values of $\beta$ for few big massive galaxies going toward the quenching state of evolution. These objects have likely experienced in the near past episodes of gas accretion and star formation, well testified by the $\beta$ parameter.

We conclude that $\beta$ gives an indication of the stellar mass and the SFR of a galaxy. This information, coupled with the fact that $\beta$ provides the direction of motion of a galaxy in the FP space, gives this parameter the status of a quantitative proxy of galaxy evolution. \\

\begin{figure*}
   \centering
   {\includegraphics[scale=0.45, angle=0]{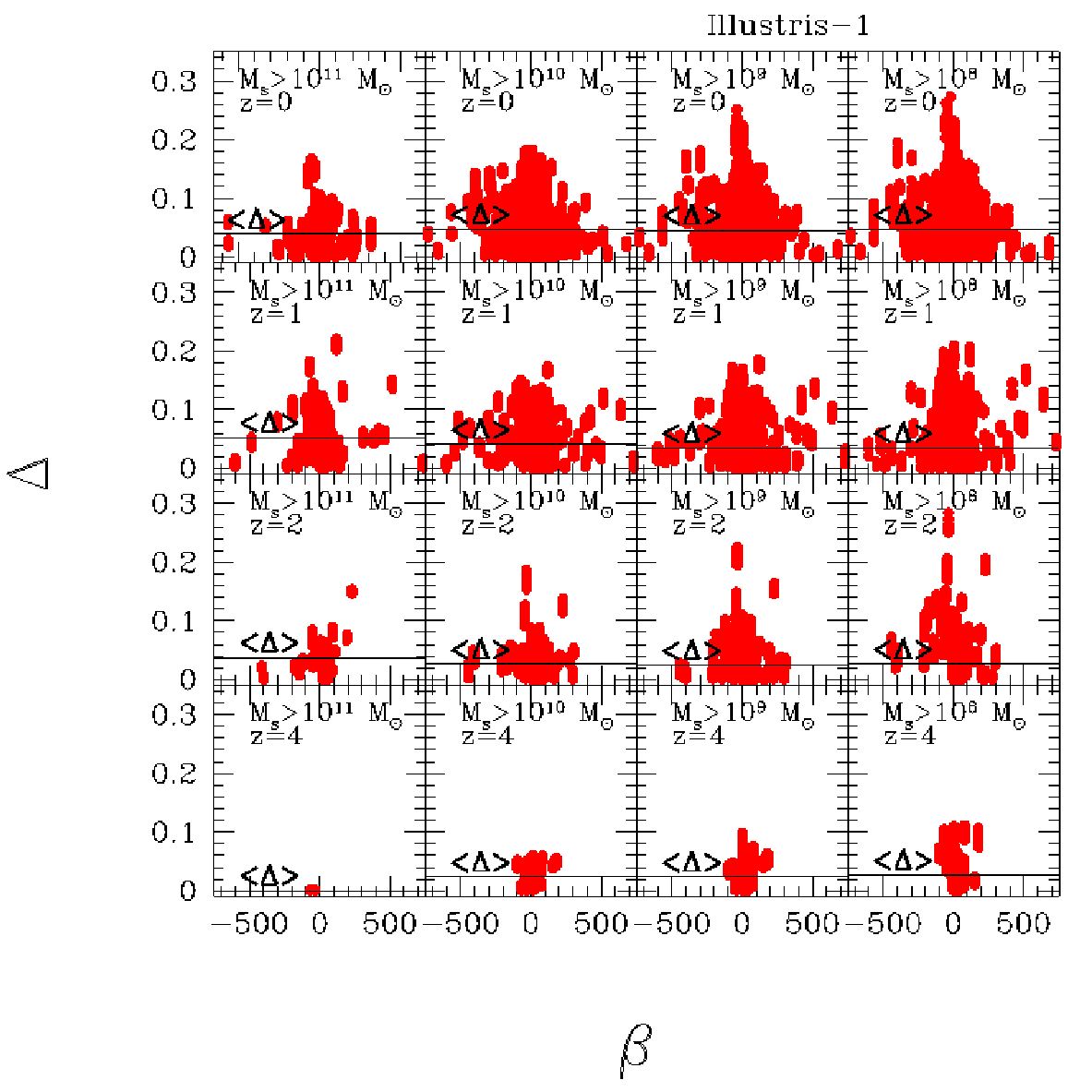}
    \includegraphics[scale=0.45, angle=0]{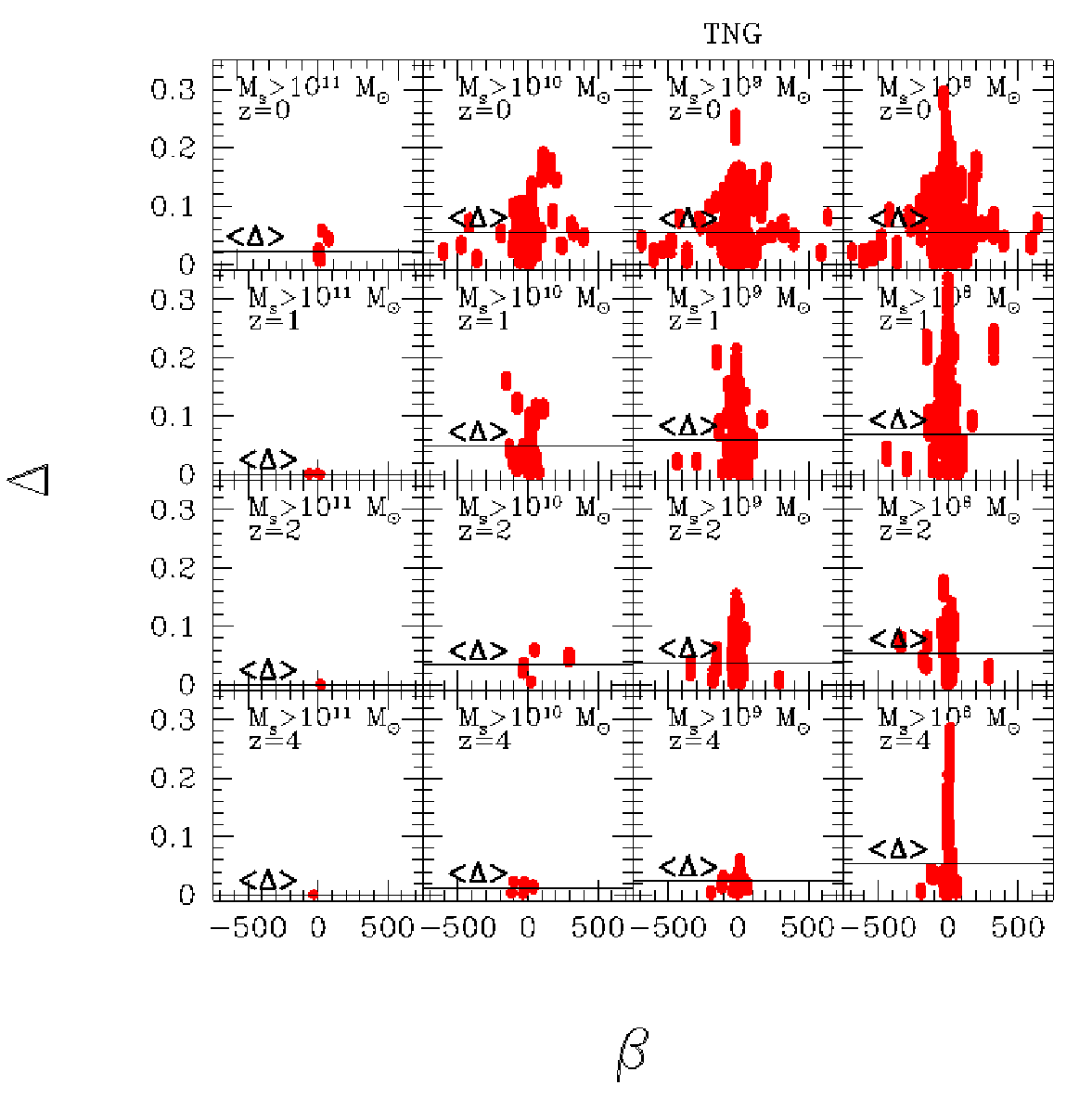}  } 
   \caption{The scatter around the FP as a function of the $\beta$ parameter for different intervals of mass and at different redshift epochs. \textsf{Left Panel}: the Illustris-1 data. 
   \textsf{Right Panel}: the IllustrisTNG-100 data. In the panels the redshft increases from the top to the bottom, while the low mass limit  of galaxies  decreases from left to right. The red little squares show the scatter around the FP, while the solid lines show the mean value of it. }
    \label{fig:11}
    \end{figure*}

\textsf{Scatter around the FP}.  Figure \ref{fig:11} shows the scatter $\Delta$ around the FP against $\beta$ for objects of different masses at different redshifts. The figure displays two groups of sixteen panels. The group at the left side shows the case of the Illustris-1 model galaxies, while the group at the right side does the same for the IllustrisTNG-100 data. In each panel is indicated the low mass limit of galaxies and the redshifts (z= 0, 1, 2, {\rm and} 4). The red areas visualize the spread while the horizontal, black, solid lines show the mean scatter.  In both groups, the spread of $\beta$ increases with redshift. Furthermore, while at z=4 only  small values of $\beta$ are possible, at z=0 both large positive and negative values of $\beta$ exist. The mean scatter around the plane does not change very much in all the redshift. In any case, examining the data and results shown in this figure, one should always remind  that the number of massive galaxies becomes smaller going toward high redshifts. \\

Concluding this section,  we can say that the classical FP is far from being the universal planar distribution valid for all stellar systems. 
The distribution of galaxies in the FP-space changes when objects with different physical conditions and evolution (star formation, mergers, etc.) are taken into account. Each object has its own peculiar position in the FP-space and its projection planes according to its peculiar history of mass accretion,  star formation, and luminosity evolution. 
The effects of these  processes are parameterized by the value of $\beta$ (and $L'_0$). 

The old idea of a universal FP is approximately true for the distribution of the large and massive ETGs, objects that are well virialized and may have large positive and negative values of $\beta$. The fit of these galaxies in the FP-space does not coincide however with the expectation of the theoretical virial plane, because the fit only gives a sort of average value of the possible values of $\beta$ \citep[see][for more details]{Donofrioetal2017}. The light-measured structural parameters do not match the mass-derived parameters.

The Illustris galaxy models suggest that at each redshift a nearly flat distribution in the FP-space is observed for the galaxies in all mass intervals.  In the present framework this simply reflects the fact that the global structural parameters of galaxies are not so far from those achieved by galaxies when the perfect virial equilibrium is reached. The real situation is that galaxies continuously move in the FP-space because of mass accretion, star formation and luminosity evolution, but the mechanical equilibrium is not very far. The $\beta$ parameter helps us to have an idea of the effects of such evolution \citep{Donofrio_Chiosi_2023a}.

While the high massive galaxies always follow the classical FP (they are in general very close to the virialization and have large $\beta$ values, either positive and negative), the smaller galaxies, easily deviate from the ideal virial condition, being their structural parameters strongly affected by feedback effects, mergers and all the other possible physical mechanisms at play during evolution. The most sensible parameter is the effective surface brightness \Ie\ that easily may induce a strong rotation of the FP via its coefficient $b$. Dwarfs might have very different \Ie\ and thus easily change their position in the FP-space.
 
\section{The FP-space and the SFR}
\label{sec:7}

In this section we  tried to understand whether star formation can influence the distribution of galaxies in the FP-space. The SFR of the real WINGS galaxies was measured by \citet{Fritzetal2007} and it refers to stellar activity that  occurred in the last 20 Myrs. For Illustris-1 and IllustrisTNG-100 we used the SFRs provided by the libraries of model galaxies.

    \begin{figure*}
   \centering
  { \includegraphics[scale=0.45, angle=0]{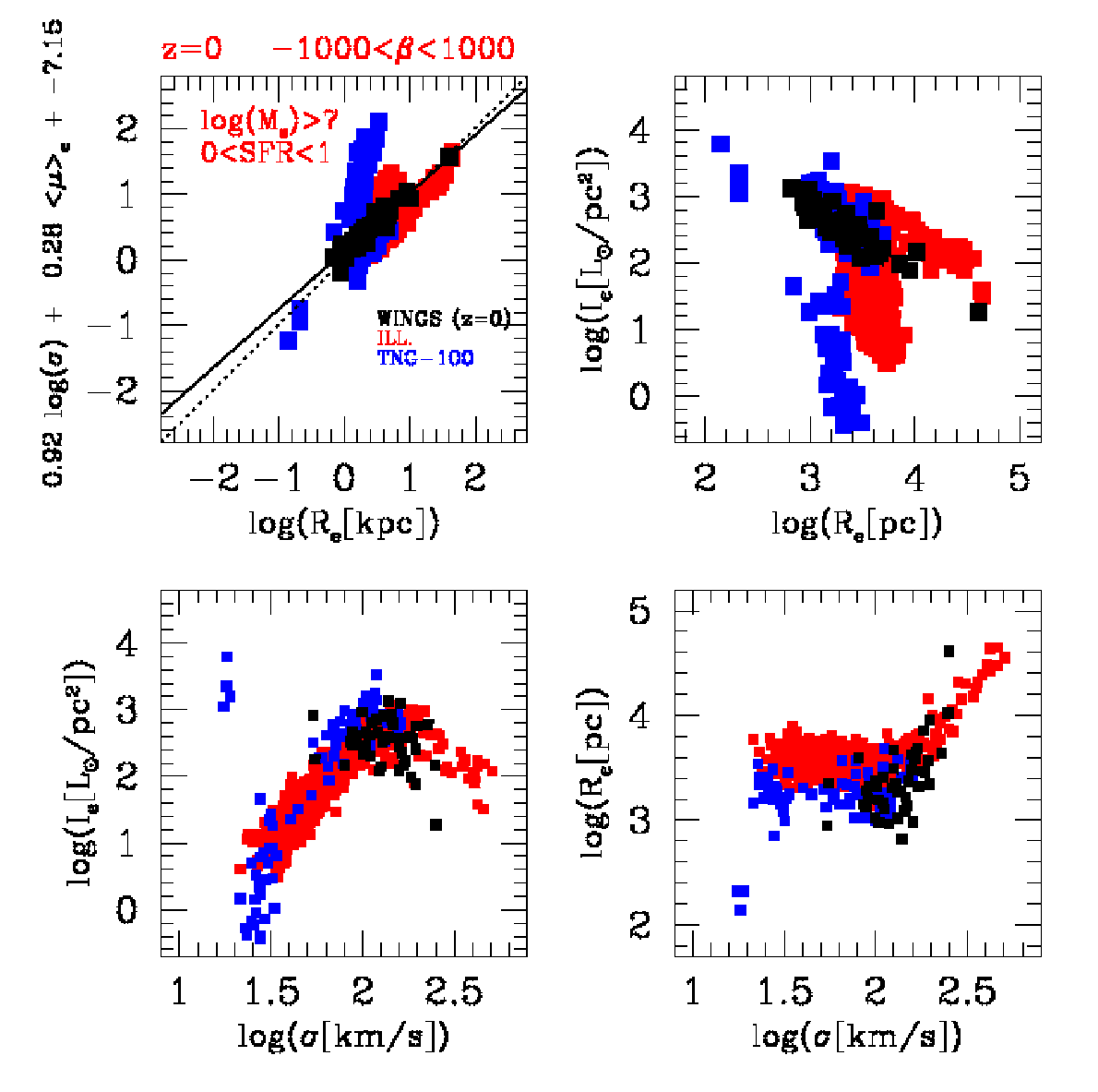}
   \includegraphics[scale=0.45, angle=0]{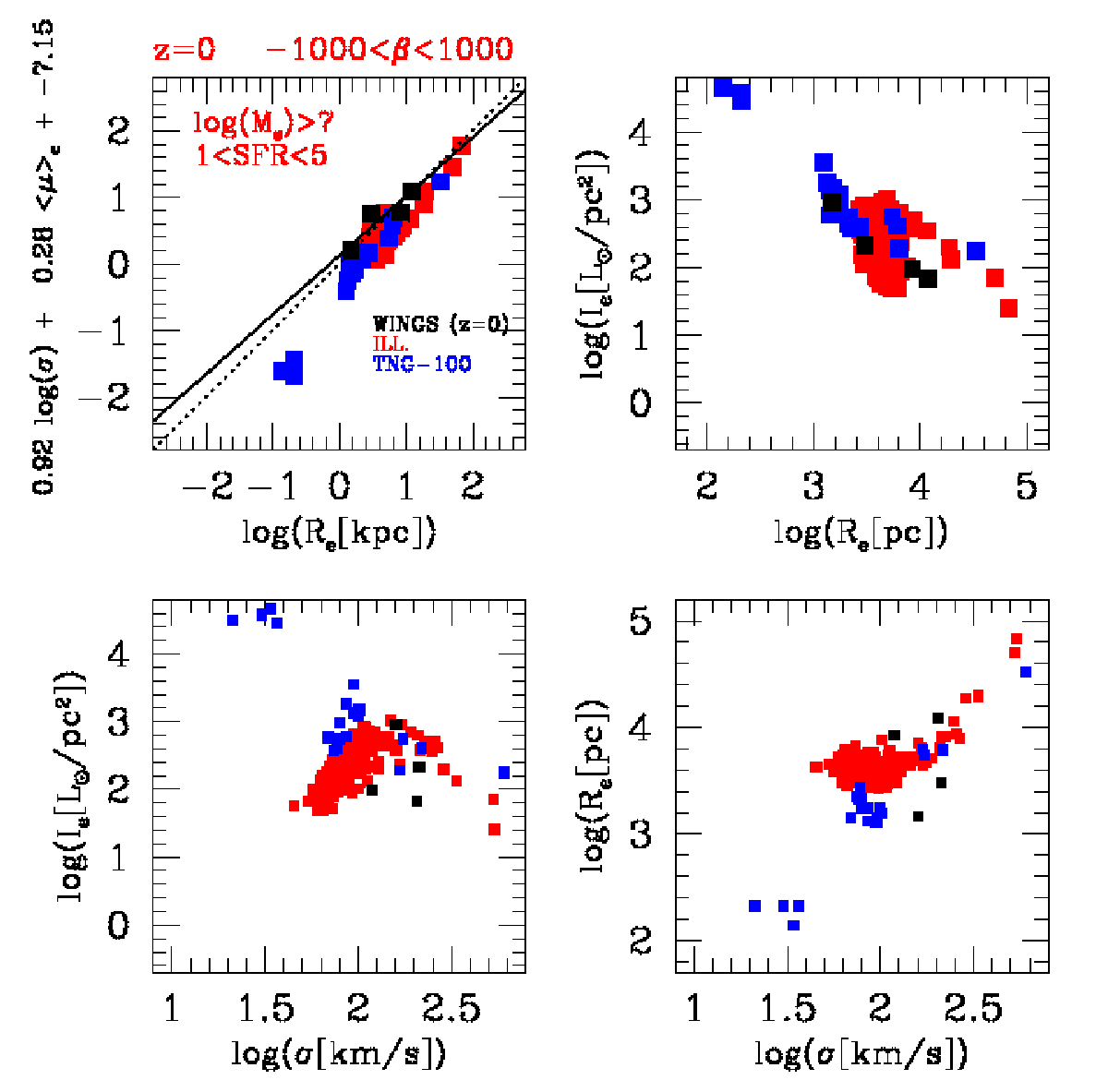}   }
   \caption{The FP-space distribution of galaxies with different star formation rates (SFR) in units of $M_\odot/yr$ and any value of $\beta$ in the interval $-1000 < \beta < 1000$. Two groups of models characterized by different values of SFR are shown.  In each group of panels are shown the FP at top left plus three projections planes as indicated. The black points are the WINGS data, the red and blue points the Illustris-1 and IllustrisTNG-100 model galaxies, respectively. The lower mass limit of galaxies is $\log (M_s/M_\odot) = 7)$. The redshift is z=0. \textsf{Left Panel}: SFR in the range  $0 < SFR < 1$. \textsf{Left Panel}: SFR in the range  $1 < SFR < 5$. }
    \label{fig:12}
    \end{figure*}
    
    \begin{figure*}
   \centering
  { \includegraphics[scale=0.45, angle=0]{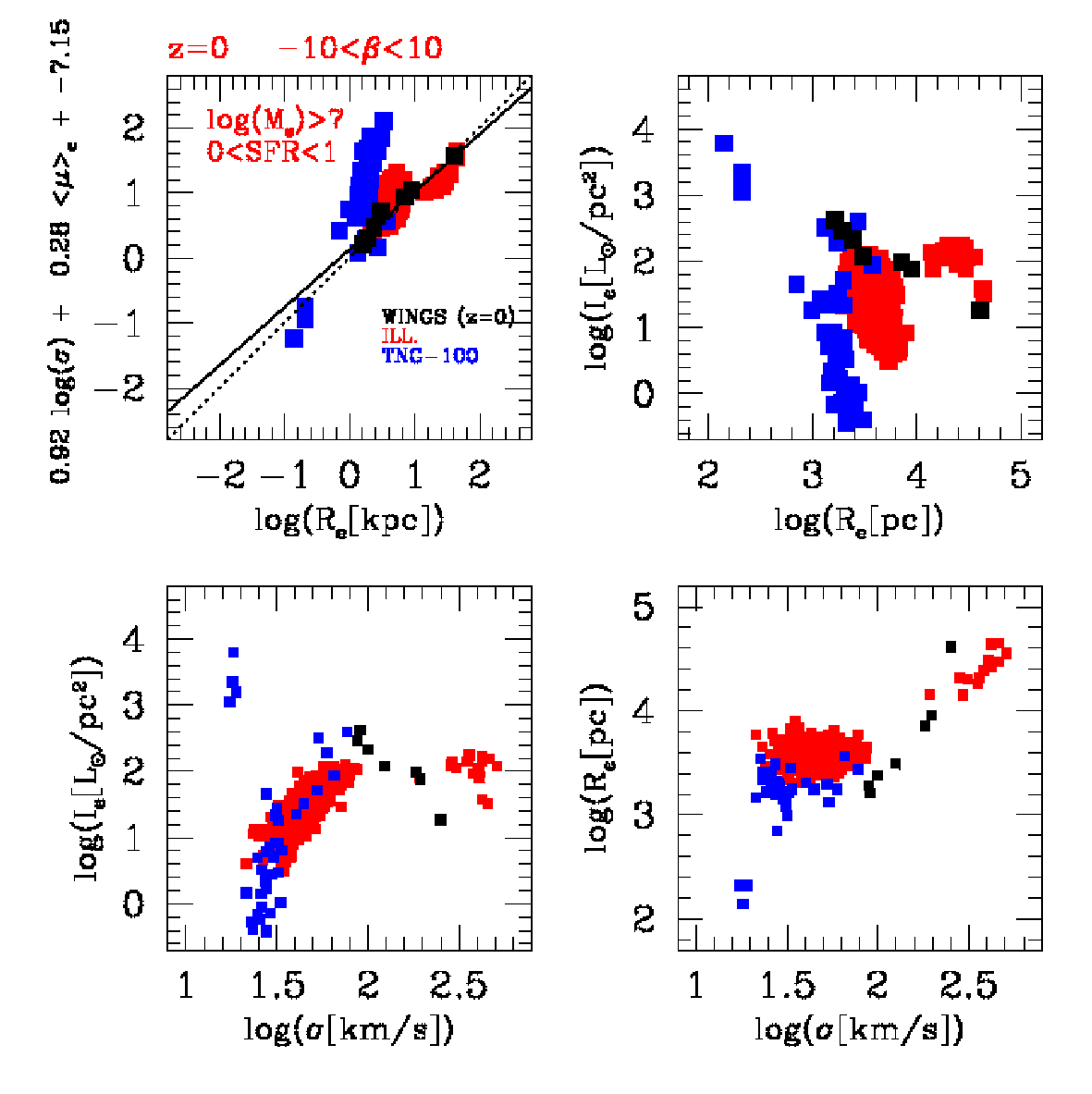}
   \includegraphics[scale=0.45, angle=0]{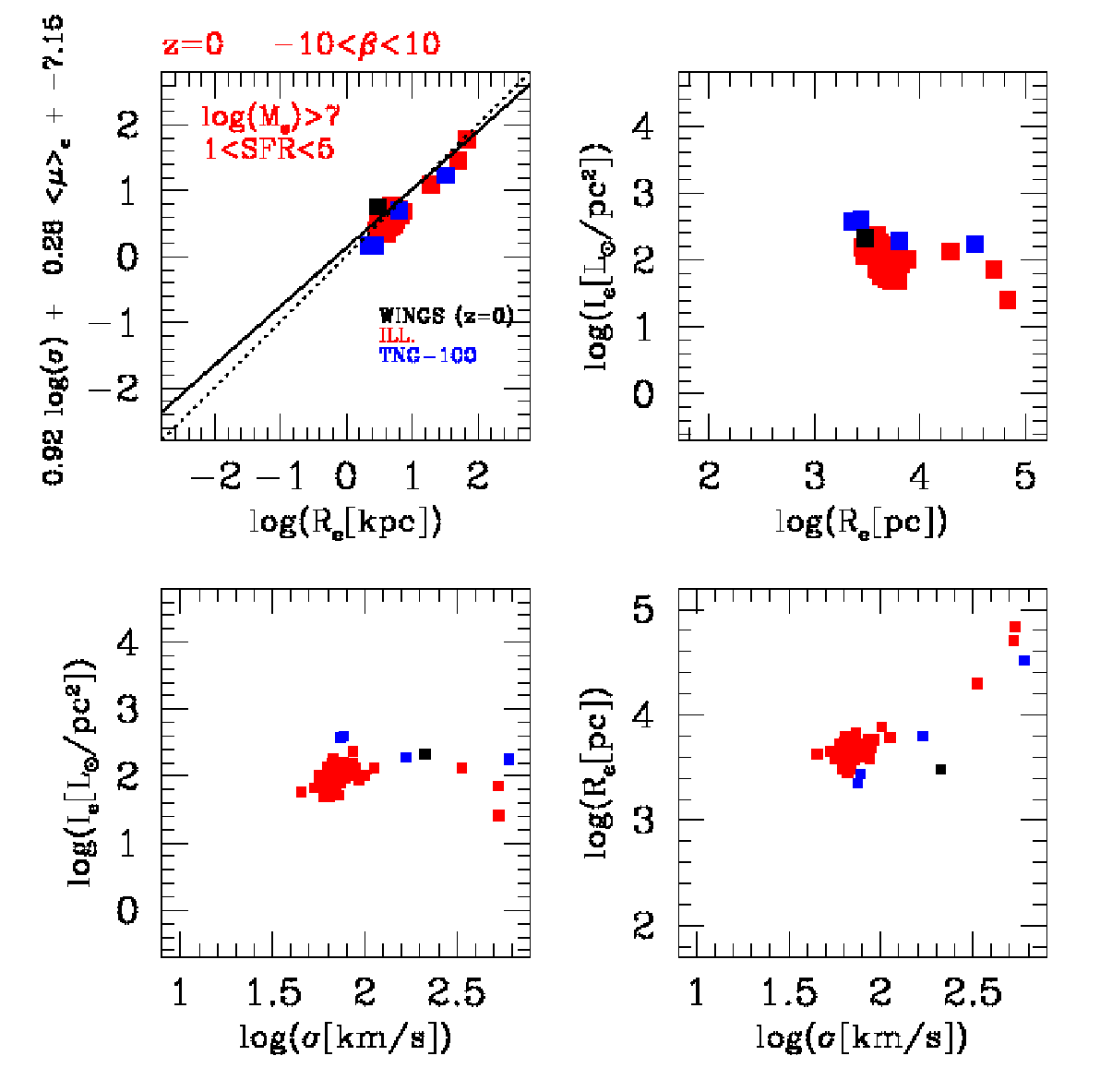}  }
   \caption{The same as in Fig. \ref{fig:12} but for  different interval of SFR and $\beta$. FP-space distribution for small values of $\beta$ and different SFRs. \textsf{Left Panel}: SFR in the range  $0 < SFR < 1$ and $\beta$ in the interval $-10 < \beta < 10$. \textsf{Left Panel}: SFR in the range  $1 < SFR < 5$ and $\beta$ in the interval $-10 < \beta < 10$.}
    \label{fig:13}
    \end{figure*}

Figure \ref{fig:12} shows the FP-space for galaxies of all possible $\beta$s ($-1000<\beta<1000$) and different SFRs. The SFR is measured in $M_\odot/yr$. The organization of the figure is similar to those already adopted in previous sections. The low mass limit of galaxies is $\log (M_s/M_\odot) = 7$ and the redshift is z=0. The left panels are for SFRs in the interval 0 to 1 $M_\odot/yr$, while in the right panels the SFR is between 1 to 5 $M_\odot/yr$. We note that at z=0 when the SFR increases (SFR > 1 $M_\odot/yr$), the upward extension characteristic of the small model galaxies disappears and only few objects with high mass are visible in the FP-space. This is a selection effect due to the small number of galaxies with high SFR at z=0. Only the galaxies with large masses have the possibility of reach SFR>1 $M_\odot/yr$.
Note that few dwarf galaxies with large \Ie\ and SFR>1 $M_\odot/yr$ are still visible below the plane of the large mass galaxies. Most likely, these objects have experienced recent bursts of star formation or mergers or both that have increased the luminosity and the effective surface brightness.

In Fig. \ref{fig:13} we restrict the interval of $\beta$ to $-10<\beta<10$.  The figure has the same layout  of Fig. \ref{fig:12}.  Also here we  note that when the SFR<1 $M_\odot/yr$  all possible masses ($>10^7 M_\odot$) are present, while when the SFR>1 $M_\odot/yr$ only galaxies with high mass are visible.  Once more, we note that in general the star forming galaxies are situated along the FP and not far from this. 

Going to higher redshifts this behaviour does not change. At any epoch the galaxies with the lowest SFR are those forming the upward tail above the edge-on FP, while those with highest SFR fall in  general below the plane.

In analogy to what already found and commented for the distribution of galaxies in the FP and its projections at varying the parameter  $\beta$, choosing the galaxy sample on the basis of the measured SFR generates different distributions of galaxies in the FP-space.
Our conclusion here is that the FP and its projections are good indicators of the evolutionary conditions of galaxies.

{There are two obvious remarks that could be made to the above analysis and conclusion. The first one is why we did not follow galaxy by galaxy of the theoretical samples the merger tree of a candidate galaxy for which the progenitors are known across the cosmic time (redshift). In such a case, all the physical variables $L$, $Re$, $I_e$, $M_s$, $\sigma$, $\beta$, $L'_0$, and SFR  would be known (albeit at discrete redshifts) so that the position on the FP at that redshift would be known. This kind of analysis was already made in 
\citet{Donofrio_Chiosi_2023b}, however limited to a small number of objects (about ten) taken from Illustris-1 sample and limited to the correlation SFR-$\beta$. The result was that $\beta$ remained small at all redshifts for most of the objects, with the exception of two of them in which $\beta$ got very large (both positive and/or negative) only at low redshift. As already anticipated, the merger three is not yet available for the IllustrisTNG-100 suite of models to our disposal (work is in progress). To cope with this, in the present study, we adopted a different strategy, that is to use all the  objects contained in the samples at each redshift, to select them according to  the mass, SFR, and $\beta$, and to plot the filtered objects on the FP at the redshft under consideration. In such a case, 
the simulations show that models with different SFR and $\beta$ occupy different regions of the FP. In this sense $\beta$ is a tracer of the mean SFR history in galaxies of a certain mass.  
The SFR history is traced on statistical sense and not object by object.

The second objection, is that the SFR taken from the simulation catalogs is an instantaneous measure of the SFR.  If the SFH is bursty because of the mergers, there is no reason to believe that this value should be related to the structural properties of the galaxies, especially at high redshifts when mergers are likely more frequent. Although this objection sounds correct, it is ill posed and may arise misunderstanding. A merger of two galaxies of given mass may cause star formation in addition to that already taking place in a galaxy due to internal reasons (induced versus spontaneous star formation). The induced star formation can be parametrized by two timescales and a specific intensity.  The first timescale is the duration of a merger (the SF trigger), this time is of the order of the dynamical timescale, or the crossing time scale, or the free-fall time scale. Current estimates set this timescale at about a few $10^8$ years. Compared to the typical evolutionary timescales of stellar populations in a galaxy, it can be neglected, The star formation trigger can be considered as an instantaneous event. The second timescale  is the time required to evolve the burst down to the typical age of the background stars already in situ. This can be estimated to be about 1 Gyr (i.e. the lifetime to pass from a turnoff mass of 20 $M_\odot$  to a turnoff mass of about 1.5 $M_\odot$). Finally, we have the intensity of the burst that can be measured by the amount of gas mass converted into stars. Since a burst of star formation is ultimately detected and measured by the variation (increase) of the luminosity of the galaxy, it is important to know how the total luminosity of a galaxy resulting from a merger  varies as a function of time. The key quantity to look at is the luminosity per unit mass of a generic event of star formation: this luminosity decreases with increasing time, it may vary by a factor of thousand  over the timescale of about 1 Gyr (see above). The fading rate of a star formation event is very rapid at the beginning and it slows down as it gets older. \citet{Chiosi_Donofrio_Piovan_2023} and \citet{Donofrio_Chiosi_2023b} have developed a simple model providing the correct final variation in the luminosity caused by a typical merger. In brief, a merger with induced star formation is thought of as three-players game the two merging galaxies of mass and luminosity $M_1$ and $L_1(t)$ and $M_2$ and $L_2(t)$, respectively,  plus  the newly born star with mass and luminosity $M_3$ and $L_3(t)$. The total mass and luminosity are {$M=\sum M_j$ and $L(t)=\sum_j L_j(t)$ } where $j=1,2,3$. Introducing the fractionary masses $m_j$ and luminosities $l_j(t)$, the definition of which is obvious, the effect of a merger with new star formation will depend on $m_j$ and $l_j(t)$. In a typical merger of two objects the  mass ratio is $M_1:M_2=10:1$ (or higher) while $M_3/M_1=0.01$ (or smaller). Apart from exceptional circumstances (e.g. very recent merger-burst), the majority of mergers would produce very little effect of the total luminosity emitted by the complex.  The only case in which a sizable effect on $L$ would be present is a merger among objects of nearly equal mass (a rare event) or just an ongoing burst of star formation. 
The Illustris simulations provide data at increasing time intervals in general longer than about 1 Gyr, so that it is extremely unlikely to catch a model galaxy at the peak of the SF burst. 
In conclusion, the SFR provided by the model simulation is the value in a situation far from ongoing mergers and/or bursts.   In any case, both continuous and/or bursting SFR change $L$, $I_e$, $M$, $R_e$ and $\sigma$,  and in turn $\beta$ and $L’_0$ and therefore the position of the galaxy on the FP.  In other words, the position of a galaxy on the FP can hint the value of the underlying $\beta$ and SFR. }

\section{The robustness of the $\beta$ parameter}
\label{sec:8}

We established that the $\beta$ parameter is a reliable indicator of a galaxy's historical star formation and evolutionary trajectory. A value of $\beta$ near zero suggests that the galaxy is significantly far from full virial equilibrium. This can be attributed to various factors such as AGN or SNe feedback effects, galaxy interactions, mergers, or environmental influences.

On the contrary, both large positive and negative values of $\beta$ imply that the galaxy is much closer to full virial equilibrium, indicating that evolutionary effects play a less significant role.

This leads us to rise question of the robustness of the $\beta$ parameter obtained from the solution of the fundamental equations of the VT and the \Lsigbtempo\ of {appendix \ref{Appendix_A} }
\citep[see also][for more details]{Donofrio_Chiosi_2023a}:
We developed  two distinct methods to estimate the uncertainty associated with $\beta$. 

\begin{figure}
   \centering
   \includegraphics[scale=0.4, angle=0]{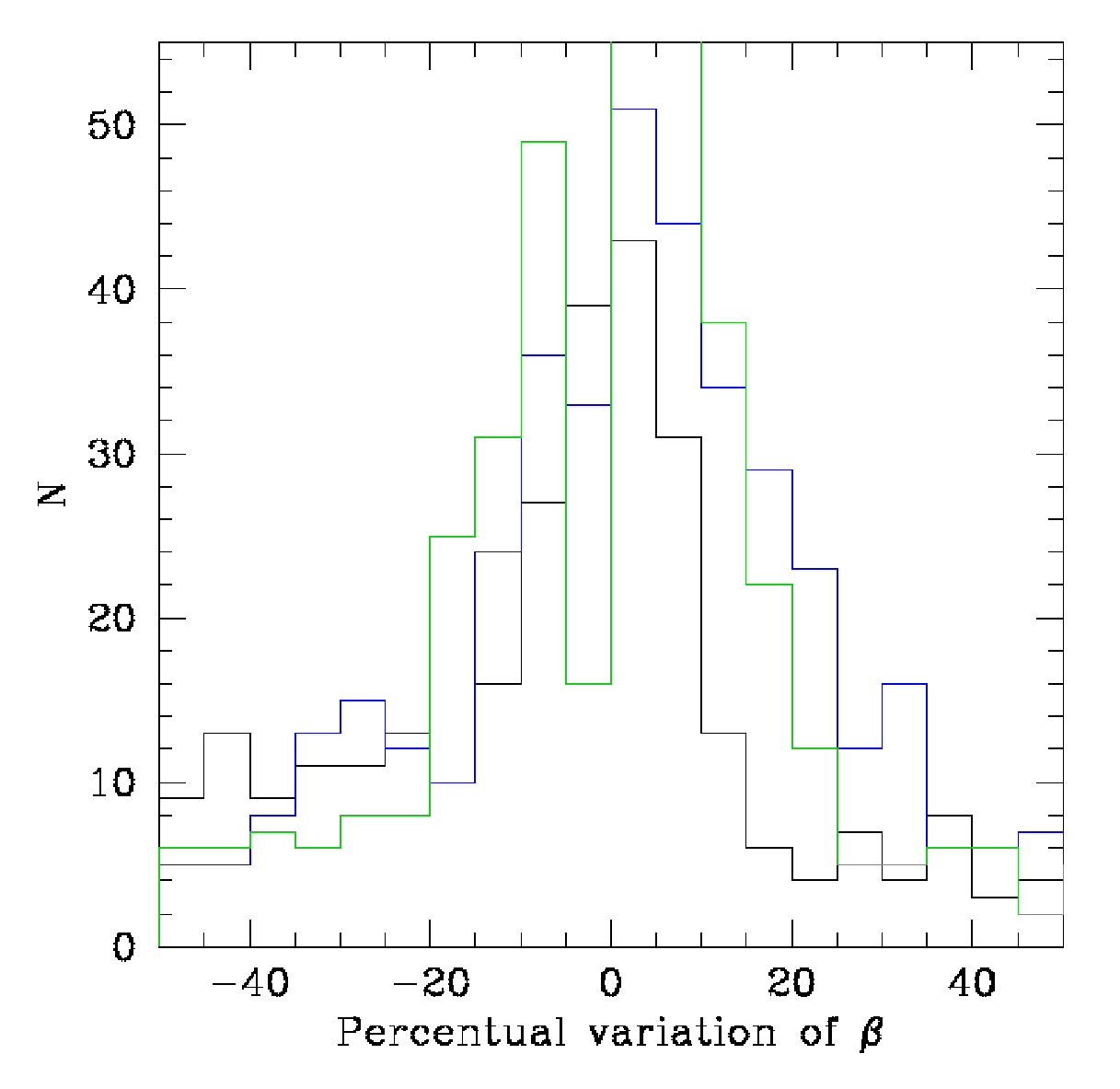} 
   \caption{The percentage  change of the $\beta$ parameter. The black histogram shows the results of our Montecarlo simulations of the errors affecting the observational parameters ($M_s,\, L,\, R_e,\, \sigma,\, Ie$ and their effects on $\log L'_0$ and $\beta$). The blue histogram shows the variations with the analytical method and Gaussian random distribution of the uncertainties affecting the characterizing parameters of each galaxy that is  described in Appendix \ref{Appendix_B} obtained from the calculus of the derivatives. The green histogram shows the results obtained using the analytical method with the flat random distribution of uncertainties (see also Appendix \ref{Appendix_B}). }
    \label{fig:14}
    \end{figure}

In the first approach, we introduced random errors in each of the parameters in eqns. (\ref{eq3}) and (\ref{eq3}), ranging from approximately 0\% to 20\%. The mathematical details on the random perturbation of the input parameters of observed galaxies (WINGS data, 479 objects) are given is the appendix \ref{Appendix_B}.  Using this technique we generated 200 simulations with these varying errors, and obtained the number frequency distribution of the $\beta$ parameter, along with its average value and standard deviation. The results are shown in Fig. \ref{fig:14} which presents the histogram of the expected percentage change of $\beta$ due to the introduction of these errors (the solid the black line). It is soon evident that $\beta$ can fluctuate within approximately a 10\% interval in both positive and negative directions, with a mean variation of around $-2$\%.

The blue and green lines in the same Fig. \ref{fig:14} shows the results of our second method, which assesses potential variations of  $\beta$ due to errors in the structural parameters. This method relies on the error propagation theory and requires the calculation of the partial derivatives of the formal relationships for $\beta$ and $\log L'_0$  with respect to the parameters $M_s$, $R_e$, $L$, $I_e$ and $\sigma$ recalling that all other parameters $L$, $R_e$ and $I_e$ are each other related by the definition of $I_e \propto L/Re^2$ and $M_s$, $R_e$ and $\sigma$ are related by the VT. The methods requires also the calculations of variances and co-variances. All details of the errors evaluation, analytical passages, and  calculations are provided in the appendix \ref{Appendix_B}. Two cases are considered for the random error distribution. In the first case we adopt a Gaussian random distribution mimicking the PSF of the photometry (blue solid line in Fig. \ref{fig:14}). In the second case, we adopt a flat random distribution of the uncertainties affecting the observational parameters (green histogram in Fig. \ref{fig:14}).  See the Appendix \ref{Appendix_B} for all details. The agreement between data and simulations is fairly good. Notably the gaussian and flat cases give approximately the same uncertainties.

\begin{figure}
   \centering
   \includegraphics[scale=0.4, angle=0]{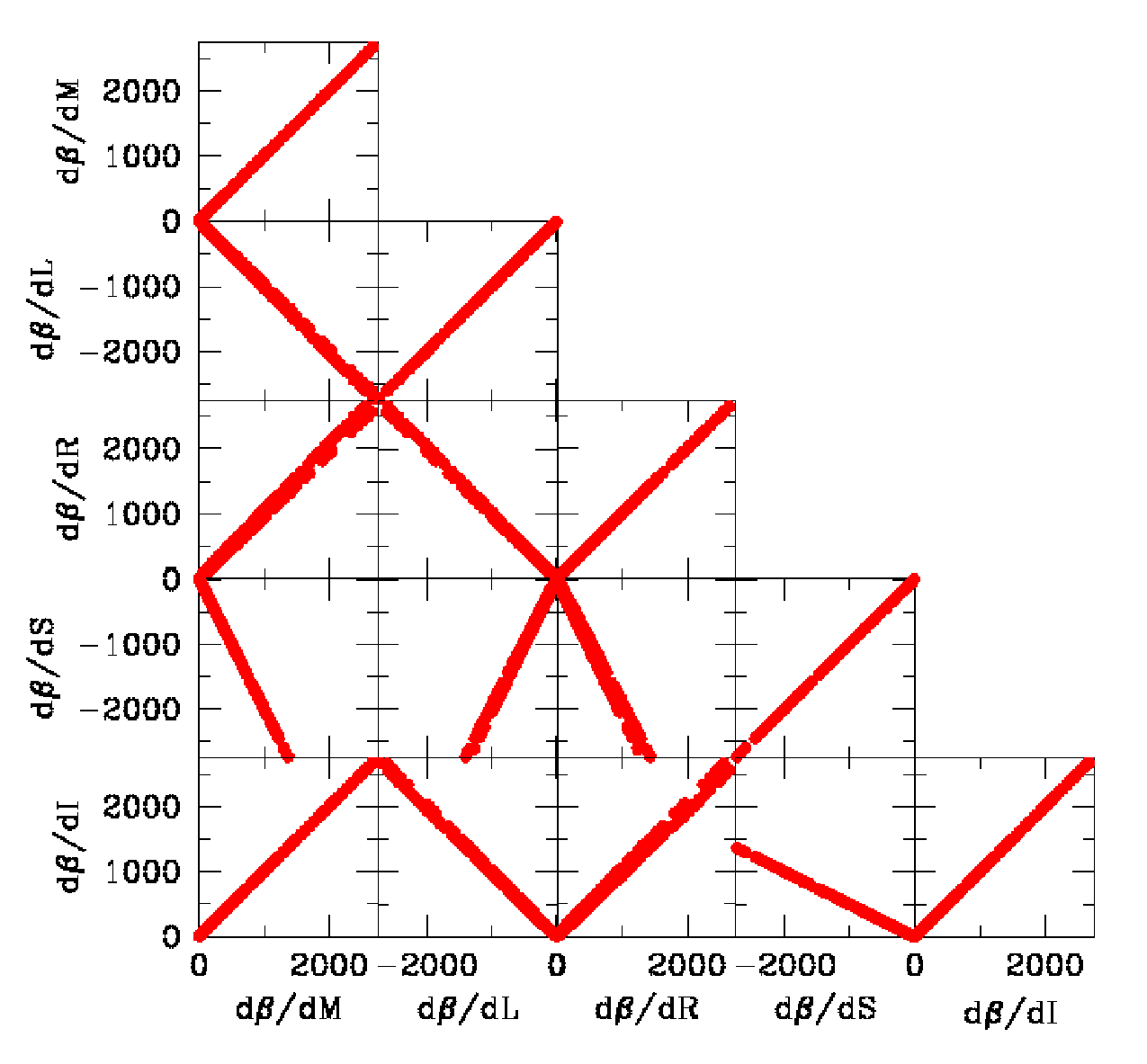}     
   \caption{The mutual dependence of the derivatives of $\beta$ with respect to all the structural parameters. }
    \label{fig:15}
    \end{figure}

Figure \ref{fig:15} highlights the expected mutual correlations  among these derivatives. The high correlation is a consequence of these derivatives being linear combinations of the various structural parameters. Furthermore, while the absolute values of these derivatives are high, they tend to cancel each other out, resulting in a percentage variation in $\beta$ that closely follows  the simulation-based results. This secures that the expected percentage frequencies of the values for $\beta$ are on solid ground. 

\section{Discussion and conclusions}
\label{sec:9}
In this study we investigated the evolution of the Fundamental Plane and its projections on the $I_e$-$R_e$, $I_e$-$\sigma$, and $R_e$-$\sigma$ planes across cosmic times (redshifts). The analysis has used the Illustris-1 and IllustrisTNG-100 libraries together with the observational WINGS data of real ETGs for comparison. Finally, the observational and theoretical data $L$, $R_e$, $M_s$, $I_e$ and $\sigma$ have been used to derive the theoretical parameters that encode the past evolution of a galaxy, namely $L'_0$ and $\beta$ that are derived from coupling the VT with the generalised luminosity-$\sigma$ relation \Lsigbtempo. 

{Before passing to conclusions, we call attention on a few points of this study that could be subject to criticism: 

i) In real galaxies, the light emitted by their stellar populations can be attenuated by the dust component of the their gas content. The same may happen to the light emitted by galaxies traveling across the cosmic medium toward the observer. In other words, internal and external dust can attenuate the light emitted by a galaxy thus affecting the luminosity and all other variables that are luminosity dependent (i.e. $I_e$, $R_e$, $M_s$, SFR, etc.) thus affecting the determination of the FP and $\beta $ in turn. Unfortunately, the current observational and theoretical data do not take dust into account. This indeed is a point of uncertainty that cannot be easily cured at the present time.

ii) The observational data we are using to compare our theoretical predictions are somewhat old and limited to redshift $z\simeq$0. In literature, there are observational data exploring the FP at z$>$0 that could be used in our analysis, see for instance \citet{Luetal2020}, \citet[][LEGA-C Survey]{deGraafetal2021}, \citet[][SDSS-IV MaNGA]{Sanchezetal2022}, \cite{vanDeSande_2013}, and \citet{Stockmann_etal_2020}. However, in this paper, first we used the WINGS data for the sake of continuity with previous papers of this series, second a tight comparison with  observational data at all redshifts was somehow beyond our aims as the main focus was to highlight the potential ability of the  $\beta$-$L'_0$ method in tracing the SFR history of galaxies.

iii) To derive $\beta$ and $L'_0$ for the Illustris-1 and IllustrisTNG-100 simulations, we kept the virial factor $k_v$ constant; see the equation of appendix \ref{Appendix_A}.
This quantity depends on the Sersic index $n$ that is not given in the {Illustris-1 and IllustrisTNG-100 databases.}  Considering the typical range spanned by the Sersic index  in real galaxies we have taken the mean value of n = 4. The maximum uncertainty introduced by $k_v$ is about 0.5 in log units.  In any case uncertainties of this order on $\beta$ are not enough to explain the large range of values spanned by $\beta$. Furthermore,  the analysis of the uncertainties on $\beta$ and $L'_0$  caused by  uncertainties on all other parameters ($L$, $I_e$, $R_e$, etc.) shows that the overall effects on  $\beta$ is rather small. Therefore, the large dispersion shown by $\beta$ ought to be real and due to physical causes ($I_e$ is likely the main driver). 

Given these premises, the main results of this study can be summarized as follows: }

   \begin{enumerate}
      \item At any cosmic time there is a plane best fitting the distribution of galaxies in the FP-space;
      
      \item The best fitting plane, i.e. the FP,  is not the same for galaxies of different masses: low mass galaxies do not share the same plane of massive galaxies;
      
      \item The coefficients of the best fitting plane, i.e. the FP,   change with the mass of the considered objects  and  the redshift;
      
      \item The best fitting plane obtained at each epoch is  the FP holding at that epoch, the analog of the FP plane inferred from objects at z=0; 
      
      \item The scatter around the FP is approximately the same for both simulations and does not change significantly with the cosmic epoch;
      
      \item The scatter around the plane of the large massive galaxies is in general smaller than the scatter of the smaller low mass galaxies;
      
      \item The distribution of galaxies in the FP-space changes with the mass of the galaxies considered and with the values of $\beta$, the parameter that gives a quantitative idea of the evolutionary stage and physical condition reached by a galaxy.
      
      \item There is a tight correlation between the rate of star formation and the existence and position of galaxies in the edge-on FP. When the star formation is low (roughly lower than $1\, M_\odot/yr$) galaxies of any mass (above $\log (M_s/M_\odot) < 7$) are visible in the FP-space, while for  $SFR > 1\, M_\odot/yr$ only the massive galaxies are visible along the edge-on FP. In any case, the scatter around the plane is small.
      
      \item { About 50\% of galaxies in the sample have the $\beta$ parameter confined in the interval $-20 < \beta < 20$, all the rest falls outside this range (on both positive and negative side).} Since galaxies are supposed to be in perfect virial equilibrium (i.e. their $I_e$, $R_e$, $L$, $M_S$ are compatible with the velocity dispersion), they should have very large values of $\pm |\beta|$. Our results suggest instead that this fraction of objects is still far from such condition. Since the classical mechanical equilibrium expressed by the VT is likely secured for most of time, this implies that the variations in luminosity (and hence $I_e$) that follow recent episodes of star formation (in which $M_s$ and likely $R_e$ have also changed) are still on the way back to recover the equilibrium condition. In other words the velocity dispersion derived from the combination of the photometric parameters is not exactly equal to the value derived from the Virial Theorem.
 
      \item
    { Full understanding of the relationship between  $L'_0$, $\beta$, and fundamental physical phenomena taking place in a galaxy, e.g. mergers, star formation both spontaneous and induced, natural evolution of the stellar content, and others is still missing despite the many available hints. In brief we know that: (i) $\beta$ depends on the galaxy mass and evolutionary status; (ii) $\beta$ spans an ample range of values going from say -1000 to +1000, preference goes to the interval -20 to 20; (iii) The variation of $\beta$ among galaxies is mirrored by their position on the FP. In general  the position of galaxies with $\beta < 0$ is different from those with $\beta > 0$ (this is in general true at all redshifts); (iv) Furthermore, also the relation between the value of $\beta$ (sign and absolute value) and position on the FP seems to change with the redshift; (v) More precisely, typically  $\beta \simeq 0$  at high redshift (z=4) nearly all over the range of $I_e$, there is an important tail toward large positive beta for $\log I_e > 2$ at z=2; there is a growth of the positive tail and onset of  an important tail of large negative values for $\log I_e > 1.5$ at z=1; finally, we note an important tails of negative and   positive values of $\beta$ all over the range of $\log I_e$ from 0 to 3  at z=0. (vi) $\beta$ is strongly related to the star formation rate; (vii) Last, in a galaxy the value of $\beta$ expressed by its stellar content, mass, radius, etc. should also   change with time as indicated by simple galaxy models \citep{Donofrio_Chiosi_2022,Donofrio_Chiosi_2023a,Donofrio_Chiosi_2023b}; (viii) Finally, the specific intensity $I_e$ seems to drive the value of $\beta$. Despite all these hints, clear understanding of the rules determining the value of $\beta$  (and $L'_0$ in turn) is still missing. Work is progress to hit the target.   }  
      
      \end{enumerate}

In conclusion, large scale cosmological simulations of galaxy formation in hierarchical scheme indicate that galaxies  at any epoch are quite close to the mechanical equilibrium and distribute on a nearly planar surface in the FP-space. This surface is well approximated by a plane when the galaxy sample satisfy the condition of ``full  virialization'', that in general is achieved by the massive galaxies at the present epoch.

The variation of the FP-space with redshift and with the $\beta$ parameter gives to these scaling relations the role of cosmological tools. Indeed, the peculiar distributions of galaxies observed at any epoch in this space might now be linked with the physical mechanisms affecting galaxies in their evolution.

\begin{acknowledgements}
      We thank the anonymous referee for his suggestions {and remarks}.
\end{acknowledgements}

   \bibliographystyle{aa} 
   \bibliography{New_FP.bib} 
 
\begin{appendix} 

\section{The basic equations of the new theory}
\label{Appendix_A}

It may be useful to refresh here the formalism and key results reached by   \cite{Donofrioetal2017,Donofrioetal2019,Donofrioetal2020} and \cite{DonofrioChiosi2021,Donofrio_Chiosi_2022,Donofrio_Chiosi_2023a,Donofrio_Chiosi_2023b}. In the following we drop the explicit notation of time dependence for the sake of  simplicity. The two equations representing the VT and the \Lsigb\ law are:

\begin{eqnarray}
 \sigma^2 &= & \frac{G}{k_v} \frac{M_s}{R_e}   \\
 \sigma^\beta &= & \frac{L}{L'_0} = \frac{2\pi I_e R^2_e}{L'_0}. 
\label{eqsig}
\end{eqnarray}

The first equation is the Virial Theorem, where $G$ is the gravitational constant, and $k_v$ a term that gives the degree of structural and dynamical non-homology. The presence of $k_v$ allows us to write $M_s$ (stellar mass) instead of the total mass $M_T$. $k_v$ is  a function of the S\'ersic index $n$, that is $k_v=((73.32/(10.465+(n-0.94)^2))+0.954))$ \citep[see][for all details]{Bertinetal1992, Donofrioetal2008}. In the two equations, all other symbols have their usual meaning. 
In these equations, $\beta$ and $L'_0$ are time-dependent parameters that depend on the peculiar history of each object. 

\textsf{Slopes of some basic FP-projections}.
From these equations one can derive all the mutual relationships existing among the parameters $M_s$, $R_e$, $ L$, $I_e$, $\sigma$ characterizing a galaxy. We find:

\noindent
(i) The \IeRe\ plane:

\begin{equation}
I_e  = \Pi R_e^{\gamma}
\label{eqIeRe}
\end{equation}
\noindent
where 

$$\gamma=\frac{(2/\beta)-(1/2)}{(1/2)-(1/\beta)}$$

\noindent
and $\Pi$ is a factor that depends on $k_v$, $M/L$, $\beta$, and $L'_0$. It is given by

$$
\Pi  = \left [ \left (\frac{2\pi}{L'_0}\right )^{1/\beta} \left (\frac{L}{M_s} \right )^{(1/2)} \left (\frac{k_v}{2\pi G} \right )^{(1/2)} \right ]^{\frac{1}{1/2-1/\beta}} .
\label{eq4}
$$

\noindent
(ii) The \IeSig\ relation:

\begin{equation}
    I_e = \left [ \frac{G}{k_v}\frac{L'_0}{2\pi}M_s \Pi^{3/\gamma} \right ]^{\frac{\beta-2}{1+3/\gamma}} \sigma^{\frac{\beta-2}{1+3/\gamma}}  .
    \label{eqIeSig}
\end{equation}

\noindent
(iii) The \Rsigma\ relation:

\begin{equation}
    R_e = \left [ \frac{G}{k_v}\frac{L'_0}{2\pi}\frac{M_s}{\Pi} \right ] \sigma^{\frac{\beta-2}{3+\gamma}}
    \label{eqReSig}  .
\end{equation}

\noindent
(iv) The \MRa\ relation:

\begin{equation}
    R_e = \left [ (\frac{G}{k_v})^{\beta/2} \frac{L'_0}{2\pi} \frac{1}{\Pi} \right ]^{\frac{2(\beta-2)}{\beta^2-6\beta+12}} M_{s}^{\frac{\beta^2-2\beta}{\beta^2-6\beta+12}}  .
    \label{eqReM}
\end{equation}

It is important to note  that in all these relationships the slopes  depend only on $\beta$. This means that when a galaxy changes its luminosity $L$,  and velocity dispersion $\sigma$, and has a given value of  $\beta$ (either positive or negative), the effects of this change in the \Lsig\ plane are propagated in all other projections of the FP-space. In these planes the galaxies cannot move in whatever directions, but are forced to move only along the directions (slopes) predicted by the $\beta$ parameter in the above equations. In this sense the $\beta$ parameter is the link we are looking for between the FP and the observed distributions in the FP projections.

\noindent
\textsf{The individual FP of galaxies}. In addition, the combination of eqs. (\ref{eqsig}) gives us another important equation. It is now possible to write a FP-like equation valid for each galaxy and depending on the parameters $\beta$ and $L'_0$:

\begin{equation}
    \log R_e = a \log\sigma + b <\mu>_e + c
    \label{eqfege}
\end{equation}

\noindent
where  $<\mu_e>$ is the mean surface brightness $<I_e>$ expressed in magnitudes and the coefficients:

\begin{eqnarray}
a & = & (2+\beta)/3 \\ \nonumber
b & = & 0.26 \\ \nonumber
c & = & -10.0432+0.333*(-\log (G/k_v) - \log (M/L) \\ \nonumber
  &   & -2*\log (2\pi)-\log (L'_0)) \nonumber
\end{eqnarray}

\noindent
are written in terms of $\beta$ and $L'_0$. We note that this is the equation of a plane whose slope depends on $\beta$ and the zero-point on $L'_0$. The similarity with the FP equation is clear. The novelty is that the FP is an equation derived from the fit of a distribution of real objects, while here each galaxy independently follows an equation formally identical to the classical FP, but of profoundly different physical meaning. In this case, since $\beta$ and $L'_0$ are time dependent, the equation represents the instantaneous plane on which a generic galaxy is located in the FP-space and consequently in all its projections.

\noindent
\textsf{The solution for $\beta$ and $L'_0$}. Finally, the equation system (\ref{eqsig}) allows us  to determine the values of $\beta$ and $L'_0$, the two basic evolutionary parameters. Let us  write the above equations in the following way:

\begin{eqnarray}
\beta [\log(I_e)+\log(G/k_v)+\log(M_s/L)+\log(2\pi)+\log(R_e)] + \\ \nonumber
    + 2\log(L'_0) - 2\log(2\pi) - 4\log(R_e) = 0 \\ 
 \beta\log(\sigma) + \log(L'_0) + 2\log(\sigma) + \log(k_v/G) - \log(M_s) + \\ \nonumber
 - \log(2\pi) - \log(I_e) - \log(R_e) = 0. 
\label{eqbet}
\end{eqnarray}
\noindent

Posing now: 

\begin{eqnarray}
A  & = & \log(I_e)+\log(G/k_v)+\log(M_s/L)+\log(2\pi)+ \\ \nonumber
   &   & \log(R_e)  \\ \nonumber
B  & = & - 2\log(2\pi) - 4\log(R_e)  \\ \nonumber
A' & = &  \log(\sigma)  \\ \nonumber
B' & = & 2\log(\sigma) - \log(G/k_v) - \log(M_s) - \log(2\pi) -  \\ \nonumber 
   &   & \log(I_e) - \log(R_e)  
   \label{eq3}
\end{eqnarray}
we obtain the following system:

\begin{eqnarray}
A\beta + 2\log(L'_0) + B = 0 \\
A'\beta + \log(L'_0) + B'= 0
\label{eqsyst}
\end{eqnarray}
\noindent 
with solutions:

\begin{eqnarray}
 \beta & = & \frac{-2\log(L'_0) - B}{A} \\ 
 \log(L'_0) & = &\frac{A'B/A - B'}{1-2A'/A}.
 \label{eq2}
\end{eqnarray}

The key result is that the  parameters $L$, $M_s$, \re, \Ie\ and $\sigma$ of a galaxy fully determine its evolution in FP-space that is encoded in the parameters $\beta$ and $L'_0$.
Given this premise,  we proceed now to show the basic scale relationships at  increasing redshift, that is at decreasing time.  

\section{The errors of $\beta$}\label{Appendix_B}

 \subsection{Useful definitions} 
Suppose we have a sample on $N_{max}$ objects (galaxies), each of which is characterized by the following parameters: luminosity $L$ in some passbands and solar units, stellar mass $M_s$ in solar units, effective radius $R_e$ in kpc or pc, velocity dispersion $\sigma$ in km/s, and specific intensity $I_e = L/2\pi R_{e}^2$ (where  by convention $R_e$ is expressed in pc). 
 
First of all,  for the sake of convenience, all these parameters are expressed in logarithmic form and in order to simplify the notation they are replaced by  $M  = \log(M_s)$, $R  = \log(R_e)$, $ S  = \log(\sigma)$,  $L  = \log(L)$, $I  = \log(I_e)$. Finally, a generic object of the sample is indicated by the index $i=1,2,......N_{max}$.     
 
With the aid of these parameters, the VT yielding the velocity dispersion as a function of the mass $M_s$ and effective radius $R_e$, and the relation $L = L'_0 \sigma^\beta$, we set up a system of equations, the solution of which is represented by equation (\ref{eq2}) and (\ref{eq3}) providing the values of  $\beta$ and $L'_0$. It is obvious that $\beta$ and $L'_0$ vary from galaxy to galaxy, and for each galaxy with time. Furthermore, since the parameters $L$, $M_s$, and so forth,  are surely affected by some observational uncertainty, the natural question arises:  ``How  stable are $\beta$ and $L'_0$? ''. In addition to it, we considered the characterizing parameters as independent quantities. However this is not the case,  for instance the total luminosity depends on the mass, the determination of the effective radius $R_e$ depends on the luminosity (and mass) profile across the galaxy, the velocity dispersion $\sigma$, derived from the virial theorem, depends on the mass and radius, finally the specific intensity $I_e$ depends on the luminosity and the radius. 

\subsection{Simulations of the errors for the galaxies in the sample}

The $N_{max}$ galaxies in the sample have characterizing parameters $L$, $M_s$, $R_e$, $\sigma$, $I_e$ whose value has been measured only once and is surely affected by some uncertainty that is difficult to assess for each object. To cope with it and be able to apply the error propagation theory on  $\beta$ and $L'_0$ of each object, we simulate artificial errors for each quantity by artificially generating for each galaxies of the list a number of virtual objects whose parameters have been slightly varied with respect to the original ones by some random correction in this way mimicking different sets of measurements. The number of moke measurements is $N_{err}$, whose value has to be suitably fixed (say around 50). The range of uncertainty for each value is of the order of 10 to 20\% at maximum. Since we are using variables expressed in the logarithmic scale, uncertainties of 10 to 20\% correspond to uncertainties of 0.042 to 0.079 in the logarithms, indicated here by  $\theta $. Looking at the mass as an example (we used here the logarithmic variables in compact notation), the procedure is a follows:  given any mass $M_0$  this is perturbed  by $ \Delta M$,  $M = M_0 + \Delta M$ so that  

\begin{equation}
M = M_0  + \Delta M =  M_0 + (\mathcal{R} - 0.5) \theta 
\end{equation}

\noindent 
where 
$\mathcal{R}$ is a random number from 0 to 1, and $\theta$ the adopted maximum uncertainty. In this way the correction $\Delta M$ can be positive or negative and the total maximum range for $\Delta$ is $\pm \theta M_0$.  The same procedure is repeated for $R$ and $L$. Recalling that the velocity dispersion is derived from the VT and the specific luminosity $I_e \propto L/R_e^2$ (the variables are correlated each others), we can simply write

\begin{equation}
 \Delta S = (\Delta M - \Delta R)/2  \qquad {\rm and} \qquad  \Delta I = \Delta L - 2 \Delta R
\end{equation}

\noindent
With new perturbed values we calculate $\log L'_0$ and $\beta$. The same procedure  is repeated a number of times ($N_{err}$) for each galaxy of the sample and for all galaxies of the list. The resulting frequency (or percentage) distribution of the $\beta's$ was already shown in Fig.\ref{fig:14}.  

Another alternative formulation for the errors affecting the parameters is a gaussian distribution mimicking the mathematical behaviour of the point spread function (PSF) used to measure the luminosity profile of the source galaxy out of which the other variables $R_e$ and $M_s$ are derived.
To this aim we adopt  

\begin{equation}
 F(x) = \frac {1}{\sigma \sqrt{2\pi} } exp[- \frac{(x-\mu)^2}{2\sigma^2}]
\end{equation}

\noindent
where $F(x)$ is the normalized distribution relative to the expected ideal case $\mu$ and $\sigma$ the standard deviation (although the same symbol is used, it should not be mistaken with the velocity dispersion), and finally $x$ is the distance both positive or negative with respect to the position of the maximum of $F(x)$. In our case, the obvious choice for $\mu$ and $\sigma$ are $\mu=0$ and $\sigma \simeq \theta $.  In order to take the observational uncertainty into account, $\sigma \simeq 0.5$ or $0.1$. With these choices  $F(x) \simeq 0$ at $x=\pm 0.5$ (actually it is zero also at smaller values of $x$. Any contribution beyond these limits can be ignored for practical purposes.  Therefore, the range of interest is $-0.5 \leq x \leq 0.5$. The maximum $F(x)$ for $x=0$ is $F(0)= \frac {1}{\sigma \sqrt{2\pi} }$.  To determine $x$, we start from the relation

                          $$x = (\mathcal{R} - 0.5)$$

\noindent
where $\mathcal{R}$ is the random number in the interval 0 to 1,  derive $F(x)$,  use the assigned difference $F(0) - F(x)$ divided by $F(0)$ as the fraction at uncertainty with respect to its maximum value $\theta$, and finally multiply all this by $\theta$ to get the real uncertainty at $x$. Again taking the mass as an example, we alter its value (in logarithmic scale) as follows 

\begin{equation}
M = M_0  + \Delta M =  M_0 + \frac{x}{|x|} \times \left(\frac{(F(0) - F(x')}{F(0)}\right) \times \theta
\end{equation}

\noindent
the uncertainty increases going toward the edges of the interval permitted to $x$ and is zero at the center of the gaussian distribution.  The definition of $\Delta M$ is straightforward.  The same procedure is repeated for $R$ and $L$. Also here, uncertainties for the velocity dispersion sigma $\Delta S$ and the specific intensity $\Delta I$ are derived from $\Delta M$, $\Delta R$ and $\Delta L$. Finally, the whole procedure  is repeated a number of times ($N_{err})$ for each galaxy  of the sample and for all galaxies of the list.

\subsection{Elements of the error propagation theory}

Let start defining the following auxiliary quantities:

(1) $\mathcal{S}_L = \sum_i L_i$, $\mathcal{S}_M = \sum_i M_i$, $\mathcal{S}_R = \sum_i R_i$, $\mathcal{S}_\sigma = \sum_i \sigma_i$, and $\mathcal{S}_I = \sum_i I_i$, where $i=1,2,......N_{err}$; \\

(2) $\mathcal{SS}_L = \sum_i (L_i)^2$, $\mathcal{SS}_M = \sum_i (M_i)^2$, $\mathcal{SS}_R = \sum_i (R_i)^2$, $\mathcal{SS}_\sigma = \sum_i (\sigma_i)^2$, $\mathcal{SS}_I = \sum_i (I_i)^2$, where $i=1,2,......N_{err}$; \\

(3) $\mathcal{SSS}_{ML} = \sum_i (M_i\times L_i)$,   $\mathcal{SSS}_{MR} = \sum_i (M_i\times R_i)$,
    $\mathcal{SSS}_{MS} = \sum_i (M_i\times S_i)$,   $\mathcal{SSS}_{MI} = \sum_i (M_i\times I_i)$,
    $\mathcal{SSS}_{LR} = \sum_i (L_i\times R_i)$,   $\mathcal{SSS}_{LS} = \sum_i (L_i\times S_i)$,
    $\mathcal{SSS}_{LI} = \sum_i (L_i\times I_i)$,   $\mathcal{SSS}_{RS} = \sum_i (R_i\times S_i)$,
    $\mathcal{SSS}_{RI} = \sum_i (R_i\times I_i)$,   $\mathcal{SSS}_{SI} = \sum_i (S_i\times I_i)$, where $i=1,2,......N_{err}$; \\

(4) Averages:   $\mathcal{M}_{L} = \sum_i L_i/N_{err} = \mathcal{SS}_{L}/N_{err}$, 
                $\mathcal{M}_{M} = \sum_i M_i/N_{err} = \mathcal{SS}_{M}/N_{err}$,
                $\mathcal{M}_{R} = \sum_i R_i/N_{err} = \mathcal{SS}_{R}/N_{err}$,  
                $\mathcal{M}_{S} = \sum_i S_i/N_{err} = \mathcal{SS}_{S}/N_{err}$, and
                $\mathcal{M}_{I} = \sum_i M_i/N_{err} = \mathcal{SS}_{I}/N_{err}$, 
                where $i=1,2,......N_{err}$;  \\

 (5) Variances: $VA_{L} = \mathcal{SS}_{L}/N_{err} - \mathcal{M}_{L}$,
 $VA_{M} = \mathcal{SS}_{M}/N_{err} - \mathcal{M}_{M}$,
 $VA_{R} = \mathcal{SS}_{R}/N_{err} - \mathcal{M}_{R}$,
 $VA_{S} = \mathcal{SS}_{S}/N_{err} - \mathcal{M}_{S}$,  and
 $VA_{I} = \mathcal{SS}_{I}/N_{err} - \mathcal{M}_{I}$;   \\

 (6) Co-variances:  $CV_{ML} = \mathcal{SSS}_{ML}/N_{err} - \mathcal{M}_M \times \mathcal{M}_L$,
 $CV_{MR} = \mathcal{SSS}_{MR}/N_{err} - \mathcal{M}_M \times \mathcal{M}_R$,
 $CV_{MS} = \mathcal{SSS}_{MS}/N_{err} - \mathcal{M}_M \times \mathcal{M}_S$,
 $CV_{MI} = \mathcal{SSS}_{MI}/N_{err} - \mathcal{M}_M \times \mathcal{M}_I$,
 $CV_{LR} = \mathcal{SSS}_{LR}/N_{err} - \mathcal{M}_L \times \mathcal{M}_R$,
 $CV_{LS} = \mathcal{SSS}_{LS}/N_{err} - \mathcal{M}_L \times \mathcal{M}_S$,
 $CV_{LI} = \mathcal{SSS}_{LI}/N_{err} - \mathcal{M}_L \times \mathcal{M}_I$,
 $CV_{RS} = \mathcal{SSS}_{RS}/N_{err} - \mathcal{M}_R \times \mathcal{M}_S$,
 $CV_{RI} = \mathcal{SSS}_{RI}/N_{err} - \mathcal{M}_R \times \mathcal{M}_I$,
 $CV_{SI} = \mathcal{SSS}_{SI}/N_{err} - \mathcal{M}_S \times \mathcal{M}_I$.
 
\subsection{Derivatives of $\beta$ and $\log(L'_0)$ with respect to  the parameters $M_s$,  $L$, $I_e$, $R_e$,  and $\sigma$ }

We start from relations (\ref{eq2}) and  (\ref{eq3}), however recast in the compact notation keeping the same units, and in which we have introduced  $c_a = log(G/k_v)$, $c_b = \log(2\pi)$, and $c_d = c_a + c_b$. For the purposes of this analysis we keep constant the structural parameter $k_v$. Relations  (\ref{eq2}) and  (\ref{eq3}) become
	   
\begin{eqnarray}	   
\log( L'_0) &=& 
            [ (-4 R - 2 c_b) S -(I + M - L + R + c_d)\times  \nonumber \\
           && (2 S - M - I - R - c_d) ] / \nonumber \\
           && [ (I  + M - L + R + c_d) - 2 S ] 
           \label{cast_lo_1}
\end{eqnarray}

\begin{eqnarray}
 \beta &=& [1/(I + M - L + R + c_d)]\times  \nonumber \\
       &&  \{ [ 2 (4 R + 2 c_b) S + 2 (I + M - L + R + c_d) \times \nonumber \\
       &&      (2 S - M - I - R - c_d) ]/       \nonumber \\
       &&  [ (I + M - L + R + c_d) - 2S ] + (2 c_b + 4 R)  \}  
       \label{cast_beta_1}
\end{eqnarray} 
 
\noindent 
 Secondly, we calculate the partial derivatives of $\log( L'_0)$ and $\beta$ with respect to 
  $\log(M_s)$, $\log(R_e)$, $\log(\sigma)$, $\log(L)$,  and $\log(I_e)$, using the same compact notation. The derivatives are

  \begin{eqnarray}
\frac{\partial \log L'_0} { \partial M} &=&   (2 c_b S + c_d^2 + 2 c_d (M + R - 2 S - L + I) + \nonumber \\
    &&     M^2 + 2 M (R - 2 S - L + I) + R^2 - 2 R L + 2 R I + \nonumber \\
    &&     4 S^2 + 2 S L - 4 S I + L^2 - 2 L I + I^2)/  \nonumber \\
    &&     ( c_d + M + R - 2 S - L + I)^2
    \label{part_lo_1}
 \end{eqnarray}

  \begin{eqnarray}
\frac{\partial \log L'_0} {\partial R} &=& (2 c_b S + c_d^2 + 2 c_d (M + R - 4 S - L + I) + \nonumber \\
     &&     M^2 + 2 M (R - 4 S - L + I) + R^2 - 4 R S - 2 R L +           \nonumber \\
     &&     2 R I + 12 S^2 + 6 S L - 8 S I + L^2 - 2 L I + I^2)/          \nonumber \\
     &&     (c_d + M + R - 2 S - L + I)^2
     \label{part_lo_2}
\end{eqnarray}
        
\begin{eqnarray}        
\frac{\partial \log L'_0} {\partial S} &=& -(2 (c_b + 2 R - L) (c_d + M + R - L + I))/  \nonumber \\
     &&    (c_d + M + R - 2 S - L + I)^2
     \label{part_lo_3}
\end{eqnarray}	  

\begin{eqnarray}
\frac{\partial \log L'_0} {\partial L} &=& (2 S (-c_b + c_d + M - R - 2 S + I))/         \nonumber \\
     &&       (c_d + M + R - 2 S - L + I)^2
     \label{part_lo_4}
\end{eqnarray}

\begin{eqnarray}
\frac{\partial \log L'_0} {\partial I} &=& (2 c_b S + c_d^2 + 2 c_d (M + R - 2 S - L + I) + \nonumber \\
     &&    M^2 + 2 M (R - 2 S - L + I) + R^2 - 2 R L + 2 R I +      \nonumber \\
     &&    4 S^2 + 2 S L - 4 S I + L^2 - 2 L I + I^2)/              \nonumber \\
     &&    (c_d + M + R - 2 S - L + I)^2
     \label{part_lo_5}
\end{eqnarray}

and

\begin{eqnarray}
\frac{\partial \beta} {\partial M} &=&  -(2 (c_b + 2 R - L))/(c_d + M + R - 2 S - L + I)^2  
\label{part_beta_1}
\end{eqnarray}

\begin{eqnarray}
\frac{\partial \beta} {\partial R} &=& -(2 (c_b - 2 c_d - 2 M + 4 S + L - 2 I))/   \nonumber \\
     &&     (c_d + M + R - 2 S - L + I)^2
     \label{part_beta_2}
\end{eqnarray}

\begin{eqnarray}
\frac{\partial \beta} {\partial S} &=& (4 (c_b + 2 R - L))/(c_d + M + R - 2 S - L + I)^2
\label{part_beta_3}
\end{eqnarray}
	
\begin{eqnarray}
\frac{\partial \beta} {\partial L} &=& -(2 (-c_b + c_d + M - R - 2 S + I))/   \nonumber  \\
     &&      (c_d + M + R - 2 S - L + I)^2 
     \label{part_beta_4}
\end{eqnarray}

\begin{eqnarray}
\frac{\partial \beta} {\partial I} &=&  -(2 (c_b + 2 R - L))/(c_d + M + R - 2 S - L + I)^2
\label{part_beta_5}
\end{eqnarray}

Although it is not indicated in the notation, each partial derivative with respect to some variable  is calculated keeping constant all other variables. The total variations  of $\log L'_0$ and $\beta$ are expressed by 

\begin{eqnarray}
   \Delta \log L'_0 &=& (\frac{\partial \log L'_0} {\partial M}) \Delta M + 
             (\frac{\partial \log L'_0} {\partial R}) \Delta R + 
             (\frac{\partial \log L'_0} {\partial S}) \Delta S +   \nonumber \\
         &&    (\frac{\partial \log L'_0} {\partial L}) \Delta L + 
             (\frac{\partial \log L'_0} {\partial I})  \Delta I
             \label{diff_lo}
\end{eqnarray}

\begin{eqnarray}
       \Delta \beta &=& (\frac{\partial \beta} {\partial M}) \Delta M + 
             (\frac{\partial \beta} {\partial R}) \Delta R + 
             (\frac{\partial \beta} {\partial S}) \Delta S +   \nonumber  \\
         &&    (\frac{\partial \beta} {\partial L}) \Delta L + 
             (\frac{\partial \beta} {\partial I})  \Delta I
             \label{diff_beta}
\end{eqnarray}

These relations are derived for each object of the sample ($i=1,2,.....N_{max}$), and for each object $i$ for all perturbations of its parameter set to simulate the uncertainty ($j=1,2,.... N_{err}$),

Particular care is paid to evaluate the variations $\Delta M$, $\Delta L$, $\Delta R$, $\Delta S$, and $\Delta $ to be inserted in relations (\ref{diff_lo} and (\ref{diff_beta}).
Since they all are written in logarithmic scale, their absolute value  cannot be larger than the uncertain $\theta=0.04$. In reality it can be significantly smaller than this, depending on the method used to perturb the original values, { either uniform-random or gaussian-random distributions}.

\subsection{Variance and co-variance of $\log L'_0$ and $\beta$}

If for the same galaxy we had many measurements of their characterizing parameters and an estimate of their errors, we could apply the technique of variances and covariances for correlated variables and  obtain an estimate of the propagation of errors on $L'_0$ and $\beta$, the target quantities of our analyses. To this aim we introduced the simulations of errors affecting the parameters (M, L, R, S, I, compact notation is used) and derived the associated variances and co-variances for each characterizing parameter of  each object in the sample. Thanks to it variance and co-variances of  $\log L'_0$ and $\beta$ of each galaxy and its simulations can be calculated. They are

\begin{eqnarray} 
       (VA_{\log L'_0})^2 &=&  \sum_J (\frac{\partial \log L'_0} { \partial J})^2 VA_J +  
         \nonumber \\
     &&    +2 \sum_{J,K}(\frac{\partial \log L'_0} { \partial J})(\frac{\partial \log L'_0} { \partial K}) CV_{J,K} 
     \label{cov_lo}
\end{eqnarray}

\begin{eqnarray} 
       (VA_{\beta})^2 &=&  \sum_J (\frac{\partial \beta} { \partial J})^2 VA_J +  
         \nonumber \\
     &&    +2 \sum_{J,K}(\frac{\partial \beta} { \partial J})(\frac{\partial \beta} { \partial K}) CV_{J,K} 
     \label{cov_beta}
\end{eqnarray}

\noindent
where $J=M,L,R,S,I$ and $K=M,L,R,S,I$, $J \neq K$, all terms with $J,K$ and $K,J$ are equal and must counted only once. The terms  
$VA_M$ , $VA_L$, $VA_R$, $VA_S$, $VA_I$ are the variances,       and $CV_{ML}$,  $CV_{MR}$, $CV_{MS}$,  $CV_{MI}$, $CV_{LR}$, $CV_{LS}$, $CV_{LI}$, $CV_{RS}$, $CV_{RI}$, $CV_{SI}$  the co-variances. 
Finally, 
\begin{equation}
 (VA_{\log L'_0}) = \sqrt{(VA_{\log L'_0})^2} \qquad {\rm and} \qquad
          (VA_{\beta}) = \sqrt{(VA_{\beta})^2}
          \label{cov_lo_beta}
\end{equation}

Lastly, the percentage change  $\Delta \Pi$ of $\beta$ and $\log L'_o$ are given by

\begin{eqnarray}
	    \Delta \Pi_{\log L'_0} &=& \frac{((abs(\Delta \log L'_0 + \log L'_0)-abs(\log L'0))} {(abs(\log_L'0)}100  \nonumber \\
	    \Delta \Pi_{\beta}  &=& \frac{((abs({\Delta \beta} + \beta)-abs(\beta))} {(abs(\beta))}100
	    \label{percent_lo_beta}
\end{eqnarray}
	    
\subsection{Summary of the whole procedure}

Given a galaxy with characterizing parameters,  $M  = \log(M_s)$, $R  = \log(R_e)$, $ S  = \log(\sigma)$,  $L  = \log(L)$, $I  = \log(I_e)$, we  calculate 20 sub-cases whose characterizing parameters are artificially changed by random amounts always smaller than the estimated maximum uncertainty of 20\% or 0.08 in logarithmic scale, that is $\pm 0.04$ the unperturbed values. The generic parameters are here indicated by $X$ where $X$ stands for  $M$, $R$, $S$, $L$,  and $I$. For these moked objects we calculate the quantities: mean values 
    $\mathcal{M}_{X} =  \mathcal{SS}_{X}/N_{err}$,
    variances
    $VA_{X} = \mathcal{SS}_{X}/N_{err} - \mathcal{M}_{X}$,    
    and covariances
    $CV_{X X'} = \mathcal{SSS}_{X X'}/N_{err} - \mathcal{M}_X \times \mathcal{M}_X'$,
\end{appendix}
where $X'$ has the same meaning of $X'$, the combinations  $X \neq X'$ are retained, and the combinations $X X' = X' X$ are counted only once. Then we derived for the original and moked objects the formal solutions of the system of equations giving  $\log L'_0$ and $\beta$ and their partial derivatives, eqns. (\ref{part_lo_1}) through eqn.(\ref{part_lo_5}) and eqns.(\ref{part_beta_1}) through (\ref{part_beta_5}). With the aid of these, we derive the uncertainties affecting $\log L'_0$ and $\beta$  and the propagation of uncertainties (errors in the estimates) for correlated variables. Finally, we derive the variances and co-variances of the solutions $\log L'_0$ and $\beta$, eqns (\ref{cov_lo_beta}). Lastly,  we calculate the percent variations of $\log L'_0$ and $\beta$. The results were presented and commented in Sect. \ref{sec:8}.
 
\end{document}